\documentclass[reprint,amsmath,amssymb,aps,longbibliography]{revtex4-1}
\usepackage{graphicx}
\usepackage{float}
\usepackage{bibunits}
\usepackage{derivative}
\usepackage{xcolor}
\usepackage[margin=0.64in]{geometry}
\usepackage{siunitx}
\usepackage{soul}
\usepackage{minitoc}

\usepackage{tocloft}
\definecolor{dark-red}{rgb}{0.4,0.15,0.15}
\definecolor{dark-blue}{rgb}{0.15,0.15,0.4}
\definecolor{medium-blue}{rgb}{0,0,0.5}
\usepackage[hypertexnames=false]{hyperref}
\usepackage[all]{hypcap}
\hypersetup{
    colorlinks, linkcolor={dark-red},
    citecolor={dark-blue}, urlcolor={medium-blue}
}

\newcommand{\sket}[1]{{\ensuremath{\lvert#1\rangle}}}
\newcommand{\lket}[1]{{\ensuremath{\left\lvert#1\right\rangle}}}
\newcommand{\tr}[1]{{\text{Tr}}}
\newcommand{\ket}[1]{\if@display\lket{#1}\else\sket{#1}\fi}
\newcommand{\sbra}[1]{{\ensuremath{\langle#1\rvert}}}
\newcommand{\lbra}[1]{{\ensuremath{\left\langle#1\right\rvert}}}
\newcommand{\bra}[1]{\if@display\lbra{#1}\else\sbra{#1}\fi}
\newcommand{\sbraket}[2]{{\ensuremath{\langle#1\rvert#2\rangle}}}
\newcommand{\lbraket}[2]{{\ensuremath{\left\langle#1\!\left\rvert\vphantom{#1}#2\right.\!\right\rangle}}}
\newcommand{\braket}[2]{\if@display\lbraket{#1}{#2}\else\sbraket{#1}{#2}\fi}

\newcommand{\sketbra}[2]{{\ensuremath{\lvert #1\rangle\!\langle #2\rvert}}}
\newcommand{\lketbra}[2]{{\ensuremath{\left\lvert #1\right\rangle\!\!\left\langle #2\right\rvert}}}
\newcommand{\ketbra}[2]{\if@display\lketbra{#1}{#2}\else\sketbra{#1}{#2}\fi}

\newcommand{\app}{Appendix}

\newcommand{\Uc}{U_\mathrm{cycle}}
\newcommand{\Ucd}{U_\mathrm{cycle}^{d}}

\newcommand{\tc}{t_\mathrm{cycle}}
\newcommand{\Uf}{{\mathfrak U}}
\newcommand{\Ufd}{{\mathfrak U}^d}
\newcommand{\Rf}{\mathfrak R}
\newcommand{\for}{\text{for }}

\defaultbibliography{QM_Refs}
\begin{document}
\doparttoc 
\faketableofcontents 
\part{} 

\title{Accurately computing electronic properties of a quantum ring}

\author{C. Neill\textsuperscript{1}}
\author{T. McCourt\textsuperscript{1}}
\author{X. Mi\textsuperscript{1}}
\author{Z. Jiang\textsuperscript{1}}
\author{M. Y. Niu\textsuperscript{1}}
\author{W. Mruczkiewicz\textsuperscript{1}}
\author{I. Aleiner\textsuperscript{1}}
\author{F. Arute\textsuperscript{1}}
\author{K. Arya\textsuperscript{1}}
\author{J. Atalaya\textsuperscript{1}}
\author{R. Babbush\textsuperscript{1}}
\author{J. C.~Bardin\textsuperscript{1,2}}
\author{R. Barends\textsuperscript{1}}
\author{A. Bengtsson\textsuperscript{1}}
\author{A. Bourassa\textsuperscript{1,5}}
\author{M. Broughton\textsuperscript{1}}
\author{B. B.~Buckley\textsuperscript{1}}
\author{D. A.~Buell\textsuperscript{1}}
\author{B. Burkett\textsuperscript{1}}
\author{N. Bushnell\textsuperscript{1}}
\author{J. Campero\textsuperscript{1}}
\author{Z. Chen\textsuperscript{1}}
\author{B. Chiaro\textsuperscript{1}}
\author{R. Collins\textsuperscript{1}}
\author{W. Courtney\textsuperscript{1}}
\author{S. Demura\textsuperscript{1}}
\author{A. R. Derk\textsuperscript{1}}
\author{A. Dunsworth\textsuperscript{1}}
\author{D. Eppens\textsuperscript{1}} 
\author{C. Erickson\textsuperscript{1}}
\author{E. Farhi\textsuperscript{1}}
\author{A. G.~Fowler\textsuperscript{1}}
\author{B. Foxen\textsuperscript{1}}
\author{C. Gidney\textsuperscript{1}}
\author{M. Giustina\textsuperscript{1}}
\author{J. A.~Gross\textsuperscript{1}}
\author{M. P.~Harrigan\textsuperscript{1}}
\author{S. D.~Harrington\textsuperscript{1}}
\author{J. Hilton\textsuperscript{1}}
\author{A. Ho\textsuperscript{1}}
\author{S. Hong\textsuperscript{1}}
\author{T. Huang\textsuperscript{1}}
\author{W. J. Huggins\textsuperscript{1}}
\author{S. V.~Isakov\textsuperscript{1}}
\author{M. Jacob-Mitos\textsuperscript{1}}  
\author{E. Jeffrey\textsuperscript{1}}
\author{C. Jones\textsuperscript{1}}
\author{D. Kafri\textsuperscript{1}}
\author{K. Kechedzhi\textsuperscript{1}}
\author{J. Kelly\textsuperscript{1}}
\author{S. Kim\textsuperscript{1}}
\author{P. V.~Klimov\textsuperscript{1}}
\author{A. N.~Korotkov\textsuperscript{1,4}}
\author{F. Kostritsa\textsuperscript{1}}
\author{D. Landhuis\textsuperscript{1}}
\author{P. Laptev\textsuperscript{1}}
\author{E. Lucero\textsuperscript{1}}
\author{O. Martin\textsuperscript{1}}
\author{J. R.~McClean\textsuperscript{1}}
\author{M. McEwen\textsuperscript{1,3}}
\author{A. Megrant\textsuperscript{1}}
\author{K. C.~Miao\textsuperscript{1}}
\author{M. Mohseni\textsuperscript{1}}
\author{J. Mutus\textsuperscript{1}}
\author{O. Naaman\textsuperscript{1}}
\author{M. Neeley\textsuperscript{1}}
\author{M. Newman\textsuperscript{1}}
\author{T. E.~O'Brien\textsuperscript{1}}
\author{A. Opremcak\textsuperscript{1}}
\author{E. Ostby\textsuperscript{1}}
\author{B. Pató\textsuperscript{1}}
\author{A. Petukhov\textsuperscript{1}}
\author{C. Quintana\textsuperscript{1}}
\author{N. Redd\textsuperscript{1}}
\author{N. C.~Rubin\textsuperscript{1}}
\author{D. Sank\textsuperscript{1}}
\author{K. J.~Satzinger\textsuperscript{1}}
\author{V. Shvarts\textsuperscript{1}}
\author{D. Strain\textsuperscript{1}}
\author{M. Szalay\textsuperscript{1}}
\author{M. D.~Trevithick\textsuperscript{1}}
\author{B. Villalonga\textsuperscript{1}}
\author{T. C. White\textsuperscript{1}}
\author{Z. Yao\textsuperscript{1}}
\author{P. Yeh\textsuperscript{1}}
\author{A. Zalcman\textsuperscript{1}}
\author{H. Neven\textsuperscript{1}}
\author{S. Boixo\textsuperscript{1}}
\author{L.~B.~Ioffe\textsuperscript{1}}
\author{P. Roushan\textsuperscript{1}}
\author{Y. Chen\textsuperscript{1}}
\author{V. Smelyanskiy\textsuperscript{1}}

\address{\textsuperscript{1}Google Quantum AI, Mountain View, CA}
\address{\textsuperscript{2}Department of Electrical and Computer Engineering, University of Massachusetts, Amherst, MA}
\address{\textsuperscript{3}Department of Physics, University of California, Santa Barbara, CA}
\address{\textsuperscript{4}Department of Electrical and Computer Engineering, University of California, Riverside, CA}
\address{\textsuperscript{5}Pritzker School of Molecular Engineering, University of Chicago, Chicago, IL}

\date{\today}

\email[Corresponding author (V.~Smelyanskiy):]{smelyan@google.com}
\email[\\Corresponding author (P.~Roushan):]{pedramr@google.com}
\email[\\Corresponding author (Y.~Chen):]{bryanchen@google.com}

\begin{abstract}
A promising approach to study condensed-matter systems is to simulate them on an engineered quantum platform \cite{feynman1982simulating,cirac1995quantum, bloch2008many, georgescu2014quantum}. However, achieving the accuracy needed to outperform classical methods has been an outstanding challenge. Here, using eighteen superconducting qubits, we provide an experimental blueprint for an accurate condensed-matter simulator and demonstrate how to probe fundamental electronic properties. We benchmark the underlying method by reconstructing the single-particle band-structure of a one-dimensional wire.  We demonstrate nearly complete mitigation of decoherence and readout errors and arrive at an accuracy in measuring energy eigenvalues of this wire with an error of \SI{\sim 0.01}{\radian}, whereas typical energy scales are of order \SI{1}{\radian}.  Insight into this unprecedented algorithm fidelity is gained by highlighting robust properties of a Fourier transform, including the ability to resolve eigenenergies with a statistical uncertainty of \SI[product-units = single]{e-4}{\radian}.  Furthermore, we synthesize magnetic flux and disordered local potentials, two key tenets of a condensed-matter system.  When sweeping the magnetic flux, we observe avoided level crossings in the spectrum, a detailed fingerprint of the spatial distribution of local disorder. Combining these methods, we reconstruct electronic properties of the eigenstates where we observe persistent currents and a strong suppression of conductance with added disorder. Our work describes an accurate method for quantum simulation  \cite{polkovnikov83nonequilibrium, carusotto2020photonic} and paves the way to study novel quantum materials with superconducting qubits.
\end{abstract}
\maketitle

In condensed-matter systems, the interplay of symmetries, interactions and local fields give rise to intriguing many-body phases.  Insight into these phases of matter comes from both experimental and theoretical developments; however, limitations in both approaches prevent a complete physical picture from emerging \cite{Qin2010, Jiang1424}.  For example, despite enormous effort, it is still not clear which state is realized at the 5/2 filling of fractional quantum Hall and which interaction one would need to generate the desired state \cite{willett1987observation, dolev2008observation, willett2019interference}.   Generally, the difficulty arises from the fact that interesting properties of quantum materials arise from subtle interference effects of many particles and small errors can lead to large deviations in observables.  Neither numerical methods nor analytics have sufficient accuracy to predict such phenomena in realistic systems. While conventional experiments provide the most direct approach, the necessary observables, such as correlated measurements, are typically inaccessible and the lack of controllability limits the benefits of such experiments.  

To outperform conventional approaches, quantum processors need to overcome two main sources of error: errors from control (unitary) and decoherence (non-unitary).  Here, we demonstrate an experimental blueprint for achieving low control error and comprehensive mitigation of decoherence.  The key insight into this development stems from robust properties of the Fourier transform.  Consider a quantum signal that oscillates in time with an envelope that decays due to decoherence.  Taking a Fourier transform of the signal will yield peaks at the oscillation frequencies.  While decoherence (as well as readout errors) will affect the amplitude and width of the peaks, the center frequencies will remain unaffected, see Appendix \ref{sec:open_sys-ring}.  On the other hand, small errors in the control parameters will manifest as shifts in the frequency of the peaks, providing a reliable signature from which we can learn these errors. The essence of our work is that studying quantum signals in the Fourier domain enables error mitigation and provides a sensitive probe of control parameters.  

We apply this insight at both the level of individual pairs for calibration and at the system level for mitigating decoherence in algorithms.  At the level of two qubits, gates can be applied periodically and local observables can be measured as a function of circuit depth.  Small errors in the control parameters are inferred from shifts in the Fourier peaks;  these errors are then corrected for.  In addition, we show that these parameters can be inferred with a remarkable statistical precision below \SI[product-units = single]{e-5}{\radian}, see Appendix \ref{sec:two_q_floquet}.  At the system level,  a similar strategy can be used, where we apply a multi-qubit unitary periodically and monitor local observables.   Here, we focus on a simple exactly-solvable model where we demonstrate an 18-qubit algorithm consisting of over 1,400 two-qubit gates with a total error in the extracted Fourier frequencies (corresponding to energy eigenvalues) of \SI{0.01}{\radian} and a statistical precision of \SI[product-units = single]{e-4}{\radian}. The 18-qubit ring formed in this experiment can be viewed as an Aharonov-Bohm interferometer. The phases associated with single qubit operations are analogous to disorder in a wire that causes a particle to scatter. The sum of these phases realizes an Aharonov-Bohm flux, which lifts the degeneracy between clockwise and counterclockwise propagating particles. In this analogy, precision stems from the sensitivity of wave interference patterns to imperfections. The underlying physics discussed in this work is general and can be adopted by other platforms.

\begin{figure}
\label{fig:engineering_flux}
\includegraphics[width=87mm]{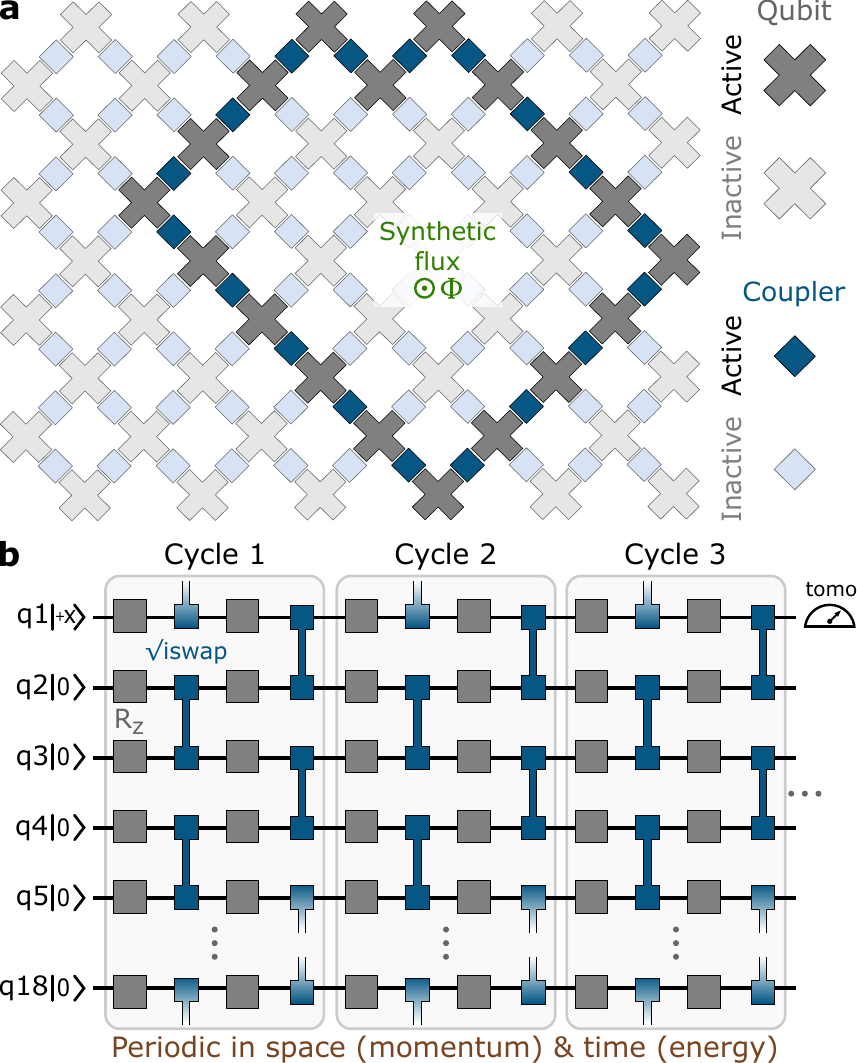}
\caption{\textbf{Engineering a 1D system with energy, momentum \& flux.  a}  Schematic of the 54-qubit processor.  Qubits are shown as gray crosses and tunable couplers as blue squares.  Eighteen of the qubits are chosen to form a one-dimensional ring.  Connecting the qubits in a ring allows us to introduce a controllable synthetic flux using single-qubit gates.  \textbf{b}  Schematic showing the control sequence used in these experiments. Each large vertical gray box indicates a cycle of evolution which we repeat many times.  Each cycle contains two sequential layers of $\surd{\text{i}}\text{SWAP}$ gates (blue), separated by single-qubit z-rotations (gray).  Periodicity in space leads to eigenstates of the cycle unitary with definite momentum.   Periodicity in time introduces conservation of energy.  Together, this realizes a digital circuit with well-defined physical properties such as energy spectrum, momentum, and flux.}
\end{figure}

\begin{figure*}
\includegraphics[width=178mm, height=60mm]{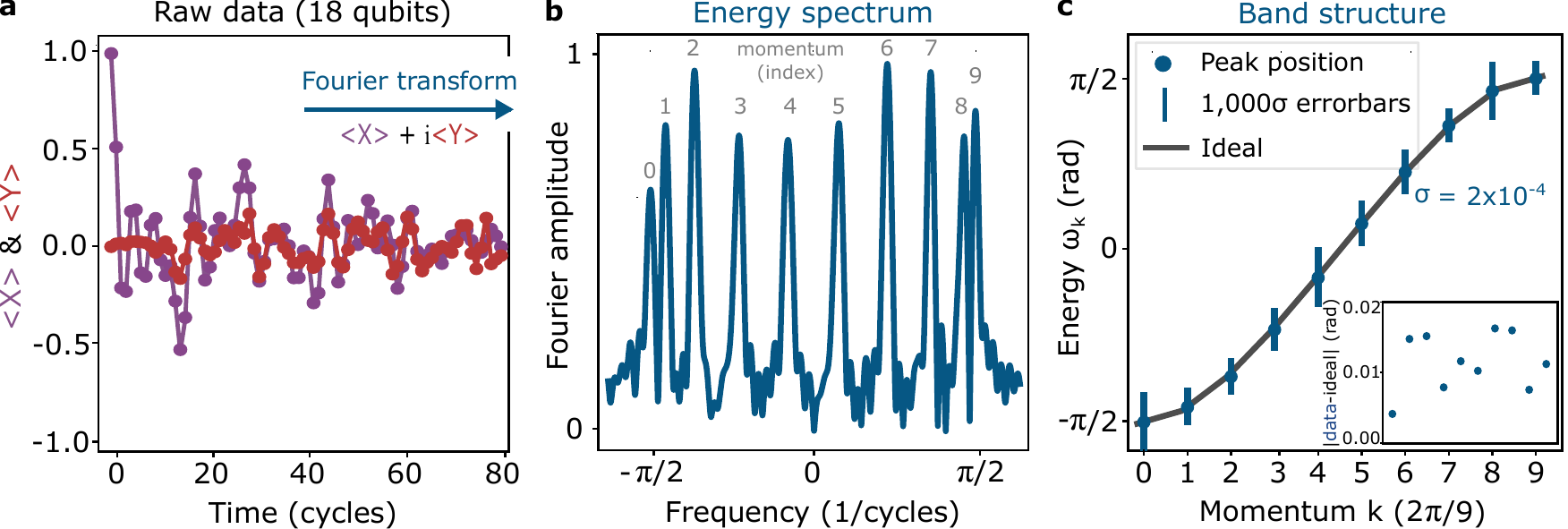}
\caption{\textbf{Measuring the single-particle band-structure.  a}  Example of typical raw-data from an 18-qubit experiment.  One qubit is initialized into the state $\left|+x\right>$, a variable number of cycles in applied, and the same qubit is measured in the $x$ and $y$ basis.  Expectation values are estimated from 100,000 repetitions of the experiment.  \textbf{b}  The Fourier transform of $\langle \text{X} \rangle + i\langle \text{Y} \rangle$ is plotted versus Fourier frequency.  Peaks in the spectrum correspond to eigenvalues of the cycle unitary.  Each eigenvalue corresponds to an eigenstate with definite momentum; the corresponding momentum can be determined by the order of the peaks.  \textbf{c}  The frequency at which each peak occurs (energy) is plotted versus the peak index (momentum) in order to recover the band structure.  Statistical error bars correspond to 1,000 standard-deviations.  The ideal single-particle band-structure is shown as a gray line and is given by Eq. \ref{eq:eigenvalues}.  The difference between the measured and ideal curves are shown inset.  This constitutes a well-defined computational problem at 18 qubits, requiring over 1,400 two-qubit gates, with a total algorithm error around \SI{0.01}{\radian}.}
\end{figure*}

We demonstrate our method using superconducting qubits with adjustable couplers as they enable control over individual frequencies, which set local fields and magnetic flux; and couplings, which set kinetic energy or hopping\,\cite{chen2014qubit, neill2017path}. A schematic of our 54-qubit processor is shown in Fig. 1a~\cite{Arute2019}.  The qubits are depicted as gray crosses and the tunable couplers as blue squares.  Eighteen of the qubits are isolated from the rest in order to form a ring.  This geometry is chosen to realize an artificial one-dimensional wire whose electrical properties we can study \cite{giamarchi2003quantum}.  While we focus on  ring geometry for simplicity, our results are sufficiently general to apply to more complex connectivities, such as a two-dimensional lattice.  

The gate sequence used in this work is shown in Fig. 1b.  Each large vertical gray box indicates a single cycle of evolution which we repeat periodically in time.  Each cycle contains two sequential layers of $\surd{\text{i}}\text{SWAP}$ gates (blue), separated by single-qubit z-rotations (gray).  Within each cycle, a two-qubit gate is applied between all possible pairs in the loop.  The $\surd{\text{i}}\text{SWAP}$ gates cause a particle (microwave excitation in this case) to hop between adjacent lattice sites (qubits).  The z-rotations are used to generate local fields and their their summation gives rise to an effective magnetic flux that threads the qubit loop~\cite{jotzu_experimental_2014, manovitz_quantum_2020}.  Here, we will focus on the dynamics of a single particle; however, our approach allows for a straightforward generalization to full many-body systems.  

The connectivity and gate sequence are chosen such that the algorithm is translationally invariant in space, resulting in a cycle unitary whose eigenstates have well-defined momentum.  Because the control sequence is periodic in time, the cycle unitary will have well-defined energies (known as quasi-energies).  
This allows us to realize a tight-binding Hamiltonian with terms of the form $(\sigma^+_m\sigma^-_{m+1}+\sigma^-_m\sigma^+_{m+1})$, 
where $\sigma^+$ ($\sigma^-$) are raising (lowering) operators that cause an excitation to propagate along the ring.
See \app~\ref{sub_sec:small_disorder} for details of the model and see \app~\ref{sec:open_sys-ring} for the embedding into qubits.  The eigenstates $\psi_k$ of this model are plane-waves and eigenvalues $\omega_k$ can be expressed in terms of the momentum $k$
\begin{gather}
\psi_k(x) \propto e^{ikx}, \,\,\,\,\,\, \cos(\omega_k) = \sin^2(k/2 - \Phi) 
\label{eq:eigenvalues}
\end{gather}
where $x$ is the position along the ring. Combined with the ability to introduce a synthetic magnetic field using z-rotations, we realize a digital quantum circuit with robust physical properties of momentum, energy and flux.

We probe the eigenspectrum of this 18-qubit ring using a many-body spectroscopy technique \cite{roushan2017spectroscopic}. Peaks in a spectroscopy experiment provide a robust signature of the underlying quantum system.  The raw data is shown in Fig. 2a where we plot the expectation values of the Pauli-$x$ and Pauli-$y$ operators on a single qubit (denoted $\langle \text{X} \rangle$ and $\langle \text{Y} \rangle$) as a function of the number of cycles in the control sequence.  While the raw data does not contain particularly intuitive features,  the complex Fourier transform of $\langle \text{X} \rangle + i\langle \text{Y} \rangle$ has the special property that peaks show up only at frequencies corresponding to the energy eigenvalues.  The Fourier transform of the time-domain data is shown in Fig. 2b where we observe clear, well-resolved peaks.

In the absence of local fields (z-rotations), the dynamics are governed entirely by the kinetic energy (or hopping), and a simple plane-wave model describes the spectrum.  This allows us to associate with each peak a corresponding value of momentum by simply noting the index of the peak, starting from 0. The momentum has units of $4\pi/N_q$, where $2\pi/N_q$ corresponds to the lattice spacing in a typical condensed matter setting and the extra factor of 2 comes the discrete evolution using gates.  
In Fig. 2c we show the measured energy as a function of the inferred momentum, realizing an experimental technique for extracting the single-particle band-structure.  The energies are inferred by fitting the data to the expression
\begin{equation} \label{eq:fitting_model}
\langle \text{X} \rangle + i \langle \text{Y} \rangle = e^{-\Gamma d}\sum_k A_k e^{-i \omega_k d}
\end{equation}
where $d$ are the measured circuit depths, $\Gamma$ is a damping rate, and $A_k$ are Fourier amplitudes. This expression is derived in  \app~\ref{sec:uniformG}.  The difference between ideal eigenvalues (given by Eq. \ref{eq:eigenvalues}) and the measured eigenvalues is shown in the inset with a typical value of around \SI{0.01}{\radian}, an unprecedented level of accuracy for an 18-qubit experiment with over 1,400 two-qubit gates.   

Extracting information from the Fourier domain has other salient features that were crucial in arriving at our results.  At large circuit depths, decoherence causes the signal to fall below the noise level of the experiment.  Maintaining a high signal-to-noise ratio is therefore key to scalable error mitigation.  Fourier transforms have the important property that the statistical uncertainty scales inversely with the length of the time-domain signal.
Consider a shallow circuit of depth $D$ where coherence can be neglected.  In this limit, the statistical uncertainty $\sigma$ scales as $1/\sqrt{N} \times 1/D$, where $N$ is the number of measurements. In the supplementary we provide the explicit relation and show that this physics is general and is not affected by the damping rate $\Gamma$.
The factor $1/D$ is expected when taking a Fourier transform and the factor $1/\sqrt{N}$ is the standard expression for finite-sampling noise.  Here, we benefit from both factors because we fit the data to Eq.~\ref{eq:fitting_model} rather than simply taking a Fourier transform. Experimentally, the statistical uncertainty in the measured eigenvalues are computed using bootstrap resampling and are shown as error-bars in Fig. 2c, multiplied by 1,000 so as to be visible.  The typical uncertainty in the measured eigenvalues is of order \SI[product-units = single]{e-4}{\radian}.  This method provides a remarkably high precision tool for probing eigenvalues in large quantum systems.  

\begin{figure*}[t]
\includegraphics[width=178mm, height=60mm]{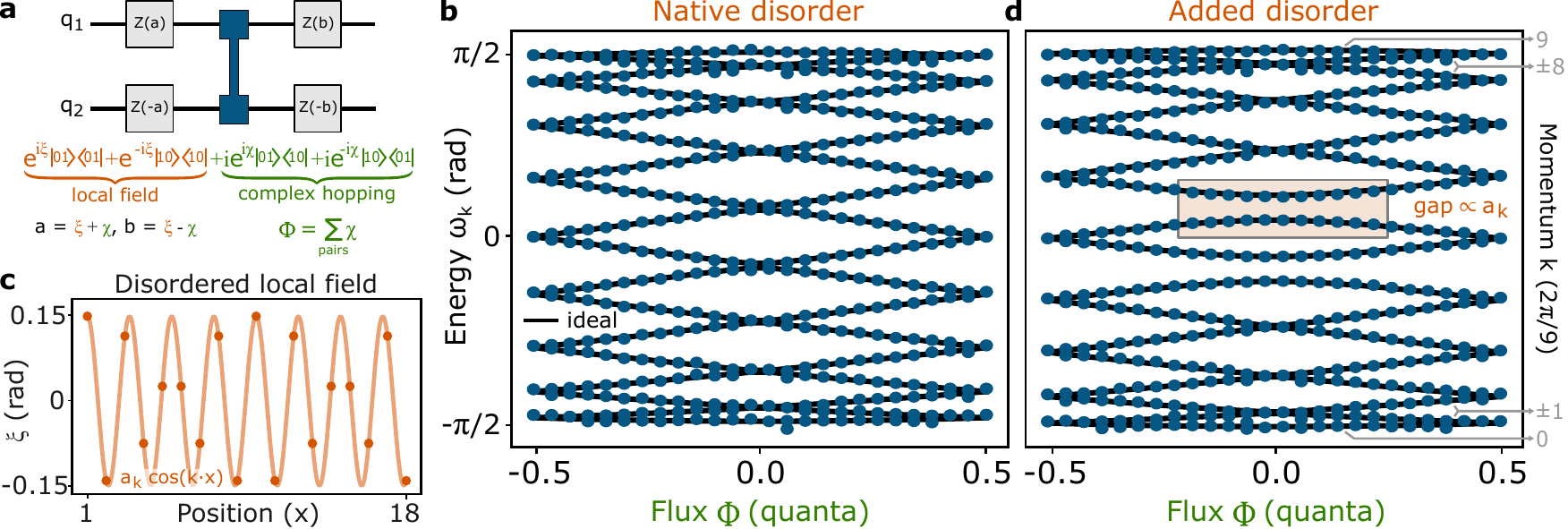}
\caption{\textbf{Synthetic flux as a probe of local disorder.  a}  Gate decomposition for producing local fields and complex hoppings.  The sum of the hopping phases over all pairs realizes a synthetic flux.  \textbf{b}  The spectrum of the cycle unitary is plotted as a function of flux in the case of nominally zero disorder.   The ideal spectrum is shown as black lines. Expectation values are estimated using 30,00 samples. \textbf{c}  The pattern of added disorder is plotted as a function of position along the ring of qubits.  \textbf{d}  The measured spectrum as a function of flux in the case of added disorder.  Only the expected transitions become gapped.  This demonstrates the correspondence between gaps in the spectrum and the spatial Fourier components of disorder in the system.  In addition, the absence of significant splitting in the native disorder case indicates that intrinsic disorder is small.}
\end{figure*}

The energy levels of atoms and materials shift in the presence of an external magnetic field, providing a simple probe of the underlying system.  In Fig. 3a, we provide a control sequence for producing a synthetic magnetic field which we will use to probe disorder in the local potentials ($\xi$).  By applying a specific pattern of z-rotations around the $\surd{\text{i}}\text{SWAP}$ gate, we can produce a complex hopping, such that a particle hopping between adjacent lattice sites accumulates the phase $\chi$; see \app~\ref{sec:gauge}.  The sum of these phases across all links produces a magnetic flux, $\Phi = \sum \chi$.  This is analogous to the Aharonov-Bohm phase~\cite{aharonov_significance_1959} that an electron accumulates when circulating in a conducting ring threaded by a magnetic flux.  In addition to flux, the z-rotations can be used to control the phase of the particle in the cases that it stays on the same site, corresponding to a dynamical phase $\xi$ that a particle would accumulate in a local potential.  
    
The measured energy eigenvalues are plotted as a function of flux in Fig. 3b.  The data (blue circles) are placed atop the exact spectrum (black lines) where we observe excellent agreement between data and theory.  At zero flux, the spectrum is highly degenerate; away from zero flux, the eigenvalues split.  This happens because in the absence of an external flux, a particle travelling clockwise and counter-clockwise have the same energy by symmetry.  The application of flux breaks this symmetry (known as chirality).  Disorder in the local potentials will also break this degeneracy and lead to gaps in the measured spectrum. These gaps enable us to infer the spatial distribution of the disorder through the relation
\begin{equation} \label{eq:gaps}
\mbox{gap}_k \propto \frac{1}{N_q} \left| \sum_{x=1}^{N_q} \xi_x e^{ikx} \right|
\end{equation}
where $\mbox{gap}_k$ is the gap at momentum $k$ and $\xi_x$ is the local field at position $x$; the right hand side of this expression is simply the Fourier transform of the local fields at spatial-frequency $k$.  See \app~\ref{sub_sec:small_disorder} for a derivation that includes over-rotations in the swap angles.  This is quite a remarkable result: gaps in the spectrum correspond one-to-one to the spatial Fourier components of disorder.  This provides a scalable metrology tool for diagnosing control errors in quantum algorithms. 

In order to better understand this effect, we controllably inject disorder into the local fields.  The pattern of local disorder is shown in Fig. 3c.  Rather than random disorder which will open gaps at all values of momentum, we have chosen to add disorder with a single spatial frequency to highlight Eq. \ref{eq:gaps}.  The resulting spectrum with added disorder is shown in Fig. 3d where we observe gaps form at the expected transitions.  The ability to systematically control the disorder enables us to explore novel condensed-matter systems, such as many-body localized phases \cite{pal2010many, ponte2015many, schreiber2015observation}.

\begin{figure*}[t]
\includegraphics[width=178mm, height=60mm]{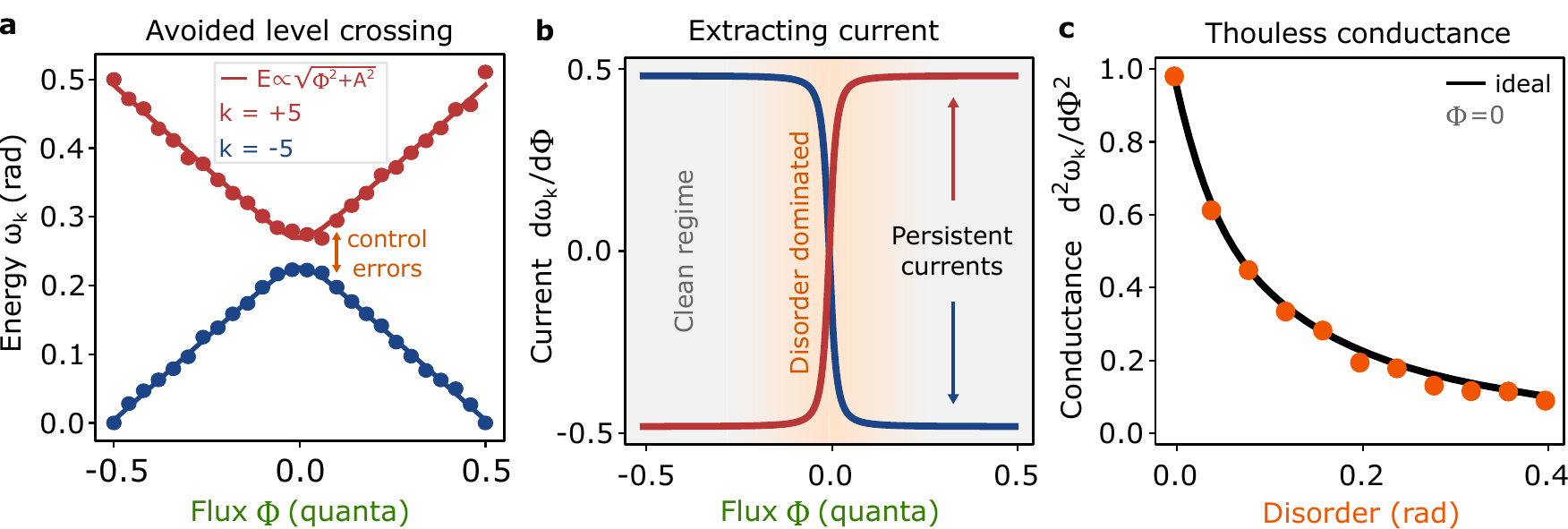}
\caption{\textbf{Inferring current and conductance from avoided level crossings.  a}  Plot of energy versus flux for a single value of momentum.  Near zero flux, we observe an avoided level crossing caused by intrinsic disorder in the local fields.  A fit to the data to a simple avoided level crossing model is shown as solid lines.  \textbf{b}  The derivative of the energy with respect to flux is shown versus flux.  This quantity corresponds to the expectation value of current in each eigenstate.  For small fluxes,  a linear response to flux is observed;  for large flux, current is independent of flux, corresponding to persistent current states.  \textbf{c}  The second derivative of energy is shown versus the amplitude of added disorder.  This quantity corresponds to the conductance of the loop.  We observe a strong suppression of conductance with added disorder. These results demonstrate that expectation values of observables in eigenstates can be measured with high precision and accuracy.}
\end{figure*}

In typical condensed-matter systems, disorder leads to scattering and is the origin of electrical resistance.  In order to study this effect, we focus on the degeneracy near zero flux, as this region of the spectrum is first-order sensitive to disorder.  Figure 4a shows a zoom in of the spectrum at $k=\pm 5$.  Disorder in the local fields cause a small gap to form between the two levels. In this region, the spectrum is well fit by a simple avoided level crossing model, shown as solid lines.  This generic model for the behavior near a level crossing will enable us to infer electrical properties of the eigenstates.

When an external magnetic field $\Phi$ give raise to a current $I$ in a wire, the Hamiltonian can be written as $H = \Phi I$.   This enables us to define the current operator as simply $\hat{I} = dH/d\Phi$.  The expectation value of the current in an eigenstate is then given by the relation
\begin{equation}
\langle \psi_k | \hat{I} | \psi_k \rangle = \frac{d \omega_k}{d\Phi}
\end{equation}
where $| \psi_k \rangle$ is the energy eigenstate at momentum k and $\omega_k$ is the corresponding eigenvalue.  In Fig. 4b we show the extracted current as a function of flux for the eigenstates at momentum $k=\pm 5$.  Near zero flux, where disorder dominates, we observe a linear dependence of current on flux, similar to that of a classical system.  Away from zero flux, we observe current that is independent of disorder - known as persistent current states, similar to that of a superconducting loop \cite{kleemans2007oscillatory, bleszynski2009persistent}. These results demonstrate that expectation values of observables in eigenstates can be extracted from the spectrum.  

The ability to measure current also enables us to infer the conductance of our one-dimensional quantum wire. We use the reasoning introduced by Thouless et al that establishes a connection between the conductance of a disordered wire and the energy levels dependence on magnetic flux~\cite{thouless_quantized_1982}. Briefly, conductance can be defined as dissipation associated with the unit voltage applied to the wire or the one resulting from flux that changes linearly with time in a ring. The single particle energy levels on a ring move as a function of the flux and experience avoided crossings. Each such crossing leads to dissipation when the level is occupied by an electron.  Therefore, one concludes that the physical conductance of a wire is proportional to the product of the slope of current versus flux near zero flux and the density of states~\cite{braun1997level}; this equation relates the non-equilibrium property conductance to the single particle spectrum. This quantity is known as Thouless conductance. In Fig. 4c we plot Thouless conductance as a function of added disorder and observe a strong suppression of the conductance with increasing disorder.  A numerical simulation is shown as a black line.  Because this quantity is computed from eigenvalues, it retains the unprecedented accuracy and precision inherent in using a Fourier transform to process the experimental data.  

Quantum processors hold the promise to solve computationally hard tasks beyond the capability of classical approaches. However, in order for these engineered platforms to be considered as serious contenders, they must offer computational accuracy beyond the current state-of-the-art classical methods. While analytical approaches occasionally provide exact solutions, they quickly lose their relevance upon small perturbations to the underlying Hamiltonian. Numerical methods, in addition to tackling groundstate problems, can handle the dynamics of highly excited states and non-equilibrium phenomena. Currently, the most powerful numerical methods, such as DMRG, have roots in renormalization group ideas and are successful in 1D and quasi-1D geometries \cite{white1992density, white1993numerical}. In dealing with higher spatial dimensions where entanglement spreads widely in space or grows rapidly in time, all numerical methods resort to approximations, where parts of the Hilbert space are truncated to make the computation feasible.  As a result of these limitations, one can safely claim that, for example, studying dynamics in an 8 by 8 spin lattice with local arbitrary interactions and predicting observables with $1\%$ accuracy is beyond any classical computational method.  With the experimental advancements presented here, going beyond this classical horizon seems within reach in the very near future.
\vspace{-3mm}
\section*{Author Contributions}
\noindent \small{C.N. designed and executed the experiment. C.N, and P.R. wrote the manuscript. C.N., T.M., and V.S wrote the supplementary material. V.S., S.B., T.M., Z.J, X.M., L.I., and C.N. provided the theoretical support and analysis techniques, theory of Floquet Calibration, and the open system model. Y. C., V. S., and H. N. led and coordinated the project. Infrastructure support was provided by the hardware team. All authors contributed to revising the manuscript and the supplementary Information.}
\textbf{Data and materials availability:} The data presented in the main text and python code for processing the data can be found in the Dryad repository located at https://doi.org/10.5061/dryad.4f4qrfj9x.

\section*{Methods}
Here, we use the Sycamore quantum processor consisting of 54 superconducting qubits and 86 tunable couplers in a two-dimensional array~\cite{Arute2019}.  This processor consists of gmon qubits (transmons with tunable coupling) with frequencies ranging from 5 to 7 GHz.  These frequencies are chosen to mitigate a variety of error mechanisms such as two-level defects.  Our coupler design allows us to quickly tune the qubit–qubit coupling from 0 to 40+ MHz.  The chip is connected to a superconducting circuit board and cooled down to below 20 mK in a dilution refrigerator.

Each qubit has a microwave control line used to drive an excitation and a flux control line to tune the frequency.  The processor is connected through filters to room-temperature electronics that synthesize the control signals.  We execute single-qubit X, Y, X/2 and Y/2 gates by driving 25ns microwave pulses resonant with the qubit transition frequency.  Single-qubit Z-rotations are implemented using 10ns flux pulses.  The qubits are connected to a resonator that is used to read out the state of the qubit.  The state of all qubits can be read simultaneously by using a frequency-multiplexing. 

Initial device calibration is performed using 'Optimus' where calibration experiments are represented as nodes in a graph~\cite{kelly2018physical}.  On top of these initial experiments, we perform a new two-qubit calibration technique known as 'Floquet Calibrations.'  The name Floquet Calibrations is based on the idea that we want to calibrate a periodic sequence of gates; the unitary describing one period of evolution is known as the Floquet unitary.  This technique is very similar to a Ramsey experiment where greater precision is achieved at long times.  Details of this procedure are presented in the supplementary materials.  

\bibliography{main}

\onecolumngrid
\newpage

\appendix

\setcounter{page}{1}
\setcounter{table}{0}
\setcounter{figure}{0}
\renewcommand{\thepage}{S\arabic{page}}  
\renewcommand{\thetable}{S\arabic{table}}  
\renewcommand{\thefigure}{S\arabic{figure}}
\part{Appendix} 
\parttoc

\section{Floquet Calibrations for two-qubit gates}\label{sec:two_q_floquet}

\begin{figure}
\includegraphics[width=178mm, height=60mm]{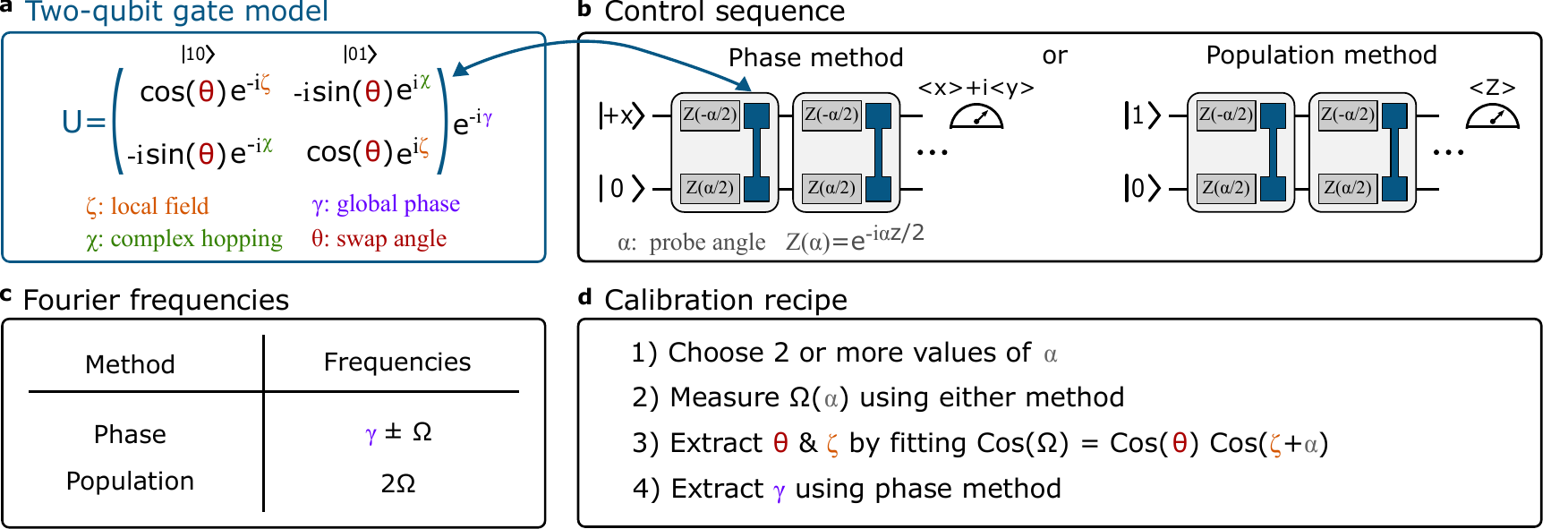}
\caption{\textbf{Two-qubit gate calibration: strategy overview.  a}  General representation of a photon-conserving two-qubit gate, truncated to the single-particle subspace.  This model has four parameters.  The parameter $\theta$ describes how much the particle hops between qubits.  The parameter $\zeta$ is the phase the particle accumulates when it stays on the same site (corresponding to a local field).  The parameter $\chi$ is the phase the particle accumulates when it hops (corresponding to a complex hopping).  The parameter $\gamma$ is a global phase.  \textbf{b}  Two methods for extracting parameters in the Fourier domain by repeated application of the two-qubit gate separated by single-qubit z-rotations.  The z-rotation provides a probe which can be varied to determine parameters.  \textbf{c}  Table showing the Fourier frequencies that each method resolves.  \textbf{d}   Calibration procedure for determining $\theta$, $\zeta$ and $\gamma$ from the measured Fourier frequencies.  The remaining parameter $\chi$ cannot be determined from frequencies at two qubits as it corresponds to a flux and thus requires a ring of qubits.}
\label{fig:2q_diag}
\end{figure}

In this work, we report an 18-qubit algorithm consisting of over 1,400 two-qubit gates with a total error in energy eigenvalues of 0.01 radians.  How is such an unprecedented accuracy possible?  As described in the main text, errors from T1, T2 and readout are mitigated by processing the data in the frequency domain and extracting peak locations.  After this mitigation, the algorithm error is dominated by miscalibrations in the single and two-qubit gates.  This places the problem of resolving small control errors at the center stage.  In this section, we describe a new procedure, called Floquet Calibrations, that allows us to resolve control errors with a remarkable precision below $10^{-5}$ radians.  The name Floquet Calibrations is based on the idea that we want to calibrate a periodic sequence of gates; the unitary describing one period of evolution is known as the Floquet unitary.  This technique is very similar to a Ramsey experiment where greater precision is achieved at long times; we now apply this principle to two-qubit gate calibration.

An overview of Floquet Calibrations is described in Fig.~\ref{fig:2q_diag}.  The general strategy is to repeat a two-qubit gate many times in order to amplify the control errors.  For example, consider a dial which is slightly over-rotated every time it is turned (e.g. 360.1 degrees instead of 360 degrees) - this small over-rotation might be hard to detect.  However, if we repeatedly rotate the dial many times, we can amplify the small error until it is large enough to easily measure.  The same idea can be applied to gates by repeating the operation many times, thus amplifying any small over-rotations.

The model that we use to describe gate parameters is shown in Fig.~\ref{fig:2q_diag} a.  This model represents the most general form of a unitary in the single-particle subspace of two qubits - this is the subspace of interest for the experiments in this paper.  Our goal is to learn the parameters of this model with high precision so that we can then correct for any offsets from the ideal values.  Each of the four parameters can be assigned intuitive physical meanings: a local field, a complex hopping, a global phase and a swap angle.  Note that at two qubits, we will not be able to learn the complex hopping, as this physically corresponds to a magnetic flux, requiring a loop in order to amplify the parameter (see main text). 

Two examples of periodic circuits that can be used to infer the control parameters is shown in Fig.~\ref{fig:2q_diag} b. In both cases, we repeat the two-qubit gate periodically in time.  Between each gate we apply a variable z-rotation by the angle $\alpha$  which we will use as a probe of the gate parameters ($\alpha$ is the same at each depth).  The key difference between the two methods is the initialization and measurement basis.  The first method is similar to a T1 experiment (excite one qubit, measure in the z-basis).  The second method is similar to a Ramsey experiment (intitialize along the equator, measure along the equator).  These experiments will provide access to different information about the unitary parameters.

For both methods, we process the data in the Fourier domain and extract oscillation frequencies.  This strategy has the benefit of being robust to T1, T2 and readout errors as well as amplifying the signal by going to large depths.  The Fourier frequencies extracted by either method is shown in Fig. ~\ref{fig:2q_diag} c.  The key difference is that the phase method is sensitive to the global phase whereas the population method is not.  Generally, for any number of qubits in the circuit, the phase method has frequencies at the eigenvalues of the unitary (as we show in the main text);  the population method has frequencies at all possible differences between eigenvalues.

Putting all of these ideas together enables a simple, robust and accurate method for calibrating two-qubit gates.  An overview of this procedure is shown in Fig. ~\ref{fig:2q_diag} d.   Only two values of the probe angle $\alpha$ are needed to extract all of the parameters; more values may be included as a consistency check of the model.  The eigenvalues of the unitary $\Omega$ can be measured using either method and the expression
\begin{equation}
\cos(\Omega) = \cos(\theta) \cos(\xi + \alpha)
\end{equation}
can be used to extract the control parameters $\theta$ and $\xi$.  The last parameter $\gamma$ can be extracted using the phase method by finding the center point between the two Fourier peaks.

\begin{figure*}[!t]
\includegraphics[width=178mm, height=60mm]{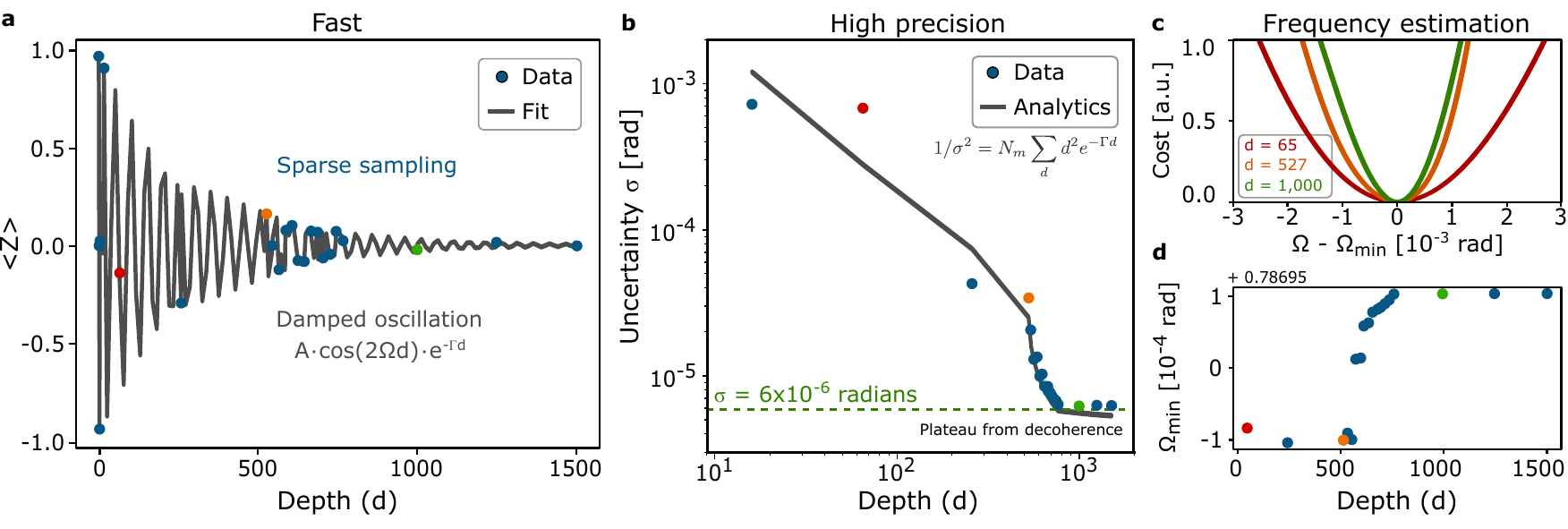}
\caption{\textbf{Two-qubit gate calibration: Fast, high-precision frequency estimation.  a}  Raw time-domain data used to extract Fourier frequencies for two-qubit gate calibration.  Depths are spaced logarithmically in order to enable fast acquisition of the signal.  The data is fit for an oscillation frequency $\omega$ and damping rate $\Gamma$.  \textbf{b}  Statistical uncertainty in the oscillation frequency as a function of depth, computed using bootstrap resampling.  The ideal analytic curve is shown in gray where $N_m=50,000$ is the number of measurements taken at each depth.  At large depth decoherence dominates and uncertainty plateaus at around $6 \times 10^{-6}$ radians.  \textbf{c}  The mean-squared fitting cost is shown as a function of $\omega$ for three values of depth.  Deeper circuits correspond to a sharper minima around the optimal frequency, corresponding to an improved accuracy with increasing depth.  \textbf{d}  Optimal frequency as function of depth.  We observe variations in the oscillation frequency on the order of $10^{-4}$ radians. 
}
\label{fig:2q_var}
\end{figure*}

Our calibration scheme is based on extracting Fourier frequencies from measurements of single-qubit observables.  Fig.~\ref{fig:2q_var} a shows a typical dataset used in this procedure where we observe a rather simple damped oscillation in $\langle z \rangle$ as a function of depth.  Sparse sampling in depth is used in order to enable fast acquisition of the signal.  It takes around one minute to measure 23 separate depths (d) out to $d=1,500$ with 50,000 repetitions at each depth.  A fit to the data is shown in gray which enables us to extract an oscillation frequency and a damping rate.  The oscillation frequency can be used to infer control parameters and the damping rate provides a metric of decoherence.  

The longer we measure the signal, the more precisely we can infer the Fourier frequencies.  Fig.~\ref{fig:2q_var} b shows the statistical uncertainty in the extracted oscillation frequency as a function of the maximum depth considered, computed using bootstrap resampling.  The data shows orders of magnitude reduction in the uncertainty with increasing depth, reaching a plateau of $6\times 10^{-6}$ radians.  The expected behavior is shown as a gray line and depends only on the maximum depth (d), the number of measurements $N_m$ and the damping rate $\Gamma$.  Note that the corresponding error (1 - fidelity) is quadratic in the over-rotation, corresponding to an error as low as $4 \times 10^{-11}$.

Intuition into this precision can be gained by looking into the optimization landscape for learning the oscillation frequency.  Fig.~\ref{fig:2q_var} c shows the cost function around the optimal frequency for three values of maximum circuit depth.  Deeper circuits correspond to a sharper minima, leading to more precision in estimating the optimal value. Fig.~\ref{fig:2q_var} d shows the extracted frequency as a function of depth where we observe small drift at the level of $10^{-4}$ radians.  Potential sources of this include fitting-bias at short depth, time-correlated errors in the control signals, and low-frequency noise.  

\section{Periodic circuit on a qubit ring: Unitary evolution}
\begin{figure}[ht]
 \includegraphics[width= 5.5in]{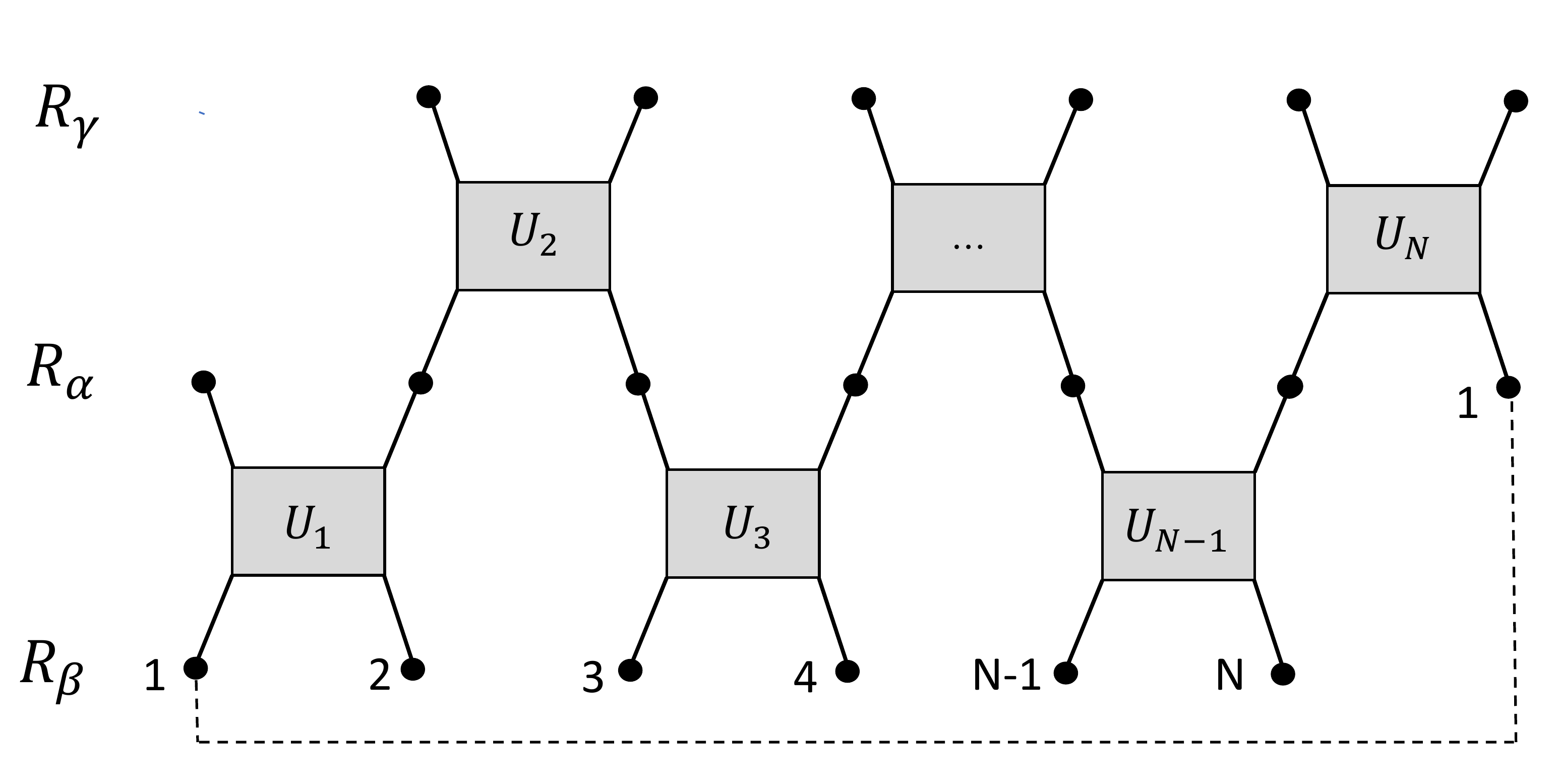}
  \caption{\textbf{Linear chain circuit structure} a periodic circuit on a linear chain with even number of qubits $N=2L$ (cf. Eq.~(\ref{eq:Ud})).   Gate $U_N$ is applied between $N^{\rm th}$ and $1^{\rm st}$ qubits.  \label{fig:ring}}
\end{figure}

Let us consider a periodic circuit composed of 2-qubit gates acting over a linear array of $N$ qubits, where the gate $U_j$ is applied on qubits $j$ and $j+1$.  The circuit unitary depicted in the Fig.~\ref{fig:ring} is equal to
\begin{equation}
U(d)=\Ucd, \quad U_{\rm cycle}=\prod_{j=1}^{N/2}U_{2j}  \prod_{j=1}^{N/2}  U_{2j-1} \;.\label{eq:Ud}
\end{equation}
Here $U_{\rm cycle}$ is a cycle unitary composed of two layers of gates. In the first layer the gates are applied between odd and even qubits, and in the second layer the gates are applied between even and odd qubits.  We shall assume that the number of qubits $N$  is even.

The circuit unitary can be expressed in terms of it's Floquet eigenstates and eigenvalues
\begin{equation}\label{eq:eigen}
U(d)=\Ucd=\sum_{\alpha}e^{-i \,d \,\omega_\alpha}\ket{\psi^\alpha}\bra{\psi^\alpha}\;,
\end{equation}
where $\omega_\alpha$ are usually referred to as quasi-energies. Here, we will focus on  unitaries $\Uc$ that preserve the total number of excitations in a chain and consider the case with 1 and 0 excitations (the vacuum state $\ket{0}$ is an eigenstate of $\Uc$ with  
$\omega_0=0$). On a short time scale when decoherence can  be neglected observable quantities are expressed in terms of the matrix elements of the superoperator
\begin{equation}
U(d)\otimes U^{\dag}(d)=\sum_{\alpha,\beta}e^{-i d\, \omega_{\alpha\beta}}F_{\alpha\beta},\quad \omega_{\alpha\beta}=\omega_\alpha-\omega_\beta\;,\label{eq:super}
\end{equation}
where $F_{\alpha\beta}  =\ket{\psi_\alpha}\otimes \ket{\psi_\beta}\bra{\psi_\alpha}\otimes \bra{ \psi_\beta}$.  In the limit $d \gg 1$, small changes in the gate parameters will lead to large changes in the phase factors $e^{-i d \omega_{\alpha\beta}}$. Therefore from stand point of Floquet calibration the  dependence of quasi-energy differences $\omega_{\alpha\beta}$  on the gate parameters are of a predominant importance. 

In what following we will consider the transformation  of the periodic circuit unitaries to the form 
$\Ucd=\Rf\, \Ufd\,{\mathfrak R}^{\dag}$ 
where the unitary $\Rf$ does not affect the quasi-energies while  $\Uf$ represents the  reduced (canonical) form of the cycle unitary that depends on a smaller number of gate parameters than the original circuit unitary $\Ucd$.

Formally, each two-qubit gate $U_j$ is  defined by 5 parameters:  3 single qubit phases $\chi_j, \zeta_j, \gamma_j$, swap angle $\theta_j$ and CZ phase $\varphi_j$. The cycle unitary  $U_{\rm cycle}$ contains $N$ gates and the total number of the  parameters is $5N$.  As will be shown below the number of parameters on which the quasi-energy differences $\omega_{\alpha\beta}$ depend upon  will be much smaller.
 To determine this we shall perform the transformation of the unitary $U_{\rm cycle}^{d}$ to the simplified ``canonical"  form as described below.

\subsection{Excitation conserving 2-qubit gates}

Throughout the paper the basic element of the circuit is an excitation-conserving two-qubit gate. Its most general form is given below with the basis states in the order $\ket{00},\ket{01},\ket{10},\ket{11}$
 \begin{equation}\label{eq:Uf}
\begin{split}
\begin{gathered}
\mathfrak{U}=\left(
\begin{array}{cccc}
 1 & 0 & 0 & 0 \\
 0 & e^{-i (\gamma +\zeta )} \cos (\theta ) & -i e^{-i (\gamma -\chi )} \sin (\theta ) & 0 \\
 0 & -i e^{-i (\gamma +\chi )} \sin (\theta ) & e^{-i (\gamma -\zeta )} \cos (\theta ) & 0 \\
 0 & 0 & 0 & e^{-i (2 \gamma +\varphi )} \\
\end{array}
\right)
\end{gathered}
\end{split}
\end{equation}
We denote this gate applied between the qubits  $j$ and $j$+1
as  $\mathfrak{U}=U_j$. Its expression in terms of the Pauli   matrices of the qubits is
  \begin{equation}
U_j=e^{-i \left(n_j+n_{j+1}\right) \gamma _j}\,e^{i \left(n_j-n_{j+1}\right)  \left(\frac{\zeta _j-\chi _j}{2}\right)} u_j(\theta_j,\varphi_j) e^{i \left(n_j-n_{j+1}\right) \left(\frac{\zeta _j+\chi _j}{2}\right)}\;,\label{eq:Uj}
\end{equation}
 where
 \begin{equation}
 n_j=\frac{ 1-\sigma_j^z }{2}\;,\label{eq:nj}
 \end{equation}
 is a number operator and the $u_j$ are the ``bare" two-qubit unitaries of the iSWAP type 
\begin{equation}
u_j(\theta,\varphi)=\exp \left(-i \theta \left(\sigma _j^+\sigma _{j+1}^-+\sigma _j^-\sigma _{j+1}^+\right)-i \varphi \,n_j
n_{j+1}\right)\;.\label{eq:uj}
\end{equation}

In the single excitation subspace the gate $U_j$ can be written in the form (up to an overall phase factor  $e^{-i \gamma _j}$)
 \begin{equation}\label{eqn:gate_unitary}
U_j=e^{i \zeta _j}|j\rangle \langle j| \cos  \theta _j-i e^{i \chi _j}|j+1\rangle \langle j| \sin\theta_j-i e^{-i \chi
_j} |j\rangle \langle j+1|\sin\theta_j+e^{-i \zeta _j}|j+1\rangle \langle j+1| \sin\theta_j
\end{equation}
where we introduced the basis states 
 \begin{equation}
|j\rangle =|0_10_2\text{$\ldots $0}_{j-1}1_j0_{j+1}\text{$\ldots $0}_n\rangle\;,\label{eq:basis}
\end{equation}
where $|j\rangle $ corresponds to the qubit $j$ in the state 1 and the rest of the qubits in the state 0.
 In analogy to the tight binding model describing the motion of a charge in a magnetic field \cite{kohn1959theory},  $\chi_j$ is a  Peierls  phase corresponding to the integral of the vector potential along the hopping path. Therefore one might expect that the physical properties of the system will depend on the magnetic flux  through the ring $\sum_{j=1}^{N}\chi_j$. To reveal this property we study the gauge transformation in the next section.
 
\subsection{Local gauge transformation of the circuit  unitary to the canonical form}
\label{sec:gauge}

We start by writing the cycle unitary in a form that separates the single-qubit $z$-rotations from the two-qubit gates $V_j$
 \begin{equation}
U_{\rm cycle}=R_{\lambda }V_e\, R_{\alpha }\, V_o\, R_{\beta }\;,  \label{eq:cycle}
\end{equation}
\begin{equation}
V_e=\prod_{k=1}^{L}u_{2j},\quad V_o=\prod_{k=1}^{L-1}u_{2j+1},\quad L=\frac{N}{2}\;.\label{eq:VoVe}
\end{equation}
where $u_k\equiv u_k(\theta_k,\varphi_k)$ and  $V_o$,  $V_e$ correspond to the first and second layer of the two-qubit unitaries $u_j$ (\ref{eq:uj}) and we used the fact that in the circuits we consider the number of qubits $N=2L$ in a chain is even.

In Eq.~(\ref{eq:cycle}) above the   matrices of single-qubit rotations  appear at the beginning of first layer of gates ($R_{\beta }$), between the layers ($R_{\alpha }$) and after the second layer
 ($R_{\lambda }$)
 \begin{align}
R_{\lambda }=\exp\left(-i \sum _{k=1}^{L-1} \left(n_{2k}+n_{2k+1}\right) \gamma _{2k}+\frac{i}{2}\sum _{k=1}^{L-1}  \left(n_{2k}-n_{2k+1}\right)
 \left(\zeta _{2k}-\chi _{2k}\right)\right),\label{eq:Rlambda}\\
R_{\alpha }= \exp \left( \frac{i}{2} \sum _{k=1}^{L-1} \left(n_{2k}-n_{2k+1}\right)  \left(\zeta _{2k}+\chi _{2k}\right)+\frac{i}{2}\sum _{k=1}^L  \left(n_{2k-1}-n_{2k}\right)
\left(\zeta _{2k-1}-\chi _{2k-1}\right)\right),\label{eq:Ralpha} \\
R_{\beta }= \exp \left( \frac{i}{2} \sum _{k = 1}^{L} \left(n_{2k-1}-n_{2k}\right) \left(\zeta _{2k-1}+\chi _{2k-1}\right)-i\sum _{k = 1}^{L} \left(n_{2k-1}+n_{2k}\right)
\gamma _{2k-1}\right)\;.\label{eq:Rbeta}
\end{align}

Let us now consider the  circuit unitary $U_{\rm cycle}^{d}$ after $d$ cycles. We push  $R_{\lambda }$ to the beginning of the next cycle, yeilding
 \begin{equation*}
U_{\rm cycle}^{d}=R_{\lambda }\,U^{d }\,R_{\lambda }^{-1}
\qquad U=\,V_e\,R_{\alpha }\,V_o\,R_{\beta }\,R_{\lambda }
\end{equation*}
 We further split $R_{\alpha }$ into two parts 
  \begin{equation*}
R_{\alpha }=R_{\alpha }^{(-) }\,R_{\alpha }^{(+)}
\end{equation*}
 \begin{equation*}
R_{\alpha }^{(\pm)}=\exp\left(-i \sum _{k=1}^L \left( n_{2k+1}\pm n_{2k}\right)a_k^{\pm} \right)
\end{equation*}
where we imply the periodicity condition
\begin{equation}
n_{N+1}=n_1\;.\label{eq:per}
\end{equation}
The unitary $R_{\alpha }^{(+) }$ can be commuted through  the second layers of iSWAP gates 
  \begin{equation}
\left[V_e,R_{\alpha }^{(+)}\right]=0,\label{q:commZaVe}
\end{equation}
while the unitary $R_{\alpha }^{(-) }$ cannot. The explicit form of the coefficients $a_k^-$ is 
 \begin{equation}
a_k^-=\frac{\zeta _{2k}+\chi _{2k}}{2}-\frac{\zeta _{2k+1}+\zeta _{2k-1}-\chi _{2k+1}-\chi _{2k-1}}{4},\quad k\in (1,L-1)\label{eq:am}
\end{equation}
\begin{equation}
a_N^-=\frac{\zeta _1-\chi _1+\zeta _{N-1}-\chi _{N-1}-2\left(\zeta _N+\chi _N\right)}{4}\label{eq:anm}
\end{equation}
The coefficients $a_k^+$ are not very important and will be given later.
Using the above commutation relation we can write after $d$ cycles
 \begin{equation}
U_{\rm cycle}^{d}=R_{\lambda }\,R_{\alpha }^{(+)}\,\tilde U^{d}_{\rm cycle}\,\left( R_{\alpha }^{(+)}\,R_{\lambda
}\right){}^{-1}\;,
\end{equation}
where the new cycle unitary has the form
\begin{equation}
\tilde U_{\rm cycle}= V_e\,R_{\alpha }^{(-)} V_o\,\ R_{\alpha }^{(-)\,\dagger}\,R\;.\label{eq:Uc1}
\end{equation}
and
\begin{equation}
R=R_{\alpha }\,R_{\beta }\,R_{\lambda }\,\label{eq:R}
\end{equation}
is simply a product of all single-qubit phase gates of the original cycle unitary $U_{\rm cycle}$ (\ref{eq:cycle}).

We seek to  simplify the factor  $R_{\alpha }^{(-)} V_o\, R_{\alpha }^{(-)\,\dagger}$ in (\ref{eq:Uc1}) by making the  local {\it  gauge}  transformation of   the  unitary $\tilde U_{\rm cycle}$
 \begin{equation*}
\mathfrak{U}=S\,\tilde U_{\rm cycle}\,S^{-1}=\,V_e \left(R_{\alpha }^{(-)} S\right)V_o\left( R_{\alpha }^{(-)} S\right){}^{-1} R
\end{equation*}
where $ S$ has the form
 \begin{equation}
S=\prod_{j=1}^{L} \exp\left(- i   \left(n_{2j}+n_{2j+1}\right)\varrho _j\right), \quad [ V_e,S]=0\;,\label{eq:Rfr}
\end{equation}
and we used the fact that $[V_e,S]=0$.

 Our strategy is  to choose the coefficients $\varrho _j$ in such a way that the phase gate  $R_{\alpha }^{(-)} S$ has the form where  all factors commute with $V_o$ except for a single link 
  \begin{equation}
[R_{\alpha }^{(-)} S,V_o]=[e^{-i\,\frac{1}{2}\left(n_1-n_N\right) \phi},V_o]\;.\label{eq:coimm}
\end{equation}
for some value of the gauge field  $\phi$. This can be achieved if the unitary $R_{\alpha }^{(-)} S$ has  the form
\begin{equation}
R_{\alpha }^{(-)} S=e^{ -i \sum _{j=1}^{N/2}  \left(n_{2j-1}+n_{2j}\right)b_j}e^{-i\,\frac{1}{2}\left(n_1-n_N\right) \phi}\;,\label{eq:RR}
\end{equation}
 with some coefficients $b_j$. Under this condition  the  expression  for the circuit unitary after $d$ cycles can be written in the form
 \begin{equation}
U_{\text{\rm cycle}}^d=\mathfrak{R}\,\mathfrak{U}^{d }\mathfrak{R}^{-1},\label{eq:Uc2}
\end{equation}

  \begin{equation}
\mathfrak{U}= V_e \, e^{-i\,\frac{1}{2}\left(n_1-n_N\right) \phi }\,V_o\,e^{i\,\frac{1}{2}\left(n_1-n_N\right) \phi } R\;.\label{eq:Ufr}
\end{equation}
where $V_e$, $V_o$ are given in (\ref{eq:VoVe}). Phase $\phi$ plays a role of the gauge field on a ring and the local gauge transformation $\mathfrak{R}$ equals
\begin{equation}
\mathfrak{R}=S^{-1} R_{\lambda }\,R_{\alpha}^{(+)}\;.\label{eq:Rf}
\end{equation}
\subsubsection{Gauge field  $\phi$}
Equating right and left hand sides in (\ref{eq:RR}) we can readily obtain the set of
 $\varrho_j$ and $b_j$ coefficients. In particular, one can show that 
 \begin{equation*}
\phi=2\sum _{j=1}^L a_j^-,\quad b_k\,=a_1^-+2\sum _{j=2}^{k-1} a_j^-,\quad k\in (1,L)
\end{equation*}
   \begin{equation*}
\varrho_1=0,\quad   \varrho _k=a_1^-+a_k^-+2\sum _{j=2}^{k-1} a_j^-\quad k\in(2,L-1),\quad
\varrho _L=\sum _{j=2}^{L-1} a_j^-
\end{equation*}
 
Using explicit form of the coefficients $a_j^-$ from (\ref{eq:am}),(\ref{eq:anm}) we get
\begin{equation}
\phi=\sum _{k=1}^N (\chi _k+(-1)^k\zeta _k)\;.\label{eq:B}
\end{equation}
\subsubsection{Derivation of the phase gate unitary  $R$}
Also one can readily obtain the explicit form of the phase gate operator in (\ref{eq:R}) 
 \begin{equation}
R =e^{-i \sum _{j=1}^N \nu _jn_j}\label{eq:Rnu}
\end{equation}
 where the parameters $\nu _j$  do  not depend  on the angles $\chi_j$ 
  \begin{align}
\nu _{2m}&=\zeta _{2m-1}-\zeta _{2m} +\gamma _{2m}+\gamma _{2m-1}\label{eq:nu} \\
\nonumber \\
\nu _{2m+1}&=\zeta _{2m}-\zeta _{2m+1} +\,\gamma _{2m}+\,\gamma _{2m+1}\nonumber 
\end{align}
 where we implied cycle conditions $\nu_j=\nu_{j+N}$. We fix that the global phase to be zero hence 
 \begin{equation}
\sum _{j=1}^n \nu_j=0\;.\label{eq:cond}
\end{equation}
One can see that the dependence on the angles $\zeta_j$ and $\gamma_j$ comes only in terms of the combinations (\ref{eq:nu}). Therefore without loss of generality we can set $\gamma_j=0$.

\subsubsection{Local gauge transformation $\mathfrak{R}$}
Using the equations (\ref{eq:Rlambda}), (\ref{eq:Rfr}) and (\ref{eq:Rf}) we obtain
\begin{equation}
\mathfrak{R}=e^{i \mathfrak{r}}\;,\label{eq:Rf1}
\end{equation}
\begin{equation}
\mathfrak{r}=\frac{1}{2}\sum _{k=1}^L \left[\left(n_{2k}+n_{2k+1}\right)f_k+\left(n_{2k}-n_{2k+1}\right)\left(\zeta _{2k}-\chi _{2k}\right)\right]\;,\label{eq:rf}
\end{equation}
where site number operators $n_j$ are given in (\ref{eq:nj}) and quantities $f_k$ expressed in terms of the combinations of single-qubit phases $x_k$
\begin{equation*}
x_k\equiv \chi _k+(-1)^k\zeta _k
\end{equation*}
\begin{equation*}
f_1=\frac{x_1-x_3}{2}\hspace{8 mm}
\end{equation*}
\begin{equation*}
f_2=\frac{x_1+2x_2+3 x_3+2 x_4}{2}\hspace{5 mm}
\end{equation*}
\begin{equation*}
\hspace{3 mm}f_k=\frac{x_1+2x_2+3x_3}{2}+x_{2k}+2\sum _{j=4}^{2k-1} x_j\hspace{9 mm}
\end{equation*}
\begin{equation*}
f_{n/2}=\frac{x_3-x_1}{2}+\sum _{j=4}^{n-1} x_j
\end{equation*}

\subsubsection{Number of independent parameters}\label{sec:param_counting}

The equations (\ref{eq:Uc2})  express  the circuit unitary    $U_{\rm cycle}^{d}$ in  terms of the  canonical form of the cycle 
unitary $\Uf$  that has the same set of eigenvalues as $U_{\rm cycle}$.   It depends on the $N$ swap angles $\theta_j$ and CZ phases $\varphi_j$, also $N-1$ independent single qubit phases  $\nu_j$  (cf.  (\ref{eq:cond})) and a gauge field   $\phi$. The eigenvalues of $\Uc$ and $\Uf$ are the same. Therefore the quasi-energy differences $\omega_{\alpha\beta}$ depend on $2N$ parameters as shown in Table \ref{eq:tab-param}. In a single excitation subspace  the CZ phases are not important  and the number of independent parameters is 2$N$. We also note that cycle unitary $\Uf$ depends on the angles $\chi_j$ only via the gauge field $\phi$.

  \begingroup
\setlength{\tabcolsep}{10pt} 
\renewcommand{\arraystretch}{1.8} 
\begin{table}[h]
\begin{tabular}{|c|c|c|c|c|}
\hline
Swap angles  and CZ phases & Single-qubit  phases   & Gauge field &All parameters \\
 $\{\theta_j,\,\varphi_j\}_{j=1}^{N}$ &  $\{\nu_j\}_{j=1}^{N} $ &  $\phi=\sum_{j=1}^{N}(\chi_j+(-1)^j\zeta_j)$&\\
  \hline
  $2N$& $N-1$ &  1 &3$N$\\
 \hline
\end{tabular}
\caption{Parameters that determine the quasi-energy differences for a periodic circuit on a ring}
   \label{eq:tab-param}
  \end{table}
  \endgroup

\subsubsection{Uniformly Distributed Flux}
 
For future purposes it is desirable to modify Eq.~ (\ref{eq:Ufr}) such that the flux is distributed evenly over the gates instead of being concentrated on one gate (involving qubits $N$ and $1$). Let us define an average flux per gate
 \begin{equation}
     \Bar{\chi} = \frac{\phi}{N},\label{eq:chibar}
     \end{equation} 
and consider a cycle unitary $\Uc$ where each gate $U_j=U_j(\Bar{\chi},0,\theta_j)$ corresponds to the same value of single-qubit phases, $\chi_j=\Bar{\chi}$ and $\zeta_j=0$ 
 \begin{equation}
  U_{\rm cycle}(\Bar{\chi},0,\vec{\theta}\,) \equiv \prod_{j\in {\rm even}} U_j(\Bar{\chi},0,\theta_j)  \prod_{j\in {\rm odd}} U_j(\Bar{\chi},0,\theta_j)\;.\label{eq:Uc0}
 \end{equation}
 Here we used a  vector notation for the set of parameters,  $\vec \theta=\{\theta_1,\ldots,\theta_N\}$. The corresponding 
 local gauge transformation $\mathfrak{R}$ in (\ref{eq:Uc2}) can be  obtained from (\ref{eq:Rf1})  by setting $\chi_j = \Bar{\chi}$ and $\zeta_j=0$.

In the general case of non-uniform single qubit angles the circuit  unitary can be expressed in terms of $U_{\rm cycle}(\Bar{\chi},0,\vec{\theta})$ as follows
\begin{equation}
    U_{\rm cycle}^{d}(\vec{\chi},\vec{\zeta},\vec{\theta})=\Rf(\vec{\delta\chi} ,\vec{\zeta})\,\,
    \Ufd(\Bar{\chi},\vec\zeta,\vec\theta)\,\,
    \Rf^{-1}(\vec{\delta\chi},\vec{\zeta})\;,\label{eq:Uc4}
\end{equation}
\begin{equation}
   \Uf(\,\Bar{\chi},\vec\zeta,\vec\theta\,)=
    U_{\rm cycle}(\,\Bar{\chi},0,\vec{\theta}\,)\,R(\,\vec{\zeta}\,)\;,\label{eq:Uc4a}
\end{equation}
 where $\vec{\delta\chi}\equiv \{\chi_1-\Bar{\chi},\ldots,\chi_N-\Bar{\chi}\}$,
 the phase gate unitary (\ref{eq:R}) $R(\,\vec{\zeta}\,)$ is given in (\ref{eq:Rnu}) and we explicitly indicated all the arguments in the equations above. 
 
 \subsection{``Reference" Circuit with Identical Gates}
 \label{sub_sec:reference_circuit}
 Here we study the ``reference" circuit corresponding to the case where all gates are identical  
 \begin{equation}
     \chi_j=\chi,\quad \zeta_j=0,\quad \theta_j=\frac{\pi}{4}\;.\label{eq:ref}
 \end{equation}
 We want to solve the eigen-problem for the nominal circuit $\Uc(\chi) \equiv \Uc(\,\chi,0,\pi/4)$ as given by
 \begin{equation}
    \Uc(\chi) \ket{\psi} = e^{-i \omega}\ket{\psi}\;.\label{eq:eig-ref}
 \end{equation}
 We introduce the nomenclature of the site basis states
 \begin{equation}
 \{\,\, \ket{\gamma, m}, \quad \gamma=1,2,\quad m=1,\ldots,N/2\,\,\}\;,\label{eq:basis-block}   \end{equation}
 where qubits at subsequent odd ($\gamma=1$) and even $(\gamma=2$) positions  form blocks enumerated by the index $m=1:N/2$.
 In this notation the matrix elements of the cycle unitary are given by
 \begin{equation}
 U_{m\,k}^{\gamma \beta} = \bra{\beta, m} \Uc(\chi) \ket{\gamma, k},\quad \gamma, \beta \in \{1,2\},\quad m,k\in\{1,\ldots,N/2\}\;.\label{eq:Ublock} 
 \end{equation}
$U_{m\,k}^{\gamma \beta}$ has the following block-translationally invariant form
 \begin{equation}
     U_{m,m-1}^{11} = -\frac{1}{2}e^{2 i \chi}, \quad U_{m,m}^{11} = \frac{1}{2}, \quad U_{m, m-1}^{12} = -\frac{i}{2} e^{i \chi}, \quad U_{m,m}^{12} = -\frac{i}{2} e^{-i \chi}\,\label{eq:Uref}
 \end{equation}
  \begin{equation}
     U_{m,m}^{21} = -\frac{i}{2}e^{i \chi}, \quad U_{m,m+1}^{21} = -\frac{i}{2} e^{-i \chi}, \quad U_{m, m}^{22} = \frac{1}{2}, \quad U_{m,m+1}^{22} = -\frac{1}{2} e^{-2 i \chi}\;.\nonumber
 \end{equation}
 Because of this translational symmetry
the components of the eigenstates $\ket{\psi^{\alpha}}$ in the basis (\ref{eq:basis-block})    $\braket{\beta, k}{\psi^{\nu}(q)} \equiv \psi_{\beta, k}^{\nu}(q)$ have the form
\begin{equation}
    \psi_{\gamma, m}^{\nu} (q) = \frac{1}{\sqrt{L}} V_{\gamma}^{\nu}(q) e^{i q m},\quad \nu=\pm,\quad q\in{\cal Q}\;,
\end{equation}
and are characterized by the values of the quasi-momentum  $q$ from the set $\mathfrak{Q}$ (to be given  below) and the branch index $\nu=\pm $. For each eigenstate the values of $V_{\gamma}^{\nu}(q)$ form a two-component spinor
 \begin{equation}
\begin{split}
\begin{gathered}
V^{\nu}(q)=\frac{1}{\sqrt{2}}\left(
\begin{array}{c}
 \nu  \left(1-\nu \sin (p/2)\left( \sin ^2(p/2)+1\right)^{-1/2}\right)^{1/2} \\
 e^{i \left(\frac{p}{2}+\chi \right)} \left(1+\nu \sin (p/2)\left( \sin ^2(p/2)+1\right)^{-1/2}\right)^{1/2} \\
\end{array}
\right),\quad p=q-2\chi\;\label{eq:Vpm}
\end{gathered}
\end{split}
\end{equation}
 The corresponding quasi-energies are given by 
 \begin{align}
   \omega_{\pm}(q) = \pm \Omega(q-2\chi), \label{eq:mupm}
   \end{align}
   where
  \begin{equation}
    \Omega(p) = \arccos\left( \sin^2\left( p/2\right)\right)\;.\label{eq:ring_omega}
 \end{equation}
 
\begin{figure}[H]
    \centering
    \includegraphics[width=0.95\textwidth]{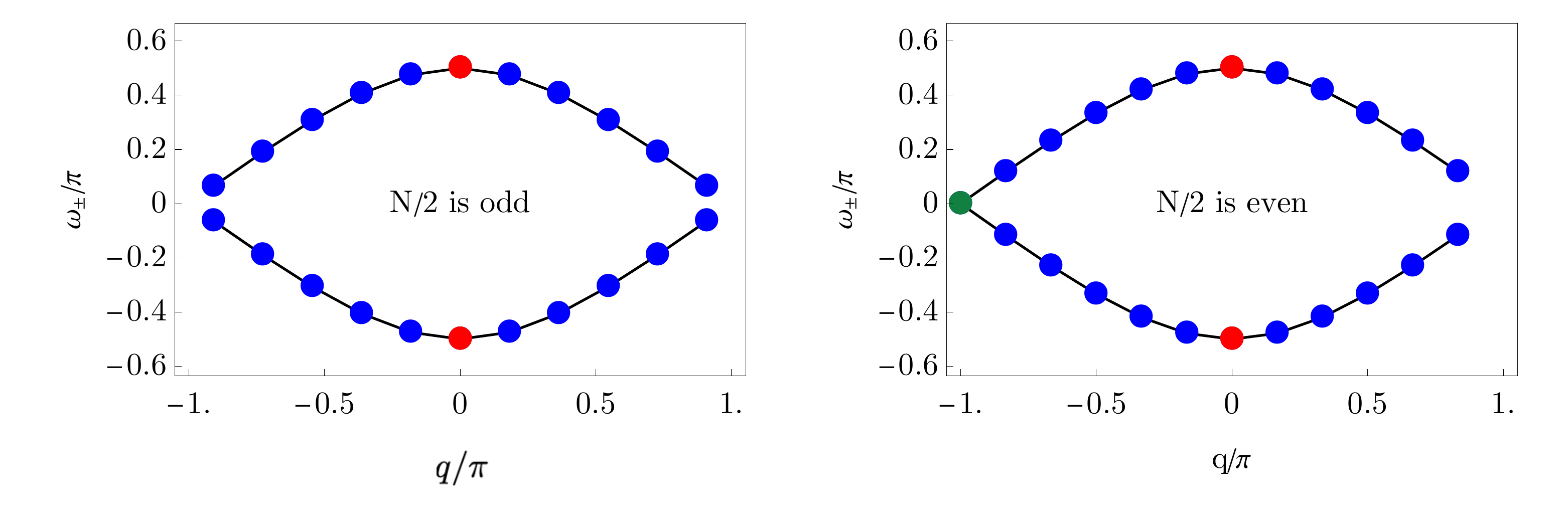}
    \caption{\textbf{Study of the spectrum of a ring} Quasi-energies $\omega_{\pm}(q)$ of the reference   circuit (\ref{eq:ref}), (\ref{eq:Uref}) {\it vs} quasi-momentum $q$ at zero flux  $\chi=0$. Left plot corresponds to the ring with $N=$22 qubits ($N/2$ is odd). Thick points correspond to the values of the quasi-momentum (\ref{eq:qodd}). Pairs of blue points with $q=\pm |q|$, $|q|\neq 0$ correspond to  the doubly-degenerate quasi-energy values (doublets). There are $N-2$ doublets in total.  Red points correspond to quasi-energy singlets with $q=0$. They  correspond to maximum quasi-energy values $\omega_\pm(0)=\pm\pi/2$. Right plot corresponds to the ring with $N=$24 qubits ($N/2$ is even). Thick points correspond to the values of the quasi-momentum (\ref{eq:qeven}). Unlike the case with odd values of $N/2$ there is a value of momentum $q=-\pi$ corresponding to the pair of degenerate quasi-energy levels $\omega_{\pm}(-\pi)=0$. There are $N-4$ doublets  corresponding each to the opposite nonzero quasi-momenta $q$.}
    \label{fig:mu-q}
\end{figure}

 For odd values of $L$, the quantized  values of quasi-momentum are given by the following
 \begin{equation}
     q\equiv q_m = -\pi + \frac{\pi \left(2m + 1\right)}{L}, \quad  m \in {0, L-1},\quad L=\frac{N}{2}\,\,({\rm odd}),\label{eq:qodd}
 \end{equation}
 and for even values of $L$
 \begin{equation}
     q\equiv q_m = -\pi + \frac{2 \pi m}{L}, \quad m \in {0, L-1},\quad L=\frac{N}{2}\,\,({\rm even}),\label{eq:qeven}
 \end{equation}
 Fig.~\ref{fig:mu-q} shows the quasi-energies $\omega_\pm(q)$ as a function of the quasi-momentum $q$. 
\begin{figure}[H]
    \centering
    \includegraphics[width=0.65\textwidth]{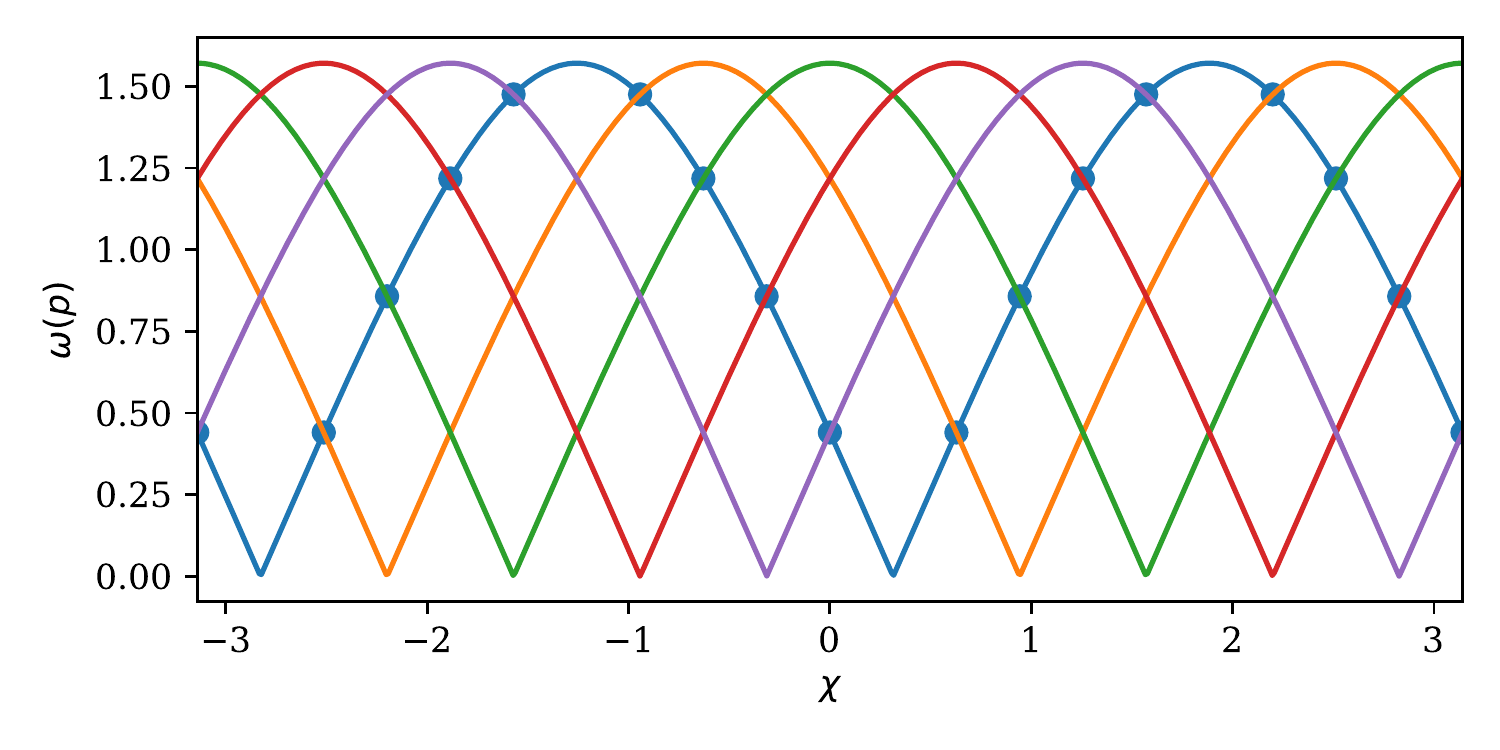}
    \caption{\textbf{Level crossings in the ring spectrum} Quasi-energies $\omega_+$ of the reference   circuit on a ring of qubits (\ref{eq:ref}), (\ref{eq:Uref}) {\it vs} flux $\chi$. The plots correspond to $N=10$.   The solid points show one set of level-crossings $\chi=\chi_{0j}$ as predicted by Eq.~\ref{eq:level_cr}}
    \label{fig:ring_quasi_spectrum}
\end{figure}

Fig.~\ref{fig:ring_quasi_spectrum} shows the positive quasi-energies $\omega_+$ as a function of $\chi$. 
Based on Eq.~\ref{eq:ring_omega}, level-crossings are expected between $\omega_{+}(q_i)$ and $\omega_{+}(q_j)$ for the values of $\chi = \chi_{ij}$ given below
\begin{equation}
    \chi_{ij} = \frac{q_i}{2} - \left(n + \frac{\kappa}{2}\right) \pi + \frac{q_j - q_i}{4}, \: n \in \mathbb{I}, \: \kappa \in \{0,1\}\;,\label{eq:level_cr}
\end{equation}
\begin{equation}
    \omega_\nu(q_i)=\omega_\nu(q_j)\quad {\rm for}\quad \chi=\chi_{ij}\;.\label{eq:doublet}
\end{equation}
Fig.~\ref{fig:ring_quasi_spectrum} shows the level crossings between $\omega_{+}(q_0)$ and $\omega_{+}(q_j)$ for all $j$. The expression for $\chi_{ij}$ is given below as a function of $q_i$ and $q_j$

\subsection{Circuit with small disorder in gate parameters}
\label{sub_sec:small_disorder}
We now consider a  circuit unitary $U_{\rm cycle}(\,\vec{\chi},\vec{\zeta},\vec{\theta}\,)$ for a  $N$-qubit ring whose parameters are  sufficiently  close to those in a reference circuit $U_{\rm cycle}(\chi_{i\,j})$. 
The  degeneracy of the quasi-energy levels  (\ref{eq:level_cr}),(\ref{eq:doublet}) corresponding to the single-excitation spectrum of the  reference circuit  is lifted when the average flux per gate deviates from its value $\chi=\chi_{i\,j}$ and the  rest of the circuit parameters distributed non-uniformly along the qubit chain around their reference values (\ref{eq:ref}).

We study the splitting of the quasi-energy levels in a doublet corresponding to one of the values of $q=q_m>0$ given by  Eqs.~(\ref{eq:qodd}) or (\ref{eq:qeven}). The solution of the   eigenproblem
\begin{equation}
U_{\rm cycle}(\,\vec{\chi},\vec{\zeta},\vec{\theta}\,)\ket{ \psi^{\alpha}}=e^{-i \omega_\alpha}\ket{\psi^{\alpha}},
\label{eq:eigUd}
\end{equation}
depends on the gate parameters $\zeta_j,\chi_j,\theta_j$ via the following 3  quantities
     \begin{equation}
\delta _{\zeta }(q)=\frac{1}{N}\sum _{k=1}^N \zeta _ke^{- i k\,q},\hspace{8 mm}\delta _{\theta }(q)=\frac{1}{N}\sum _{k=1}^N \left(\theta
_k-\pi /4\right)e^{- i k\,q} ,\hspace{8 mm}\Bar{\chi} =\frac{1}{N}\sum _{k=1}^N \chi _k\;.\label{eq:fur}
\end{equation} 
Here $\Bar{\chi}$ is an average flux per gate discussed above (\ref{eq:chibar}) and $\delta _{\zeta }(q)$, $\delta _{\theta }(q)$  are Fourier  transforms of the disorder in $\zeta_j$ and $\theta_j$.

When the disorder parameters are much smaller than the separation $4\pi |\omega^\prime(q)|/N$ between neighbouring quasi-energies of the reference circuit with different values of $|q|$ (cf. (\ref{eq:ring_omega}))
\begin{equation}
  |\delta _{\zeta }(q)|,\,|\delta _{\theta }(q)|,\,|\Bar{\chi}|\ll \frac{2\pi  }{N}\frac{ \sin (q)}{\sqrt{1-\sin (q)^4}}\;,\label{eq:cond}
\end{equation}
the solution of matrix eigenvalue problem (\ref{eq:eigUd}) 
can be obtained using the degenerate perturbation theory of quantum mechanics.
A pair of   zeroth-order eigenstates $V_{\gamma }^{(\nu )}(\pm q)e^{\pm i q m}$ of the reference circuit unitary form two linear superpositions
\begin{equation}
\psi_{\gamma , m}^{\pm,\nu ,q}=\frac{1}{\sqrt{N/2}}\sum _{\sigma =\pm } u_{\sigma }^{\pm}(\nu,q) V_{\gamma }^{(\nu )}(\sigma  q)\hspace{2 mm}e^{i \sigma q m}\;,\label{eq:psi}
\end{equation}  
each corresponding to the eigenstate $\ket{\psi^{f,\nu ,q}}$ of the cycle unitary 
\begin{equation}
    U_{\rm cycle}(\,\vec{\chi},\vec{\zeta},\vec{\theta})\,\ket{\psi^{f,\nu ,q}}=e^{-i \omega_\nu^f(q)}\,\ket{\psi^{f,\nu ,q}},\quad f=\pm 1\;.\label{eq:fnuq}
\end{equation}
Therefore in the limit of weak disorder (\ref{eq:cond}) the eigenstates of $\Uc$ are defined by a triple of  quantum numbers $\alpha=(f,\nu,q)$. The coefficients $u_{\pm}^{f}(\nu,q)$ above  form  spinors that are  eigenstates of the   $2\times2$ "Floquet Hamiltonian" $h(\nu,q)$
 \begin{equation}
     h(\nu,q) 
\left(
\begin{array}{c}
 u_{+}^{f}(\nu,q) \\
 u_{-}^{f}(\nu,q)
\end{array}
\right)=\delta\omega_\nu^f(q) \left(
\begin{array}{c}
u_{+}^{f}(\nu,q) \\
 u_{-}^{f}(\nu,q)
\end{array}
\right),\quad f=\pm 1\;.\label{eq:eigF}
\end{equation} 
 where 
  \begin{equation}
\begin{split}
\begin{gathered}
h=\nu\,  B  \left(
\begin{array}{cc}
 1 & 0 \\
 0 & 1 \\
\end{array}
\right)+\nu\, \delta\omega \left(
\begin{array}{cc}
 \sin  \gamma  &  e^{i \delta } \text{cos$\gamma $} \\
  e^{-i \delta }\cos  \gamma  & - \sin  \gamma  \\
\end{array}
\right)\;.\label{eq:h}
\end{gathered}
\end{split}
\end{equation}
(above  we omitted the argument $(\nu,q)$ for brevity). The expressions for the eigenstates are
 \begin{equation}
\begin{split}
\begin{gathered}
\hspace{8 mm}\left(
\begin{array}{c}
 u_+^{f }(\nu ,q) \\
 u_-^{f }(\nu ,q) \\
\end{array}
\right)=\left(
\begin{array}{c}
 f  e^{-i \delta } \sin \left(\frac{\pi }{4}+f \frac{\gamma }{2}\right) \\
 \sin \left(\frac{\pi }{4}-f \frac{\gamma }{2}\right) \\
\end{array}
\right),\hspace{4 mm}f =\pm 1\;\label{eq:uupm}
\end{gathered}
\end{split}
\end{equation}

The parameters $B\equiv B(q)$ and $\delta\omega\equiv \delta\Omega(\nu,q)$ determine the disorder-induced uniform shift and splitting, respectively,  of the  quasi-energy levels in a doublet relative  to their unperturbed value $\nu\Omega(q)$
 \begin{equation}
     \omega_\nu^\pm(q)=\nu\Omega(q)+\nu B(q)\pm \nu \delta\Omega(\nu,q)\;.\label{eq:muDis}
     \end{equation}
where
  \begin{equation}
B(q) =\frac{\cos (q/2)}{\sqrt{1+\sin (q/2)^2}}\delta _{\theta }(0)
\end{equation}\newline 
 and 
  \begin{equation}
\delta\Omega(\nu ,q)=\left(\frac{4\hspace{2 mm}\bar{\chi }^2}{1+\sin \left(\frac{q}{2}\right)^{-2}}\hspace{2 mm}+| \xi (\nu,q)| ^2\right)^{1/2}\;.\label{eq:A}
\end{equation}
The parameter $\xi(\nu,q)$ equals
 \begin{equation}
\xi (\nu ,q)= e^{-\frac{3i q}{2}}\left(\delta _{\theta }(q)-2i \nu  \frac{\sin (q/2)}{\sqrt{1+\sin (q/2)^2}} \delta _{\zeta }(q)\right)\,\label{eq:xi}
\end{equation}
 The angles $\gamma,\,\delta$ equal
  \begin{equation}
\gamma (\nu ,q)=\arcsin \left(\frac{2 \bar{\chi }}{\delta\Omega(\nu ,q) \sqrt{1+\sin (q/2)^{-2}}}\right)\;,\label{eq:gamma}
\end{equation}
 \begin{equation}
\delta (\nu ,q)=\arg  \xi (\nu ,q)\;.\label{eq:delta}
\end{equation}
 
We note that the  doublet levels $\omega_{\nu}^{\pm}(q)$ varying with the average flux $\Bar{\chi}$ undergo avoided crossing at $\Bar{\chi}=0$. The level splitting at the avoided crossing is $2|\xi(\nu,q)|$. It is of interest to consider the case where  $\Bar{\chi}$   is a sum of the flux $\chi$ from the reference circuit and some systematic errors 
\begin{equation}
    \chi_j=\chi+\delta\chi_j\;.
    \end{equation}
    
 The value of $\chi=\chi^{(c)}$ corresponding to  the avoided crossing point is
    \begin{equation}
    \chi^{(c)}=-\sum_{j=1}^{N}\delta\chi_j\;,\label{eq:chiC}
    \end{equation}
For $N\gg1$ and in the case where $\delta\chi_j$ are zero-mean i.i.d. random numbers
$|\chi^{(c)}|\sim N^{1/2}$ is much greater than the typical value of   $|\chi_j|$.
 
\subsection{Measuring the spectrum of quasi-energies}
\label{sec:measure-q}

The most direct way to obtain the quasi-energies in a single-excitation spectrum of the cycle unitary $\Uc$  is to measure for a given qubit $m$ the spectral decomposition of the expectation value $\langle \sigma^{+}_{m}(t)\rangle$ dependence on  the cycle number $d$.  We define  the basis states  with 0 and 1 excitation in terms of computational basis states for $N$-qubit system \begin{equation}
    \ket{j}=\ket{0_10_2\ldots1_j\ldots0_N},\quad \ket{0}=\ket{0_10_2\ldots 0_N},\quad j\in(1,N)\;. \label{eq:basis1}
    \end{equation}
In this basis the Pauli matrices have the form 
\begin{equation}\sigma_m^z = \ket{0}\bra{0} + \sum_{k \neq m} \ket{k}\bra{k} - \ket{m}\bra{m}, \quad \sigma_m^- = \ket{m}\bra{0},\quad   \sigma_m^+ = \ket{0}\bra{m}\;,\label{eq:sigma-D}\end{equation}
we note that because a single qubit state $\ket{0_m} \equiv \begin{pmatrix}1 \\0 \end{pmatrix}_m$
 corresponds to a spin-1/2 state  $\ket{\uparrow_m}$ at the site $m$ the operator $\sigma_m^-$  creates the excitation  and $\sigma_m^+$ annihilates the excitation at that site.

We start from the vacuum state $\ket{0}$ and apply $\pi/2$ pulse $\exp(i \pi\sigma_r^x/4)$  to the $r$th qubit, $r\in (1,N)$ 
\begin{equation}
  \ket{\Phi_1}\equiv e^{i \pi\sigma_r^x/4}  \ket{0}=\frac{\ket{r}+\ket{0}}{\sqrt{2}}\;.\label{eq:Phi1}
\end{equation}
We then apply periodic circuit $\Ucd$ to the state $\ket{\Phi_1}$ and obtain the quantum state after $d$ steps
 \begin{equation}
|\Phi_{1+d}\rangle =\Ucd|\Phi_1\rangle =\frac{\Ucd|r\rangle +|0\rangle }{\sqrt{2}}\;,\label{eq:Phid}
\end{equation}
 where we used $\Uc\ket{0}=\ket{0}$.  The expectation value of the operator $\sigma_r^\dagger$ equals
  \begin{equation}
\langle \Phi_{1+d} | \sigma _r^+|\Phi_{ 1+d}\rangle =\frac{1}{2}\sum _{\alpha =1}^{N} e^{-i d\, \omega _{\alpha
}}|\left\langle r\left|\psi ^{\alpha }\right.\right\rangle|^2 \;.\label{eq:cmd}
\end{equation}
The spectral function
 $\sum_{d=0}^{d_{\rm max}-1}\langle \Phi_{1+d} | \sigma _r^+|\Phi_{ 1+d}\rangle\, e^{-2\pi i k d/d_{\rm max}}$ (\,$k=0,\ldots,d_{\rm max}-1$)  gives the quasi-energy spectrum $\{\omega_\alpha\}$.

\section{Persistent current for periodic circuit on a qubit ring \label{sec:spin_cur}}
The gauge field $\phi$ (\ref{eq:B}) corresponds to the total flux through the ring over one  cycle. It determines the integral of the  persistent current in the ring over the cycle duration.
To obtain the persistent current we write a time-periodic control  Hamiltonian $H(t)$  that acts on qubit system
   \begin{equation}
H(t)=\sum _{m=1}^N \epsilon _m(t)\sigma _m^z +\sum _{m=1}^N g_m(t)\left(\sigma_m^{+}\sigma _{m+1}^{-} e^{i \vartheta}+\sigma _{m+1}^{-}\sigma _m^{+}e^{-i \vartheta}\right),\hspace{6 mm}\sigma^{k} _{N+1}\equiv \sigma^{k} _N\;.\label{eq:LH}
\end{equation} 
where $\vartheta$ is a fiducius  twist angle that will be eventually set to zero.

 The quantum circuit $\Ucd$ (\ref{eq:Ud}) is defined by specifying the  time-dependence  of the frequency detunings between qubits $\epsilon_m(t)$ and coupling coefficients $g_m(t)$. They are periodic in time 
 \begin{equation}
     H(t)=H(t+t_{\rm cycle})\;,
 \end{equation}
 with the period $t_{\rm cycle}$  equal to the physical duration of the cycle of gates $\Uc$ (\ref{eq:Ud}). From (\ref{eq:LH})  the current operator is equal to
 \begin{equation}
      {\mathcal J}(t)=\frac{\partial H(t)}{\partial\vartheta}\;.\label{eq:calJ}
 \end{equation}
 
 In a Heisenberg picture the current operator equals $U^\dagger(t,0){\mathcal J}(t) U(t,0)$
 where $U(t,0)$ is the quantum propagator
  \begin{equation}
U(t,0)=T e^{-i \int_0^t H(s,\vartheta ) \, ds}\;.\label{eq:U0t}
\end{equation}
 One can obtain the expression for the integral of the current  over the interval of time $(0,t)$
  \begin{equation}
J(t)\equiv \int _0^tU^{\dagger }(s,0)\mathcal{J}(s)U(s,0)ds=i U^{\dagger }(t,0) \frac{\partial U(t,0)}{\partial \vartheta }\;,\label{eq:UJU}
\end{equation}
 
In the  cycle unitary $U(\tc,0)\equiv \Uc$ the twist angle $\vartheta$ shifts the  values of the Peierls phases (cf. (\ref{eqn:gate_unitary}))
\[\chi_k\rightarrow \chi_k+\vartheta,\quad k=1:N\;.\]
 Therefore  we have
 \begin{equation}
     \frac{\partial \Uc}{\partial \vartheta}=\sum_{k=1}^{N} \frac{\partial \Uc}{\partial \chi_k}
 \end{equation}
In Eq.~(\ref{eq:UJU})  we set the fiducius  twist angle to zero $\vartheta=0$ and for $t=d \tc$   obtain for the integral of the spin current over $d$ cycles
  \begin{equation}
J(d\,\tc)=i (U_{\rm cycle}^{d})^{\dagger} \sum_{k=1}^{N}\frac{\partial U_{\rm cycle}^{d}}{\partial \chi_k }\;\label{eq:Jc}
\end{equation}\newline 
 
 Using the expression (\ref{eq:Ud}) for  $\Uc$ in terms of the  product of the
 gate unitaries we obtain 
  \begin{equation}
J(\tc)=\sum _{k=1}^{N/2} \left(J_{2k-1}+ J_{2k}^{\prime } \right),\hspace{11 mm}J_{2k}^{\prime }=U_{2k+1}^{\dagger }U_{2k-1}^{\dagger
} J_{2k}U_{2k+1}U_{2k-1}\;,\label{eq:Jc1}
\end{equation} 
 where $J_k$ \begin{equation}
J_k=i U_k^{\dagger }\frac{\partial U_k}{\partial  \chi_k }\,\label{eq:Jm}
\end{equation}\newline 
 corresponds to the spin current density operator for the magnetization transport from the site  $k$ to $k+1$ over the duration of the gate $U_k$ (\ref{eqn:gate_unitary}). In the single excitation subspace it has the form
 \begin{equation}
 J_k=i\left( |k+1\rangle \langle k|e^{i \left(\chi _k+\zeta _k\right)}- |k\rangle  \langle k+1| e^{-i \left(\chi _k+\zeta _k\right)}\right)\sin\theta_k\sin\theta_{k+1}+ (|k+1\rangle  \langle k+1|- |k\rangle \langle k|)\sin ^2\left(\theta _k\right) 
\end{equation}
 The first term corresponds to the standard  form of the  spin operator while the second term $\propto (|m+1\rangle  \langle m+1|- |m\rangle \langle m|)$ is due to the two-layered form of the cycle of gates. 
 We note in passing that the local spin currents in (\ref{eq:Jc1}) and (\ref{eq:Jm}) obey the two continuity equations, separately for odd and even  sites that was obtained in \cite{PhysRevLett.122.150605} by a different method.
 
 \subsection{Average persistent current}

 We use the expression for the  circuit unitary (\ref{eq:eigen}) in terms of Floquet eigenstates and eigenvalues of the cycle unitary to obtain the operator for a time-average of the  persistent current $J(t)/t$ (\ref{eq:UJU})  over the duration of the quantum circuit $U(d)$ (\ref{eq:Ud}) with $d$ cycles, $t=d\tc$ 
  \begin{equation}
\frac{J\left(d \,t_{\text{cycle}}\right)}{d\, t_{\text{cycle}}}=\sum_{k=1}^{N}\sum _{\alpha =1}^n \frac{\partial \omega _{\alpha }}{\partial \chi_k }|\psi ^{\alpha }\rangle
\langle \psi ^{\alpha }|+\frac{1}{d\, t_{\text{cycle}}} \sum _{\alpha =1}^N \sum _{\gamma \neq  \alpha } e^{-i d t_{\text{cycle}} \left.\left(\omega _{\alpha
}-\omega _{\gamma }\right)\right/2}F_{\alpha \gamma }(d)|\psi ^{\gamma }\rangle \langle \psi ^{\alpha }|\;,\label{eq:Jav}
\end{equation}\newline 
where
 \begin{equation*}
F_{\alpha \gamma }(d)=2\pi  i \delta _d\left(\omega _{\alpha }-\omega _{\gamma }\right)e^{i\left.\left(\omega _{\alpha }+\omega _{\gamma }\right)\right/2}\sum_{k=1}^{N}\langle
\psi ^{\gamma }| \frac{dU_{\text{cycle}}}{d \chi_k }|\psi ^{\alpha }\rangle
\end{equation*}\newline 
 and
  \begin{equation}
\delta _d(x)=\frac{1}{2\pi }\sum _{l=-(d-1)/2}^{(d-1)/2} e^{i x l}\hspace{4 mm}=\frac{1}{2\pi }\frac{\sin (d x/2)}{\sin (x/2)}\;\label{eq:dd}
\end{equation}\newline 
(it is assumed that $d$ is an odd integer). In the limit of large number of cycles
\begin{equation}
    d\gg \Delta\omega^{-1},\quad \Delta\omega\equiv \min_{\alpha\neq\beta\in(1,N)}(\omega_\alpha-\omega_\beta)\;,\label{eq:dlim}
\end{equation}
the expression (\ref{eq:Jav}) for the  time-averaged persistent current operator is dominated by the first term. Because eigenvalues  of the cycle unitary $\omega_\alpha$ depend on angles $\chi_k$ only via the total flux $\phi$ (\ref{eq:Ufr}) given in (\ref{eq:B}). Then to the leading order in $1/(d\Delta\omega) $ we have
 \begin{equation}
\frac{J\left(d\, t_{\text{cycle}}\right)}{d\, t_{\text{cycle}}}\simeq  \sum _{\alpha =1}^n \frac{\partial \omega _{\alpha }}{\partial \phi }|\psi ^{\alpha
}\rangle \langle \psi ^{\alpha }|+\mathcal{O}\left(\frac{1}{d \,\Delta \omega }\right)\;.\label{eq:Javd}
\end{equation}\newline 
 
One can also derive the  expression for the Drude weight $D$ in terms of the quasi-energy level curvatures following the approach similar to that presented above
\begin{equation}
    D=\lim_{N\rightarrow \infty}N\sum_{\alpha=1}^{N} p_\alpha \frac{\partial^2 \omega_\alpha}{\partial \phi^2}
\end{equation}
where $p_\alpha=\bra{\psi^\alpha}\rho(0)\ket{\psi^\alpha}$ is the population of the Floquet state $\ket{\psi^\alpha}$. This expression for $D$  is a generalization of the Kohn's formula \cite{PhysRev.133.A171} for the case of periodically driven systems.

\section{ Periodic Circuit on a Qubit Ring: Open System Dynamics}\label{sec:open_sys-ring}

 Precise determination of the quasi-energies requires studying the periodic circuits $\Ucd$ of a large depth when  the environmental effects must be included. 
 As will be shown in the next section the non-Markovian effects of the low frequency noise and parameter drift can be neglected on the time scale of the experiment. 
Therefore we  will model open system dynamics  with Lindblad master equation
\begin{equation}\label{eqn:general_linbald}
    \frac{d \rho(t)}{dt} = -i \left[ H(t), \rho(t)\right] + L[\rho(t)]\;,
\end{equation}
where the Lindbladian operator has the form   \cite{breuer2002theory}
\begin{equation}\label{eqn:dephase_photon}
    L[\rho] = \sum_{m=1}^{n} \frac{\Gamma_{2\phi}^{m}}{2} \left( \sigma_m^z \rho \sigma_m^z - \rho \right) + \Gamma_1^m \left( \sigma_m^- \rho \sigma_m^+ - \frac{1}{2} \left( \sigma_m^+  \sigma_m^- \rho + \rho \sigma_m^+ \sigma_m^-\right)\right)
\end{equation}
Here $\Gamma_1^m, \Gamma_{2\phi}^{m}$ are decay and (intrinsic) dephasing rates for individual qubits. In (\ref{eqn:general_linbald}) $H(t)$ is time-periodic control  Hamiltonian that acts on qubit system. In its simplest form $H(t)$ is equal to
   \begin{equation}
H(t)=\sum _{m=1}^N \epsilon _m(t)\sigma _m^z +\sum _{m=1}^N g_m(t)\left(\sigma_m^{+}\sigma _{m+1}^{-}+\sigma _{m+1}^{-}\sigma _m^{+}\right),\hspace{6 mm}\sigma^{k} _{N+1}\equiv \sigma^{k} _N\;.\label{eq:H}
\end{equation} 
 The quantum circuit (\ref{eq:Ud}) is defined by specifying the  time-dependence  of the frequency detunings between qubits $\epsilon_m(t)$ and coupling coefficients $g_m(t)$. They are periodic in time 
 \begin{equation}
     H(t)=H(t+t_{\rm cycle})\;,
 \end{equation}
 with the period $t_{\rm cycle}$  equal to the physical duration of the cycle of gates $\Uc$ (\ref{eq:Ud}). The  corresponding quantum propagator can be expanded in the basis of the Floquet eigenstates of the Hamiltonian $H(t)$ 
   \begin{equation}
U(t,0)=T \exp \left(-i \int_0^t H(t ) \, dt \right)=\sum _{\alpha} |\psi^\alpha(t)\rangle \langle \psi^\alpha(0)| e^{-i \lambda _\alpha t}\;,\label{eq:Ut0}
\end{equation} 
where  
\begin{equation}
|\psi^\alpha(t)\rangle =|\psi^\alpha\left(t+t_{\text{\rm cycle}}\right)\rangle\;.\label{eq:Floq-u}
\end{equation}
At  time intervals $t= d\, t_{\rm cycle}$ that are  integers of the cycle duration  the propagator equals to the $d$th power of the cycle unitary considered above (\ref{eq:Ud})
\begin{equation}
    U( d\, t_{\rm cycle},0)=\Ucd,\quad d=0,1,2,\ldots\;.
\end{equation}
This defines the  connection between the  eigenstates and eigenvalues of 
 $U(t,0)$ and those of  $\Uc$  (\ref{eq:eigUd}),(\ref{eq:psi})
\begin{equation}\ket{\psi^\alpha(d\, t_{\rm cycle})}= \ket{\psi^\alpha},\quad \omega_\alpha=\lambda_\alpha t_{\rm cycle}\;.\end{equation} 
We move into the interaction picture
 \begin{equation}
\tilde{\rho }(t)=U^{\dagger }(t,0) \rho (t) U(t,0)\;,\label{eq:rhoI}
\end{equation} 
 and consider the density matrix $\tilde\rho_{\alpha\beta}(t)$
 in the Floquet basis 
  \begin{equation}
\hspace{2 mm}\tilde{\rho }_{\alpha \beta }(t)\equiv \langle \psi^{\alpha }| \tilde{\rho }(t)|\psi^{\beta }\rangle =e^{i \left(\lambda
_{\alpha }-\lambda _{\beta }\right)t} \langle \psi^{\alpha }(t)| \rho (t)|\psi^{\beta }(t)\rangle\,.
\end{equation}
 In the subspaces with 0 and 1 excitations  the Lindblad equation for $\tilde{\rho }_{\alpha \beta }(t)$ has the form
 \begin{equation}
\begin{split}
\begin{gathered}
\frac{\partial  \tilde{\rho }_{\alpha \delta }(t)}{\partial t}=\sum _{\beta , \gamma  =0}^N R_{\alpha \beta \gamma \delta }(t) \tilde{\rho }_{\beta
\gamma }(t)\\
R_{\alpha \beta \gamma \delta }(t) =\sum _{m=1}^N \Gamma_{2\phi}^{m}\left(2U_{m\alpha }^*(t)U_{m\beta }(t)U_{m\gamma }^*(t)U_{m\delta }(t)-\delta _{\alpha
,\beta }U_{m\gamma }^*(t)U_{m\delta }(t)-\delta _{\gamma ,\delta }U_{m\alpha }^*(t)U_{m\beta }(t)\right)\\
+\sum _{m=1}^N \Gamma _1^m\left(\delta _{\alpha ,0} \delta _{\delta ,0}U_{m\beta }(t) U_{m\gamma }^*(t)-\frac{1}{2}\left(U_{m\alpha }^*(t)U_{m\beta
}(t)\delta _{\gamma ,\delta }+\delta _{\alpha ,\beta }U_{m\gamma }^*(t)U_{m\delta }(t)\right)\right)\;,\label{eq:RI}
\end{gathered}
\end{split}
\end{equation}
 where $\delta_{\gamma,\delta}$ is a Kronecker delta, the Floquet state  $\ket{\psi^0}=\ket{0}$ corresponds to the vacuum with the quasi-energy $\lambda_0=0$ and \begin{equation}
U_{m\beta }(t)\equiv \langle \psi^m| U(t,0)|\psi ^{\beta }\rangle=U_{m\beta}(t-d(t)\tc) \,e^{-i d(t)\,\omega_\alpha},\qquad
d(t)\equiv\left \lfloor t/\tc \right\rfloor =0,1,\ldots\;.\label{eq:Umb}
\end{equation}
Here  $d(t)$ is the number of  cycles elapsed before the moment $t$ and $U_{m\beta}(t-d(t)\tc)$ is  a periodic function of time with period $\tc$. 
 
For the technique  to estimate the values of  quasi-energies from experimental data we will follow the same approach as described in Sec.~\ref{sec:measure-q}.
To compare directly with the Eq.~(\ref{eq:cmd}) we write the expectation value of the operator $\sigma_{r}^{\dagger}(d\tc)$ after $d$ cycles starting from the pure initial state 
$(\ket{0}+\ket{r})/\sqrt{2}$ (\ref{eq:Phi1})
\begin{equation}
\text{Tr}\left[\sigma _r^{\dagger }\rho (d\,\tc)\right]=\sum _{\alpha =1}^n e^{-i d \omega _{\alpha }}\tilde{\rho }_{\alpha 0}(d\,\tc)\,\left\langle
r\left|\psi ^{\alpha }\right.\right\rangle\;.\label{eq:cmd0}
\end{equation} 
 It is expressed in terms of the off-diagonal matrix elements $\tilde{\rho }_{\alpha 0 }(t)$ that connects vacuum state to a Floquet  state in a single excitation subspace, $\alpha =1,2,\ldots ,N$. 
For those matrix elements the Eq.~(\ref{eq:RI}) takes the form
\begin{equation}
\frac{\partial  \tilde{\rho }_{\alpha 0}(\tau )}{\partial \tau }=-\sum _{\beta  =1}^N \sum _{m=1}^N \left(\Gamma_{2\phi}^{m}+\frac{\Gamma _1^m}{2}\right)U_{m\alpha
}^*(t)U_{m\beta }(t)\tilde{\rho }_{\beta 0}(t),\quad
\tilde{\rho }_{\alpha 0}(0)=\frac{1}{2}\braket{\psi^\alpha}{r}\;.
\label{eq:rho0alpha}
\end{equation}

\subsection{\label{sec:Secular-O}Secular approximation}

 Dominant matrix elements in $\tilde\rho(t)$ are changing on the time scale corresponding to the typical   dephasing  and  decay  times ($1/\Gamma_2$ and $1/\Gamma_1$, respectively) 
that is  much greater   than the cycle duration. 
 Therefore we can coarse-grain Eq.~(\ref{eq:rho0alpha}) over the time  $\Delta\,t$, such that
 $\Gamma_{2\phi},1/\Gamma_1\gg \Delta\,t \gg t_{\rm cycle}$.  After the coarse-graining the sum 
 in the right hand side of the Eq.~(\ref{eq:rho0alpha})  only contains terms where
 the separation between the quasi-energies is sufficiently large, $|\omega_\alpha-\omega_\beta|\lesssim \Gamma_{1,2} \tc$.
   This corresponds to a secular approximation \cite{breuer2002theory}.

 We note that for not too long qubit chains the decay and dephasing rates are much smaller then the separation between the   quasi-energies of the  reference circuit with  nearest  values of  quasi-momenta $|q_{m+1}-q_{m}|\sim\frac{2\pi}{N}$ 
 \begin{equation}
\Gamma_1,\,\Gamma_{2\phi} \ll|\omega_\nu(q_{m+1})-\omega_\nu(q_{m}|\simeq \frac{2\pi}{N}\frac{\sin(q)}{\sqrt{1-\sin(q)^2}},\quad
\quad {\rm for}\quad  |q|\simeq\mathcal{O}(1)\;.\label{eq:condG}
\end{equation}
Therefore the above secular approximation is applicable here and the corresponding  Floquet eigenstates  are not coupled in  Eq.~(\ref{eq:rho0alpha}) after coarse-graining. However  the  quasi-energy splittings $|\omega_\nu^+(q)-\omega_\nu^-(q)|$ for the Floquet states  $\ket{\psi^{\pm,\nu,|q|}}$ (\ref{eq:psi}) that are superpositions of planes waves with the same value of $|q|$ are   limited by disorder and not by $2\pi/N$. 
Therefore for sufficiently small disorder  the condition $|\omega_\nu^+(q)-\omega_\nu^-(q)|\gg \Gamma_{1,2}\tc$ can be violated in which case the Eq.~(\ref{eq:rho0alpha}) couples the states $\ket{\psi^{\pm,\nu,|q|}}$ and the secular approximation breaks down for these transitions. This regime will be referred to a future study. Here we focus on the case
\begin{equation}
    \Gamma_1,\Gamma_{2\phi}\ll  |\delta_\zeta(q)|, |\delta_\theta(q)|,|\Bar{\chi}|\lesssim\frac{2\pi}{N}\;.\label{eq:cond2}
\end{equation}

Under the condition given in (\ref{eq:cond2})  the secular approximation (\ref{eq:wp})-(\ref{eq:Walpha}) applies for all transitions $\ket{0}-\ket{\psi^{\alpha}}$ ($\alpha=1:N$). Then the expressions  for the elements of the density matrix in  Schrodinger picture $\rho_{\alpha0}(d\,\tc)$ and the observable  (\ref{eq:cmd0}) have the form
\begin{align}
 \langle \psi ^{\alpha}|\rho (d\,\tc)|0\rangle=\frac{1}{2}\braket{\psi^\alpha}{r}\,e^{-i d\, \omega_\alpha -d \,W_\alpha}\;,\label{eq:wp}\\
\langle\sigma_r^+(d)\rangle=\frac{1}{2}\sum _{\alpha =1}^n e^{-i d \omega _{\alpha }-d\,W_\alpha}\,\left |\left\langle
r\left|\psi ^{\alpha }\right.\right\rangle\right|^2\;.\label{eq:cmd1}
\end{align}
 where
 \begin{equation}
W_\alpha=\sum _{m=1}^N\left(\Gamma _2^{m}+\frac{\Gamma _1^{m}}{2}\right)\overline{|U_{m,\alpha}(s)|^2}\;. \label{eq:Walpha}
\end{equation}
Here dephasing rate $W_\alpha$ corresponds to the eigenstate $\ket{\psi^\alpha}$. The horizontal bar above denotes averaging $t_{\text{cycle}}^{-1}\int _0^{t_{\text{cycle}}}\cdots ds$.


\subsubsection{Limit of small disorder}
In the case where disorder is small compare to level separation for different values of $|q|$ (\ref{eq:cond}) we can  use the results for the reference circuit in Eqs.~(\ref{eq:cmd1}) and (\ref{eq:Walpha}). In particular, we use Eq.~(\ref{eq:psi}) for the Floquet eigenstates $\psi_{\gamma , m}^{\pm,\nu ,q}$ 
and Eq.~(\ref{eq:muDis}) for the quasi-energies  $\omega_\nu^\pm(q)$. Then the dephasing rate 
(\ref{eq:Walpha}) corresponding to an eigenstate $\ket{\psi^{\pm,\nu,q}}$ equals
  \begin{equation}
W_{\nu }^{\pm}(q)=\sum _{l=1}^N\sum_{\gamma=1,2} \left(\Gamma _2^{\gamma , l}+\frac{\Gamma _1^{\gamma ,l}}{2}\right)\overline{|U_{\gamma,l}^{\pm,\nu,q}(s)|^2},\qquad U_{\gamma,l}^{\pm,\nu,q}(s)\equiv \langle \gamma ,l| U(s,0)|\psi
^{\pm, \nu ,q}\rangle\;. \label{eq:Wnu}
\end{equation}
 Here   $\Gamma _{2\phi}^{\gamma ,l}+\Gamma _1^{\gamma ,l}/2$  is a total dephasing rate  for a qubit at the position $m=2(l-1)+\gamma$ on a ring and 
 \begin{align}
U_{\gamma,l}^{\pm,\nu,q}(s) =\left(\frac{2}{N}\right)^{1/2} \sum _{\sigma =\pm } \sum _{\lambda =1,2}
&\left(\mathcal{U}^{3-\gamma , 1}(s)\mathcal{U}^{2, \lambda }e^{-i \frac{\sigma  q}{2}} 
+  \mathcal{U}^{3-\gamma , 2}(s)\mathcal{U}^{1,
\lambda } e^{i \frac{\sigma  q}{2}}\right)\nonumber\\
&\times V_{\lambda }^{(\nu )}(\sigma  q)u_{\sigma }^f(\nu ,q)e^{i (-1)^{\gamma } \frac{\sigma  q}{2}}e^{i \sigma q l}\;.\label{eq:Ugl}
\end{align}

Here ${\mathcal U}_{i,j}(s)$ are time-dependent matrix elements of the 
unitary ${\mathcal U}(s)=T\exp(-\int_{0}^{s}H_{\rm gate}(s')ds')$, with $s\in(0,\tc)$.   The  gate Hamiltonian $H_{\rm gate}(t)$ applies to a given pair of qubits and  implements the  gate unitary ${\mathcal U}$ (\ref{eq:Uf}) where \({\mathcal U}_{\lambda , \lambda }=i {\mathcal U}_{\lambda , 3-\lambda }=2^{-1/2 }\text{
    }(\lambda =1,2)\). The gate Hamiltonian is part of the system control Hamiltonian (\ref{eq:H}). In the case of  the reference circuit control pulses  $\epsilon_j(t)=\epsilon(t)$, $g_j(t)=g(t)$) are identical for all qubits.   
    
    The coefficients $\overline{|U_{\gamma,l}^{\pm,\nu,q}(s)|^2}$ in (\ref{eq:Wnu}) depend on the  elements of the tensor 
\begin{equation}
w_{ijkl}=\overline{{\mathcal U}_{i,j}^{*}(s ){\mathcal U}_{k,l}(s )}\;.\label{eq:cijkl}
\end{equation}
    The coefficients  $w_{ijkl}$ and therefore  the density matrix $\wp _{f\nu q}(t)$
 depend on the shape of the control pulses and not  only on the parameters of the logical gate unitary. This is a difference from the closed quantum   system evolution where final sate depends only on the logical circuit.
  This happens because the processes
 of decoherence  and decay are continuous in time. The coefficients $c_{ijkl}$  are not all independent from each other due to the 
 unitary constrains. One can show that coefficients $\Lambda$ depend only on 
 two parameters
 \begin{equation}
w_{1111}=\frac{1}{t_{\text{cycle}}}\int_0^{t_{\text{cycle}}} \left| \mathcal U_{1,1}(\tau )\right| {}^2 \, d\tau ,\hspace{4 mm}w_{1112}=\frac{1}{t_{\text{cycle}}}\int
_0^{t_{\text{cycle}}}{\mathcal U}_{1,1}^*(\tau ){\mathcal U}_{1,2}(\tau )d\tau\;.\label{eq:cI}
\end{equation}
Because of the  constraint $|{\mathcal U}_{1,1}(\tau )|^2+|{\mathcal U}_{1,2}(\tau )|^2=1$ there are in total 3 real-valued pulse-dependent parameters. 

\subsection{Case of large  rings $\mathbf{ N\gg1}$ \label{sec:uniformG} }

 Here we make an interesting observation. Under the  assumption that dephasing  and decay rates  are the same for all qubits,
\begin{equation}
    \Gamma_1^m\equiv \Gamma_1,\quad  \Gamma_{2\phi}^m\equiv \Gamma_{2\phi},\quad m\in(1,N)\;,\label{eq:Gmoe}
\end{equation} Eq.~(\ref{eq:rho0alpha})  can be  simplified using the orthonormality and completeness of Floquet basis, $\sum_{m=1}^{N}U_{m\alpha
}^*(t)U_{m\beta }(t)=\delta_{\alpha,\beta}$. From here it immediately follows that all matrix elements $\tilde{\rho }_{\alpha 0}(t)$ for $\alpha=1:N$  undergo exponential decay independently from each other with the same rate $\Gamma _{2\phi}+\Gamma _1/2$, starting from the initial value $\tilde{\rho }_{\alpha 0}(0)=\langle \psi ^{\alpha }|r\rangle/2$ where $r$ is a site number for the excitation at $t=0$ (cf. (\ref{eq:Phi1})).   The expectation value of $\sigma_r^\dagger$ after $d$ cycles from Eq.~(\ref{eq:cmd0}) is
 \begin{equation}
\left\langle \sigma _r^+(d)\right\rangle =\frac{1}{2}e^{-\Gamma d}\hspace{2 mm}\sum _{\alpha =1}^N e^{-i
d \omega _{\alpha }}A_\alpha\;.\label{eq:c-uni}
\end{equation}
where we introduced  dimensionless damping  rate $\Gamma$ and
\begin{equation}
 \Gamma=   \left(\Gamma _{2\phi}+\frac{\Gamma _1}{2}\right)\tc,\quad A_\alpha=\left|\braket{r}{\psi^\alpha}\right|^2\;.\label{eq:def}
\end{equation}
Above the $\ket{\psi^\alpha}$ and $\omega_\alpha$ and, respectively, eigenstates and eigenvalues  of the cycle unitary (\ref{eq:eigUd}) and $\ket{r}$ is a site basis state.
In the main text, the expectation value of $\sigma^+_r$ at a given  site is constructed by measuring the Pauli operators $X_r$ and $Y_r$ through the relation $\left<\sigma^+\right> = \left<X\right> + i \left<Y\right>$.  The expression (\ref{eq:c-uni}) is obtained above is given in the main text.

We now proceed by showing that  for large qubit rings $N\gg1$ the expression (\ref{eq:c-uni}) is correct under much more relaxing conditions than (\ref{eq:Gmoe}). We  re-write the expression for $W_{\nu }^{\alpha }(q)$ (\ref{eq:Walpha}) in the following form 
 \begin{equation}
W_{\nu }^{\alpha }(q)=\sum _{\sigma ,\sigma '=\pm } \sum _{\gamma =1}^2 \Gamma _{\gamma }\left(q(\sigma  -\sigma')\right) \Lambda _{\gamma ,\sigma
, \sigma '}^{\nu ,\alpha }\left(q\right)\;,\label{eq:W1}
\end{equation}
 where
  \begin{equation}
\Gamma _{\gamma }(q)=\frac{1}{N}\sum _{l=1}^{N/2} \left(\Gamma _{2\phi}^{2(l-1)+\gamma}+\frac{\Gamma _1^{2(l-1)+\gamma}}{2}\right)e^{-i q l}\;,\quad, \gamma=1,2\label{eq:Gammag}
\end{equation}
 are Fourier transforms of the qubit dephasing rates $\Gamma _{2\phi}^{m}+\Gamma _1^{m}/2$ taken over odd ($\gamma=1$) and even ($\gamma=2$) sites only. The expression for  $\Lambda _{\gamma ,\sigma
, \sigma '}^{\nu ,\alpha }\left(q\right)$ is not given here, it follows immediately from comparing (\ref{eq:W1}) with the equations (\ref{eq:Wnu}) and (\ref{eq:Ugl})  in the previous section.

In general,  dephasing  rates may fluctuate form  site to site. 
However these fluctuations are statistically independent  for super-conducing qubits. 
Therefore the dominant terms in the sum (\ref{eq:W1}) correspond to $\sigma=\sigma'$ and the result depend on the dephasing rate averages  $\Gamma_1(0),\Gamma_2(0)$ taken  over odd and even sites, respectively. 
Therefore $W_{\nu }^{\alpha }(q)$ does not depened on the individual dephasing rates but rather on their averages over odd and even sites. 
From the central limit theorem 
\begin{equation}
    \Gamma_{1,2}(0)\simeq\Gamma +{\mathcal O}(N^{-1/2})
\end{equation}
 In this case the value of $W_{\nu }^{\alpha }(q)\sim\Gamma$ and we again arrive to the expression (\ref{eq:c-uni}).
Therefore for larger qubit rings $N\gg1$ the later is correct under the  condition that fluctuations on $\Gamma _{2\phi}^{m}+\Gamma _1^{m}/2$ are bounded in magnitude and statistically independent  from site to site.

\subsection{Estimation of quasi-energy spectrum from the data: scaling considerations} \label{sec:scaling} 
In the  protocol described in the main text we apply a family of  periodic circuits to an $N$-qubit array. The circuits have the same cycle unitary $\Uc$ but differ in the number of cycles $d=1,2,\ldots,D$. We estimate the circuit parameters given in Table \ref{eq:tab-param} by measuring the expectation value  $\left\langle \sigma _r^+(d)\right\rangle$ of the operator $ \sigma _r^+$ after each cycle $d$. We then fit the analytical model given in Eq.(\ref{eq:c-uni}) to the empirical values $ \sigma _r^{\rm data}(d)$. The fitting is done using the least mean square approach via the cost function 
\begin{equation}
    {\mathcal L}=\frac{1}{2}\sum_{d=1}^{D}M_d \Bigl\vert\left\langle \sigma _r^+(d)\right\rangle- \sigma _r^{\rm data}(d)\Bigr\vert^2\;.\label{eq:Lr}
\end{equation}
Here, $M_d$ is the number of times a  circuit of the depth $r$ is repeated to obtain the expectation value.
Using (\ref{eq:c-uni}) we get
\begin{equation}
    {\mathcal L}(\omega,A,\Gamma)=\frac{1}{2}\sum_{d=1}^{D}M_d \left|\frac{1}{2}e^{-\Gamma d}\hspace{2 mm}\sum _{\alpha =1}^N e^{-i
d \omega _{\alpha }}A_\alpha -\sigma _r^{\rm data}(d)\right |^2\;.\label{eq:Lr}
\end{equation}
Minimizing the cost function $ {\mathcal L}(\omega,A,\Gamma)$ with respect to the parameters $\omega_\alpha,\, A_\alpha,\,\Gamma$ we get their estimated values.  It is instructive to compute the inverse covariance matrix with respect to the quasi-energies
\begin{equation}
    K_{\alpha,\beta}\equiv {\rm Cov}[\omega_\alpha,\omega_\beta],\quad (K^{-1})_{\alpha,\beta}\propto\frac{\partial^2\mathcal{L}}{\partial\omega_\alpha\partial \omega_\beta}\;,\label{eq:K}
\end{equation}
where $\partial ^2\mathcal{L}/\partial \omega _{\alpha }\partial \omega _{\beta }$ is evaluated at the
 minimum of the cost function ${\mathcal L}$. Its expectation value equals
  \begin{equation}
\frac{\partial ^2\mathcal{L}}{\partial \omega _{\alpha }\omega _{\beta }}=\frac{\left| F_{\alpha }(r)\right| {}^2\left| F_{\beta }(r)\right| {}^2}{N^2}\sum
_{d=1}^D M_d d^2e^{-2\Gamma  d}\cos \left[d \left(\omega _{\beta }-\omega _{\alpha }\right) \right]\;\label{eq:Cov1}
\end{equation}\newline 
where we used the relation (\ref{eq:def}) for $A_\alpha$ 
\begin{equation*}
  A_\alpha=  |\left\langle r\left|\psi _{\alpha }\right\rangle\right|^2=\frac{2 |F_\alpha(r)|^2}{N}\;. 
\end{equation*}
and $F_\alpha(r)\simeq 2^{-1/2}$ is the  rescaled eigenfunction (cf. (\ref{eq:psi}), (\ref{eq:uupm})).

We assume that the number of measurements are the same for each circuit depth, $M_d=M$. In this case the contribution to the inverse covarience from the circuits of the same depth $d$ is $\propto d^2 e^{-2\Gamma d}$.  The $d^2$ dependence is a hallmark of Heisenberg scaling in quantum metrology. Eqs.~(\ref{eq:K}) and (\ref{eq:Cov1}) correspond to the Eq.(4) in the main text.

Performing the summation over $d$ we get the explicit form of  the  matrix elements $\partial ^2\mathcal{L}/\partial \omega _{\alpha }\partial \omega _{\beta }$. One can show that for 
\begin{equation}
    D\gg N\;,\label{eq:ND}
\end{equation}
the eigenvalues of the matrix $\partial ^2\mathcal{L}/\partial \omega _{\alpha }\partial \omega _{\beta }$ are well approximated by the its  diagonal elements that determine the variances of the  individual quasi-energies, $\sigma(\omega_\alpha)\propto (\partial^2 \mathcal{L}/\partial \omega _{\alpha }^2)^{-1}$. The condition (\ref{eq:ND})  has a clear physics meaning:  to resolve individual quasi-energies we need circuit depth $D$ greater than the inverse separation between their nearest values, $2\pi/|\omega_\alpha-\omega_{\alpha+1}|\sim N$.

 Under the above condition (\ref{eq:ND}) we obtain for the inverse variance of individual quasi-energy 
  \begin{equation}
\frac{1}{{\sigma}(\omega_\alpha)}\propto \frac{\partial^2 \mathcal{L}}{\partial \omega _{\alpha }^2}=M\frac{1-e^{-2 x}\left(1+2 x +2 x^2 \right)}{4 \Gamma ^3 N^2}\left| F_{\alpha }(r)\right|
{}^4,\quad x=\Gamma D\;.\label{eq:var}
\end{equation}\newline 
 \begin{figure}[H]
    \centering
    \includegraphics[width=4.0in]{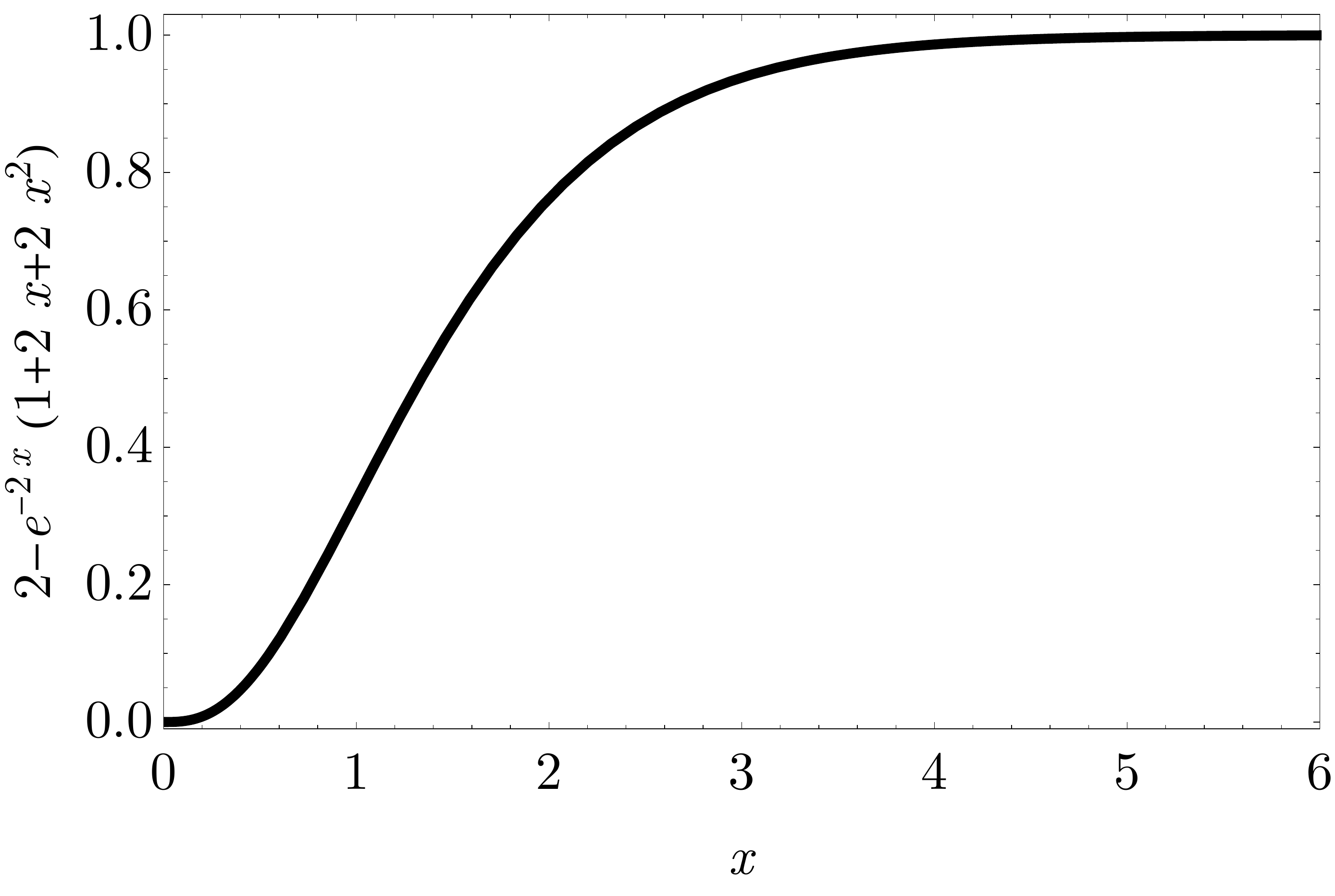}
    \caption{Plot of the coefficient in the expression for the diagonal elements of the inverse covarience  matrix (\ref{eq:var}).}
    \label{fig:var}
\end{figure}
In (\ref{eq:var}) we introduced quantity $x=\Gamma D$ that controls how much the variance of the estimated quasi-energies  scales with the maximum circuit depth $D$. As can be seen from Fig.~\ref{fig:var} the accuracy  improves with the maximum circuit depth  till $D \,\Gamma \leq 3$-$4$ and then it saturates. For a shorter circuit depth $D$  we obtain from Eq.~(\ref{eq:var}) 
\begin{equation}
    \frac{1}{{\sigma}(\omega_\alpha)}\propto M\frac{D^3}{3 N^2}\label{eq:var1}
\end{equation}
and the variance scales as $1/D^3$.   For the experimentally observed  damping rate $\Gamma\simeq 0.007$  the maximum circuit depth beyond which variance does not decrease is
$D <$500-600. Combining this with the  condition (\ref{eq:ND}) we arrive at the estimate for the number of qubits in a ring $N< 150$.

We now use (\ref{eq:var}) to investigate  how the total number of repetitions   depends on $N$ assuming that the value of the variance  $\sigma$  remains {\it fixed}.  The number of  circuits  with different
 depths (from 1 to $D$) grows linearly with $N$ and the total number of repetitions is $D M$.
According to the discussion  above (cf. (\ref{eq:ND})) we  set the circuit depth $D=c N$ with $c=$ 3-4. For $N\ll \Gamma^{-1}$ 
we use the Eq.~(\ref{eq:var1}) and obtain $M\propto N^2/D^3\propto N^{-1}$. The decrease of the required  number of repetitions with $N$ is due to the fact that the numerator in (\ref{eq:var1}) increases  as  $ D^3\propto N^3$ that more than compensates for the  factor $N^2$ in the denominator. 
Therefore the total required number of repetitions $D M$ does not change with $N$ at $N\ll \Gamma^{-1}$.
 
 On the other hand, for larger number of qubits $N> \Gamma^{-1}$ the necessary circuit depth $D$ is large so that  the expression (\ref{eq:var}) for the inverse variance  saturates at
\begin{equation}
    \frac{1}{{\sigma}(\omega_\alpha)}\propto \frac{M}{\Gamma^3 N^2}\;.\label{eq:var2}
\end{equation}
The factor $N^2$ in the denominator is due to the fact that the probability to find an excitation on a given site scales down as $1/N$ that leads to the increase in the number of repetitions at each circuit depth
 to reach the same accuracy $M\propto N^2$.  The total number of repetitions $D M$  grows rapidly as $N^3$.
 
  \begin{equation}
\frac{1}{{\sigma}(\omega_\alpha)}\propto \frac{\partial^2 \mathcal{L}}{\partial \omega _{\alpha }^2}=M\frac{1-e^{-2 x}\left(1+2 x +2 x^2 \right)}{4 \Gamma ^3 N^2}\left| F_{\alpha }(r)\right|
{}^4,\quad x=\Gamma D\;.\label{eq:var}
\end{equation}\newline 
 \begin{figure}[H]
    \centering
    \includegraphics[width=4.0in]{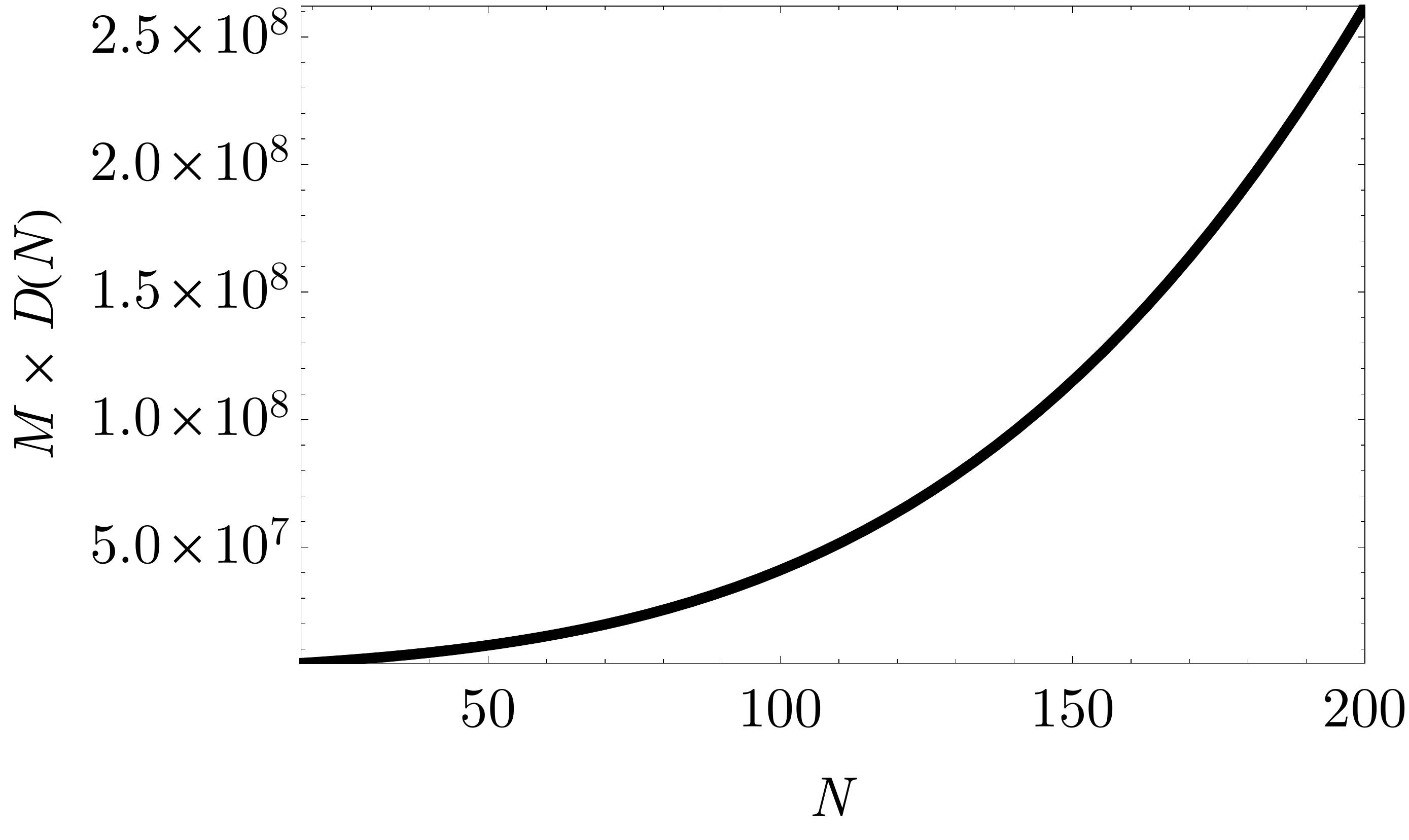}
    \caption{Plot of the total number of repetitions $M D(N)$ to reach a given accuracy from Eq.~(\ref{eq:var}) assuming that $D(N)=3N$ and  taking in account the experimental parameters described in the text.}
    \label{fig:var1}
\end{figure}

In Fig.~\ref{fig:var1} we plot the total number of repetitions $D M$ as a function of $N$ from Eq.~(\ref{eq:var}) by fixing the  value of $\sigma$.  We set $D=3N$ and use as a reference point 
our experimental parameters:   the  number of repetitions   per each circuit depth $M$=100000, maximum depth $d=80$, number of qubit $N=18$,    the accuracy of the quasi-energy estimation $\sigma^{1/2}\simeq 10^{-4}$ and $\Gamma=0.007$. One can see that at $N=100$ the total number of repetitions is 4$\times$10$^7$ while at $N=150$ it is 10$^8$.

In our experiments the set or 100000 repetitions of the circuit of the same length takes approximately 2-3 secs (this number is determined by the communication latency and not by the total circuits duration). The entire experiment takes around 3 minutes. Therefore  for $N=100$ the experiment is expected to
take about 15-20 mins and for $N=150$ between 40 to 60 mins. Those numbers are well within the stability range of our system.

\section{Periodic circuit on the open chain of qubits: Unitary Evolution }\label{sec:open_chain_unitary}
 The circuit on the open chain of qubits is obtained from the one defined on a qubit ring  in Eq.(\ref{eq:Ud}) by setting the gate $U_N$ applied between the qubits $N-1$ and $N$ to identity at each cycle. In follows from the discussion  in Sec.~\ref{sec:gauge} that  the gauge field $\phi=0$ in this case (cf. Eq.(\ref{eq:Ufr})) and therefore reduced cycle unitary $\Uf$ does not depend on the single qubit phases $\chi_j$. The  circuit unitary  can be written in the  form given  in Eq.~(\ref{eq:Uc2})
 \begin{equation*}
U_{\text{\rm cycle}}^d=\mathfrak{R}\,\mathfrak{U}^{d }\mathfrak{R}^{-1}\;,
\end{equation*}
where the reduced cycle unitary $\Uf$ equals
 \begin{equation}
\mathfrak{U}= G  R\;.\label{eq:Ufro}
\end{equation}
Here
\begin{equation}
G=\prod_{k=1}^{N/2}u_{2j} \, \prod_{k=1}^{N/2-1}u_{2j+1}\;,\label{eq:Gdef}
\end{equation}
(cf. (\ref{eq:VoVe})). The unitary $R$ is given by Eq.(\ref{eq:Rnu}) with phases $\nu_m$ given in (\ref{eq:nu}) except for  the   $\nu_1$ and $\nu_N$ given below 
\begin{equation}
\nu _1=-\zeta _1+ \gamma _1,\quad \nu _{N}=\zeta _{N-1}+\gamma _{N-1}
\end{equation}

In the site basis (\ref{eq:basis}) of a single excitation subspace  $R=\sum_{j=1}^{N}e^{i \nu_j}\ket{j}$ and  the  eigenproblem for the cycle unitary (\ref{eq:eigen})  has the matrix form
\begin{equation}
\sum_{l=1}^{N} G_{kl} \,e^{i \nu_l} \,\psi_{l}^{\alpha}=e^{-i\omega_\alpha}\psi_{k}^{\alpha},\quad \alpha,k=1:N\;,\label{eq:eigen1}
\end{equation}
  where nonzero elements of the $N\times N$ matrix $G_{kl}$ are 
  \begin{equation}
G_{1,1}=\cos \theta _1 ,\quad G_{1,2}= -i\sin  \theta _1\,\label{eq:G}
\end{equation}
\begin{equation*}
G_{2k,2k-1}=-i\sin  \theta _{2k-1}\cos  \theta _{2k},\quad G_{2k,2k}=\cos \theta _{2k-1}\cos  \theta _{2k},
\end{equation*} 
\begin{equation*}
G_{2k,2k+1}=-i
\sin  \theta _{2k}\cos \theta _{2k+1},\quad G_{2k,2k+2}=-\sin \theta _{2k}\sin \theta _{2k+1}
\end{equation*}
\begin{equation*}
G_{2k+1,2k-1}=-\sin\theta_{2k-1}\sin \theta _{2k},\quad G_{2k+1,2k}=-i \cos\theta_{2k-1}\sin \theta _{2k},
\end{equation*}
\begin{equation*}
G_{2k+1,2k+1}= \cos \theta _{2k+1}\cos \theta _{2k},\quad G_{2k+1,2k+2}=-i \sin  \theta _{2k+1}\cos  \theta _{2k}
\end{equation*}
\begin{equation*}
G_{N,N-1}=-i\sin \theta _{N-1},\quad G_{N, N}=\cos\theta_{N-1}
\end{equation*}


\subsection{Circuit with identical gates}\label{sec:chain_nominal}
Here we consider the particular example of the periodic circuit where all gates $u_j$ are identical
\begin{equation}
\zeta_j=\zeta,\quad \theta_j=\theta\;.\label{eq:uniform}
\end{equation}
The parameters $\nu_j$ are 
\begin{equation}
\nu_1=- \nu_N=\zeta,\quad \nu_{k \in (2,N-1)}=0\;.\label{eq:nu-u}
\end{equation}
One can show that for not too large values of $\zeta$
\begin{equation}
|\zeta | \leq \zeta_c(N)=\arccos\left(\frac{1}{\sqrt{2}}\frac{N-2}{N-1}\right)\;,\label{eq:bulk-cond}
\end{equation}
\begin{equation*}
\lim_{N\rightarrow \infty}\zeta_c(N)=\pi/4\;.
\end{equation*}
eigenstates of $\Uf$ form two branches of "bulk" eigenmodes each corresponding to the standing wave with momentum $q$.
Using spinor  notation for mode amplitudes on neighboring odd and even sites we can write 
 \begin{equation}
\begin{split}
\begin{gathered}
\left(
\begin{array}{c}
 \psi _{2m+2}^{\alpha } \\
 \psi _{2m+1}^{\alpha } \\
\end{array}
\right)=\frac{e^{ i m q_{\alpha }} A_+(\omega_\alpha,\zeta )+e^{ -i m q_{\alpha }}A_-(\omega_\alpha ),\zeta}{2} \psi _1^{\alpha },\quad m\in(1,N/2-2)\;,\label{eq:psi}
\end{gathered}
\end{split}
\end{equation}
 For the sites at the boundaries of the chain the mode amplitudes have the form
   \begin{equation}
\psi _2^{\alpha }=\frac{e^{i \zeta } \cos \theta -e^{-i \omega _{\alpha }}}{i\hspace{2 mm}\sin  \theta  } \psi _1^{\alpha }
\end{equation} \begin{equation}
\begin{split}
\begin{gathered}
\left(
\begin{array}{c}
 \psi _N^{(\alpha )} \\
 \psi _{N-1}^{(\alpha )} \\
\end{array}
\right)=b(\zeta )\frac{e^{ i (N/2-1) q_{\alpha }} A_+(\omega_\alpha,\zeta )+e^{ -i (N/2-1) q_{\alpha }}A_-(\omega_\alpha,\zeta)}{2}\psi _1^{\alpha }\hspace{3 mm}
\end{gathered}
\end{split}
\end{equation}
 Here 
 $A_\pm(\omega,\zeta)$ and $b(\zeta)$ are 2$\times$2 matrices of the form
   \begin{equation}
\begin{split}
\begin{gathered}
A_{\mp }(\omega ,\zeta )=\left(
\begin{array}{c}
 -i e^{i \zeta }+i \sqrt{2} e^{-i \omega }\pm \frac{i e^{i \zeta -\frac{i \omega }{2} }-\hspace{2 mm}i e^{i \zeta +\frac{i \omega }{2}} - i \sqrt{2} e^{-\frac{3
i \omega }{2}}}{\sqrt{2} \sqrt{\text{Cos}[\omega ]}} \\
 e^{i \zeta }\pm \frac{e^{i \zeta -\frac{i \omega }{2}}+e^{i \zeta +\frac{i \omega }{2}}-\sqrt{2} e^{-\frac{i \omega }{2}}}{\sqrt{2} \sqrt{\text{Cos}[\omega
]}} \\
\end{array}
\right)\;,\label{eq:Amp}
\end{gathered}
\end{split}
\end{equation}
  \begin{equation}
\begin{split}
\begin{gathered}
b(\zeta)=\left(
\begin{array}{cc}
 \hspace{1 mm}e^{i \zeta}\hspace{2 mm} & 0 \\
 0  & \hspace{1 mm}1 \\
\end{array}
\right)\;.\label{eq:b}
\end{gathered}
\end{split}
\end{equation}
 
 Two branches of the eigenmodes corresponds to the quasienergies $\omega _{\pm }(q)$ 
  \begin{equation}
\omega _{\pm }(q)=\pm  \omega (q)\;,\label{eq:mupm}
\end{equation}
\begin{equation}
\omega (q)=2\arcsin \left(\frac{\cos (q/2)}{\sqrt{2}}\right)\;.\label{eq:mu-open}
\end{equation}

The momentum $q$ is quantized taking $N/2$ distinct values that are roots  of the transcendental equation
 \begin{equation}
\tan \left(q\left(\frac{N}{2}-1\right)\right)+\tan \left(\frac{q}{2}\right)\left(1-\frac{1}{2-\sqrt{2} -\text{Cos}[q]}\right)=0\;.\label{eq:q-tran}
\end{equation}

\begin{figure}[H]
    \centering
    \includegraphics[width=\textwidth]{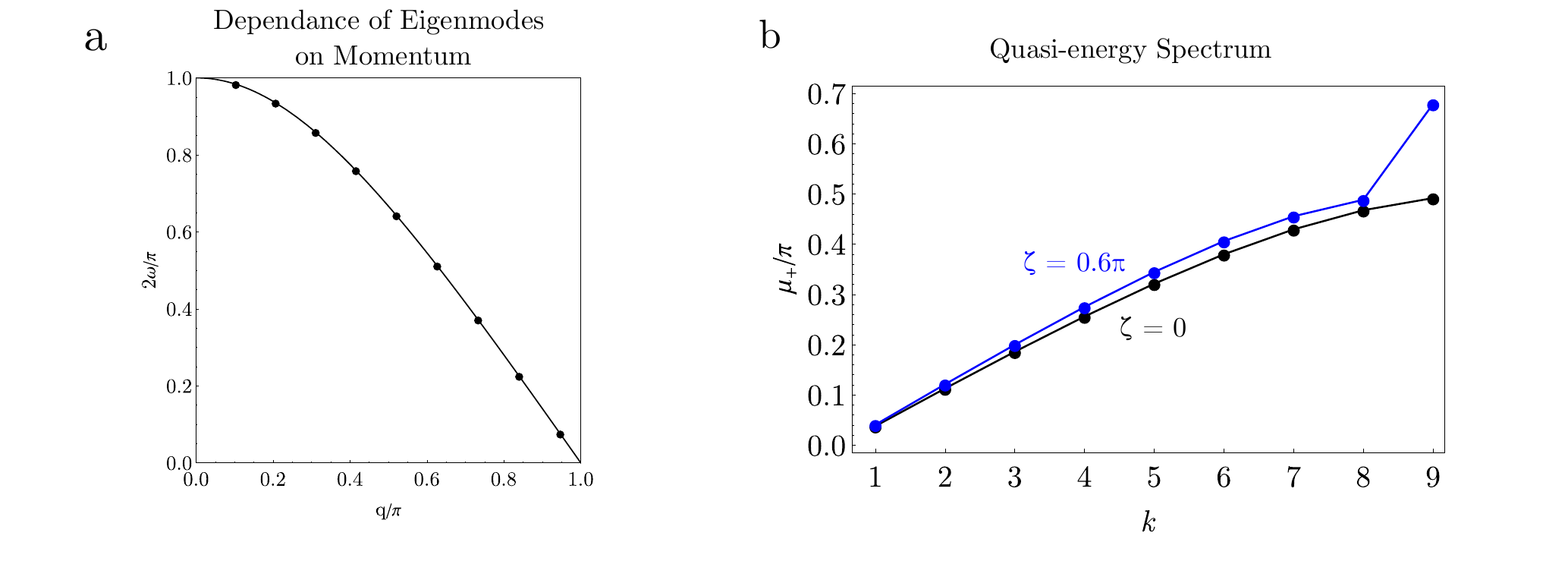}
    \caption{\textbf{Plots of the open chain spectrum} Uniform (reference) circuit eigenmodes and quasi-energy spectrum. (a) The relationship between $\Omega(q)$ and $q$. The solid line shows the continuous relationship between the two, while the points show the quantized values of momentum. (b) The plots of the sorted arrays of positive quasi-energies $\omega_+$ for $\zeta = 0$ and $\zeta = 0.6 \pi$. We see that in the latter case the largest quasi-energy splits from the bulk spectrum.}
    \label{fig:quasi_spectrum}
\end{figure}

This quantized relationship between $\omega$ and $q$ is shown in Fig.~\ref{fig:quasi_spectrum} a, and the full quasi energy spectrum for two values of $\zeta$ is shown in Fig.~\ref{fig:quasi_spectrum} b.

The function  $\Omega(q)$  is approximately quadratic for $q << 1$ 
\begin{equation}
    \Omega(q) \approxeq \frac{\pi}{2} - \frac{q^2}{4} 
\end{equation}
 \begin{equation*}
q\equiv q_k=\frac{\pi  k}{N/2-1},\quad k\ll N/2
\end{equation*}
For small values of $\Omega(q)$ it is approximately  linear 
\begin{equation}
    \Omega(q) \approxeq \frac{\pi - q}{\sqrt{2}}\ll 1
\end{equation}
 \begin{equation*}
q\equiv q_k=\pi -\frac{\pi  \left(N/2-\frac{1}{2}-k\right)}{N/2-1}
\end{equation*}

\begin{figure}
    \centering
    \includegraphics[width=1.0\textwidth]{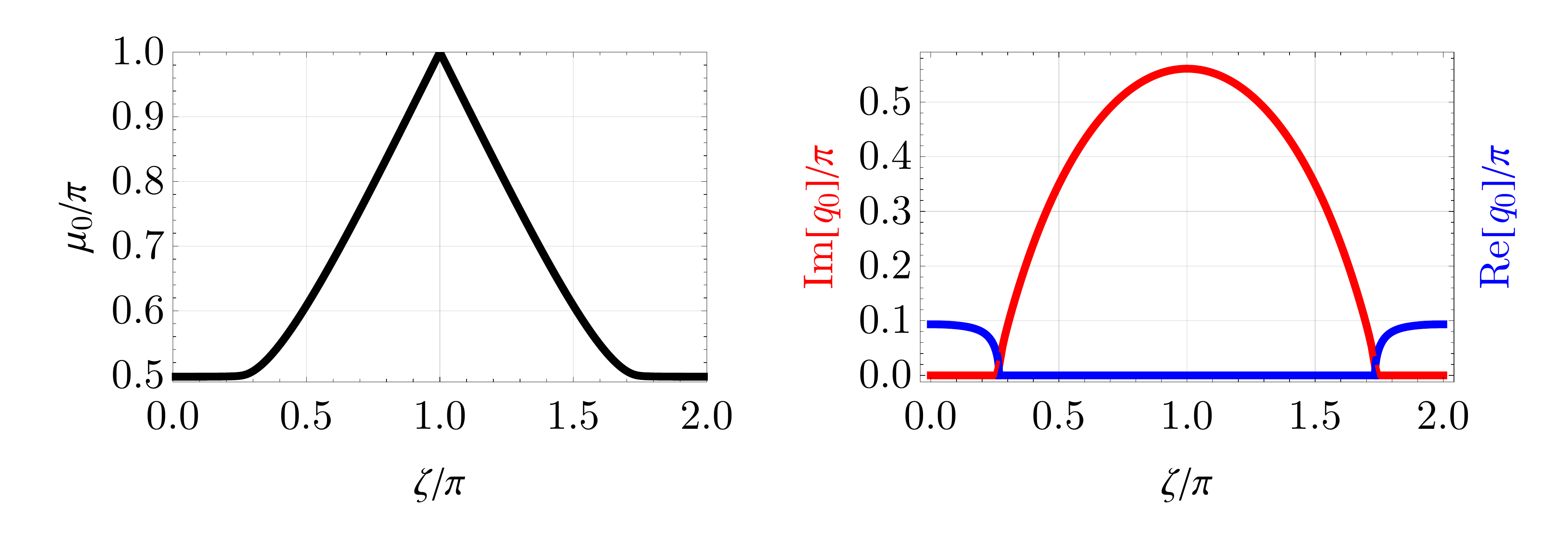}
    \caption{\textbf{Quasi-energy and quasi-momentum of the localized state {\textbf vs} {\boldmath$\zeta$}.} Right figure the plots of  real and imaginary parts of the quasi-momentum depicted with  blue and red colors respectively for the chain with $N$=10 qubits. The  localised state is formed at $\zeta=\zeta_C(N)\simeq \pi/4$. The left figure shows the dependence of the absolute value of the quasi-energies $\pm\omega_0$ of the localised states.}
    \label{fig:mu-q-loc}
\end{figure}

For sufficiently large values of $\zeta$ 
\begin{equation}
\pi\geq |\zeta | \geq \zeta_c(N)=\arccos\left(\frac{1}{\sqrt{2}}\frac{N-2}{N-1}\right)\;,\label{eq:loc-cond}
\end{equation}
 The pair of localised Floquet states are formed at the ends of the chain with quasi-energies $\pm\omega_0$, $\pi \leq \omega_0\leq \pi/2$. The dependence  of the quasi-energy and quasi-momentum of the localized state is depicted  in Fig.~\ref{fig:mu-q-loc}.
The localized state is formed from the delocalized state with zero momentum.
 Near the point  $|\zeta -  \zeta_c(N)|\ll 1$
 the quasi-momentum $q_0$ corresponding to the localized states equals
  \begin{equation}
q_0\approx c_N \left( \zeta _c(N)-\zeta \right)^{1/2}\;,\label{eq:q0} 
\end{equation}
 \begin{equation}
c_N=\frac{\sqrt{12}\left(N^2-2\right)^{1/4}}{(N (1+ N))^{1/2}}\;.\label{eq:cL}
\end{equation}
Fig.~\ref{fig:mu-q-loc} shows the quasi-momentum as a function of  $\zeta$ in the vicinity of the $\zeta_c$.
\begin{figure}
    \centering
    \includegraphics[width=0.8\textwidth]{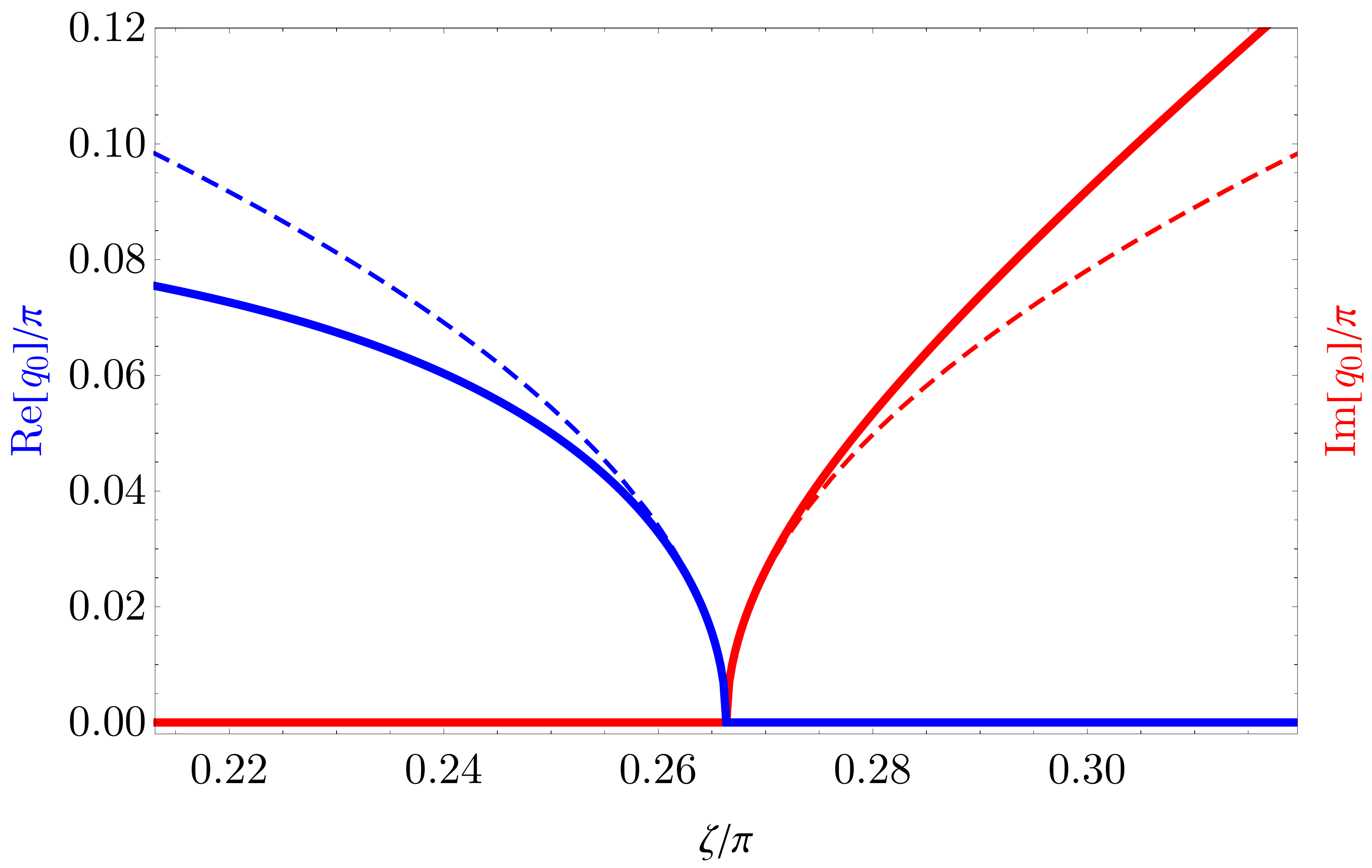}
    \caption{\textbf{Quasi-momentum vs   {$\boldsymbol\zeta$}.} Solid lines show the exact values of the real and imaginary parts of the quasi-momentum, dashed lines corresponds to the  approximate solution (\ref{eq:q0}).}
    \label{fig:q}
\end{figure}

 \subsection{\label{eq:measure-q-O}  Measuring the spectrum of quasi-energies and number of independent parameters}
 Another method to measure the spectrum of quasi-energies is to measure the two-time density-density correlator of spin excitations. In the case of a single excitation this corresponds we start with an excitation on one of the $N$ sites, $\rho (0) = \ket{k}\bra{k}$. We then let the state evolve for $d$ applications of the open chain control sequence, and then measure the probability the excitation has moved to site $m$, 

\begin{equation}
    p_{md} = \bra{m}\rho(d t_{\mathrm{cycle}}) \ket{m}, \quad \rho(0) =  \ket{k}\bra{k} \;.\label{eq:pmd}
\end{equation}
 The advantage of this method is that it does not require applying any additional gates other then the periodic circuit itself. Also  unlike the method described in Sec.~\ref{sec:measure-q} that requires microwave pulses the circuit $\Ucd$ conserves number of excitations and therefore post-selection is possible.
 
 The transition probability can be expressed in terms of the Floquet eigenstates and eigenvalues of the cycle unitary $\Uc$ (\ref{eq:eigen})   \begin{equation}\label{eq:pmks}
p_{md}=\left| \langle m|U_{\text{cycle}}^d|k\rangle \right| {}^2\equiv \left|\bra{m}\Ufd\ket{k}\right|^2 =\sum _{\alpha =1}^N \sum _{\beta =1}^N e^{i d\left(\omega _{\alpha }-\omega _{\beta
}\right)} \left\langle \left.\psi ^{\alpha }\right|m\right\rangle \left\langle m\left|\psi ^{\beta }\right.\right\rangle  \left\langle \left.\psi
^{\beta }\right|k\right\rangle \left\langle k\left|\psi ^{\alpha }\right.\right\rangle\;
\end{equation}
 Amongst the quasi-energy differences $\omega_\alpha-\omega_\beta$ there are only $N-1$ independent quantities that can be chosen, e.g.,  as $ \omega_{\alpha 1} = \omega_\alpha-\omega_1$ where $\{\omega_\alpha\}$ is a sorted array of quasi-energies. 
 
The equation (\ref{eq:Uc2})  $U_{\text{\rm cycle}}^d=\mathfrak{R}\,\mathfrak{U}^{d }\mathfrak{R}^{-1}$ express  the circuit unitary    $U_{\rm cycle}^{d}$ in  terms of the  canonical form of the cycle 
unitary $\Uf$  that has the same set of eigenvalues as $U_{\rm cycle}$.   It depends on the $N-1$ swap angles $\theta_j$, CZ phases $\phi_j$, and single qubit phases  $\nu_j$.  Therefore the quasi-energy differences $\omega_\alpha-\omega_1$ depend on $3N-3$ parameters as shown in Table \ref{eq:tab-param-O}. In a single excitation subspace  the CZ phases are not important  and the number of independent parameters is $2N-2$ (also no dependence
 angles $\chi_j$)
  \begingroup
\setlength{\tabcolsep}{10pt} 
\renewcommand{\arraystretch}{1.8} 
\begin{table}[h]
\begin{tabular}{|c|c|c|c|}
\hline
Swap angles  and CZ phases & Single-qubit  phases   &All parameters \\
 $\{\theta_j,\,\phi_j\}_{j=1}^{N}$ &  $\{\nu_j\}_{j=1}^{N} $ &\\
  \hline
  $2N-2$& $N-1$ &3$N-3$\\
 \hline
\end{tabular}
\caption{Parameters that determine the quasi-energy differences for a periodic circuit in an open chain of qubits}
  \end{table}
   \label{eq:tab-param-O}
  \endgroup

\section{Open Chain Parameter Estimation}\label{sec:fitting}

In the main text we estimate the unitary parameters of a ring  of $\sqrt{\text{iSWAP}}$ gates by fitting exponentially decaying oscillations to a time series of expectation values. Here we consider a similar problem, but on an open chain instead of a ring. Instead of fitting a generic model, we will very carefully fit to a secular approximation to the Lindblad equation (\ref{eqn:general_linbald}, \ref{eqn:dephase_photon}), which describes single qubit $T_1$ and $T_{2\phi}$ processes. This secular approximation is found by following a procedure similar to that outlined in section \ref{sec:Secular-O}, resulting in the following equations that approximately govern the open system dynamics in the Schrodinger picture, where $\rho_{\alpha \beta} = \bra{\psi^{\alpha}} \rho \ket{\psi^{\beta}}$:

\begin{equation}\label{eqn:secular_diag}
    \frac{d \rho _{\alpha \alpha} (t)}{d t} = \sum_{\beta=0}^N \rho_{\beta \beta}(t) W_{\beta \to \alpha} - \rho _{\alpha \alpha} (t) \sum_{\beta = 0}^N W_{\alpha \to \beta}
\end{equation}

\begin{equation}\label{eqn:secular_off_diag}
    \rho _{\alpha \beta} (d t_{\rm cycle}) = \rho _{\alpha \beta}(0) e^{-i \omega_{\alpha \beta} d} e^{-Y_{\alpha \beta} d t_{\rm cycle}}, \alpha \neq \beta
\end{equation}

Note that we have also assumed that the initial state is in the 1-excitation subspace. $W_{\beta \to \alpha}$ and $Y_{\alpha \beta}$ are time-independent transition and decay rates, and are given below. 

\begin{equation}\label{eqn:trans_rate}
     W_{\beta \to \alpha} =
    \begin{cases}
      \sum_{m=1}^N 2 \Gamma_{2\phi}^{m} \overline{|U_{m \alpha} (t)|^2 | U_{m \beta}(t)|^2}, & \for \alpha \neq \beta \neq 0 \\
       \sum_{m=1}^N \Gamma_1^m \overline{|U_{m \beta}(t) |^2}, & \for \alpha=0 \\
    \end{cases}
\end{equation}

\begin{equation}\label{eqn:decay_rate}
    Y_{\alpha \beta} = \sum_{m=1}^N  \left( \Gamma_{2\phi}^{m} + \frac{\Gamma_1^m}{2}\right) \left( \overline{|U_{m \alpha}(t)|^2} + \overline{|U_{m \beta}(t)|^2}\right) - 2 \Gamma_{2\phi}^m \overline{|U_{m \alpha}(t)|^2 |U_{m \beta}(t)|^2} \: ,
\end{equation}
where the overbar indicates averaging over one cycle. As can be seen in equation \ref{eqn:secular_diag}, the secular approximation solution yields a set of time-independent linear ODEs for the diagonal elements of the density matrix in the basis of the eigenvectors of the cycle unitary. This set of equations have a standard analytical solution that yeilds exponential decay according to the eigenvalues and eigenvectors of a matrix of transition rates constructed using the $W_{\beta \to \alpha}$. In section \ref{sec:drift} we will show that this Markovian model well represents the dynamics of the Sycamore device while performing this experiment on relevant timescales. 

Open chain unitary parameter estimation will be accomplished via the experiment described in section \ref{eq:measure-q-O}. We start with an excitation on one of the $N$ sites, $\rho (0) = \ket{k}\bra{k}$. We then let the state evolve for $d$ applications of the open chain control sequence, and then measure the probability the excitation has moved to site $m$, 

\begin{equation}
    p_{md} = \bra{m}\rho(d t_{\rm cycle}) \ket{m}, \quad \rho(0) =  \ket{k}\bra{k} \: .
\end{equation}

We can collect a matrix of experimental probabilities $p^{\text{data}}$ from the Sycamore device. We can then simulate this sequence using the secular approximation model and collect $p^{\text{model}}(\Vec{\theta})$, where $\Vec{\theta}$ are the model parameters, as will be outlined explicitly in section \ref{sec:open_params}. Estimating $\Vec{\theta}$ is therefore achieved by fitting $p^{\text{model}}(\Vec{\theta})$ to $p^{\text{data}}$. $p^{\text{model}}_{dk}$ in the secular approximation is given explicity below,

\begin{align}
    \begin{split}
        p^{\text{model}}_{dk} =&  \sum_{\alpha, \beta} \rho_{\alpha \beta} (d t_{\rm cycle}) \braket{k}{\psi^{\alpha}} \braket{\psi^{\beta}}{k} \\
        =& \sum_{\alpha} \rho_{\alpha \alpha} (d t_{\rm cycle}) |\braket{k}{\psi^{\alpha}}|^2 +  \sum_{\alpha \neq \beta} \rho _{\alpha \beta}(0) e^{-i \omega_{\alpha \beta} d} e^{-Y_{\alpha \beta} d t_{\rm cycle}}  \braket{k}{\psi^{\alpha}} \braket{\psi^{\beta}}{k} \\
    \end{split}\label{eqn:psim}
\end{align}

Drawing analogy to equation \ref{eq:cmd0}, we can see that within the range of applicability of the secular approximation fitting equation \ref{eqn:psim} to experimental data will allow for the quasi-energies to be extracted to the Heisenberg limit, and uncertainty will be limited by decoherence instead of shot noise. This will be shown explicitly in section \ref{sec:stat_anal}.

\begin{figure}
    \centering
    \includegraphics[width=0.4\textwidth]{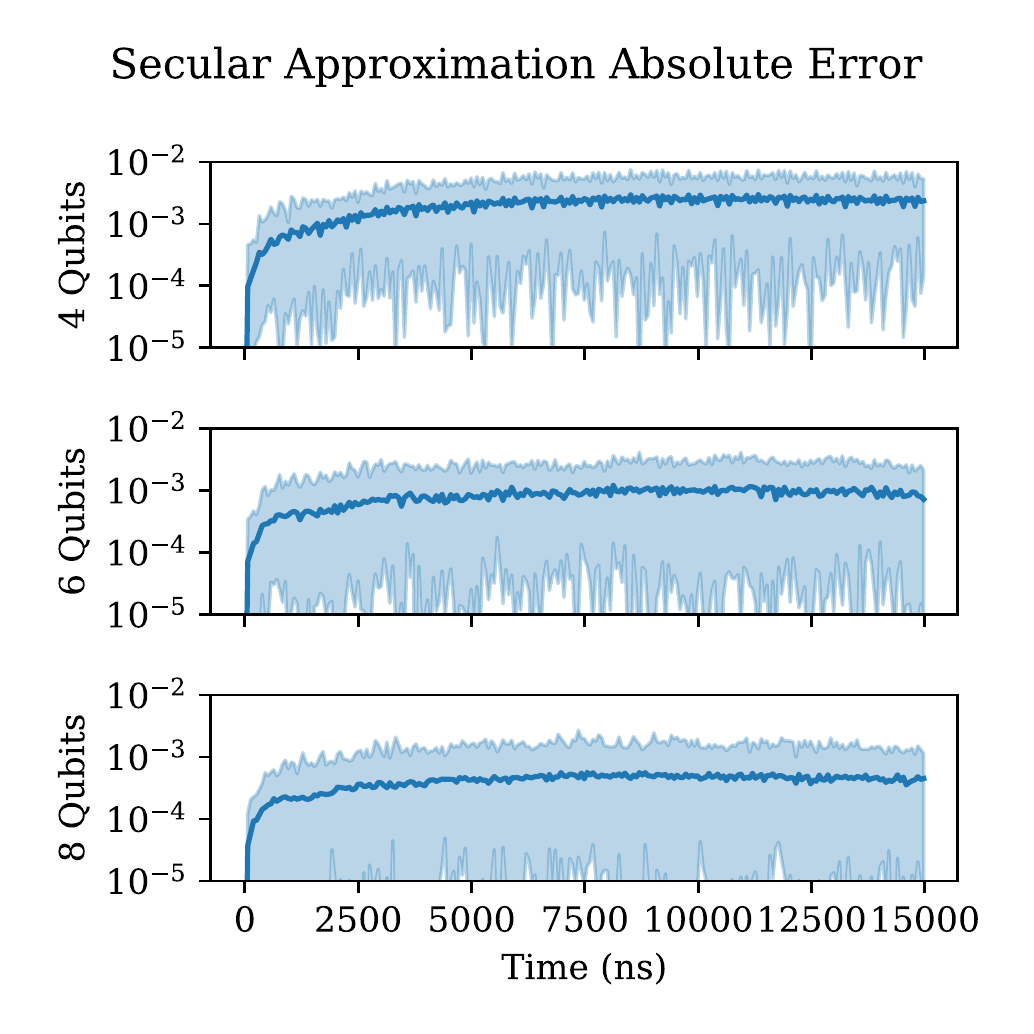}
    \caption{\textbf{Numerically validating the secular approximation} Absolute error $\epsilon = |\text{Secular} - \text{Full}|$ for $\sqrt{\mathrm{iSWAP}}$ chains on 4-8 qubits for various initializations. The shaded reigon indicates the maximum and minimum recorded error over all initializations, while the solid line indicates the mean error.}
    \label{fig:sec_error}
\end{figure}

Fig.~\ref{fig:sec_error} compares values of $p^{\text{model}}_{dk}$ computed using the secular approximation to those obtained via direct numerical simulation of the master equation for various initial conditions. The error introduced by the secular approximation for realistic values of $T_1$ and $T_{\phi}$ is on the $10^{-3}$ level, which makes it comparable to the expected deviation in experimental population measurements due to shot noise and other imperfections.  This suggests the secular approximation will be more than good enough in practice for the fitting task we wish to accomplish here.  

The secular approximation developed here is useful for fitting experimental data. The solution is simple to compute, with the added benefit that it relies only on simple mathematical primitives such as matrix inversion and eigen-decomposition, allowing one to access first and higher order derivatives of the density matrix elements with respect to the solution parameters for free via a TensorFlow \cite{abadi2016tensorflow} implementation. This is useful for both optimization and statistical analysis, as will be described in section \ref{sec:stat_anal}.

Experimentally, $\bra{k}\rho(d t_{\rm cycle}) \ket{k}$ is estimated by measuring bitstrings in the computational basis. All qubits are measured simultaneously, such that $p^{\text{data}}_{dk}$ is evaluated for all values of $k$ at a single value of $d$ simultaneously. Therefore, if there are $D$ total values of $d$, the data is collected in $D$ packages, each package containing on the order of 10000 bitstrings which can be collected in a few seconds. The exact time required  to collect these bitstrings depends on the length of the applied control sequence. 

An example of a fit to $p^{\text{data}}$ is given in Fig.~\ref{fig:var} (a), with the relationship between the standard deviation of the $N-1$ estimated ``fundamental" quasi-energy differences and total fitting depth given in Fig.~\ref{fig:var} (b). As in the case of the ring in the main text and two qubits in section \ref{sec:two_q_floquet}, the quasi-energy differences of the open chain are measured with Heisenberg scaling, with a minimum standard deviation of approximately $7 \times 10^{-5}$ being achieved before the uncertainty starts to saturate due to decoherence. 

\begin{figure}
    \centering
    \includegraphics[width=\textwidth]{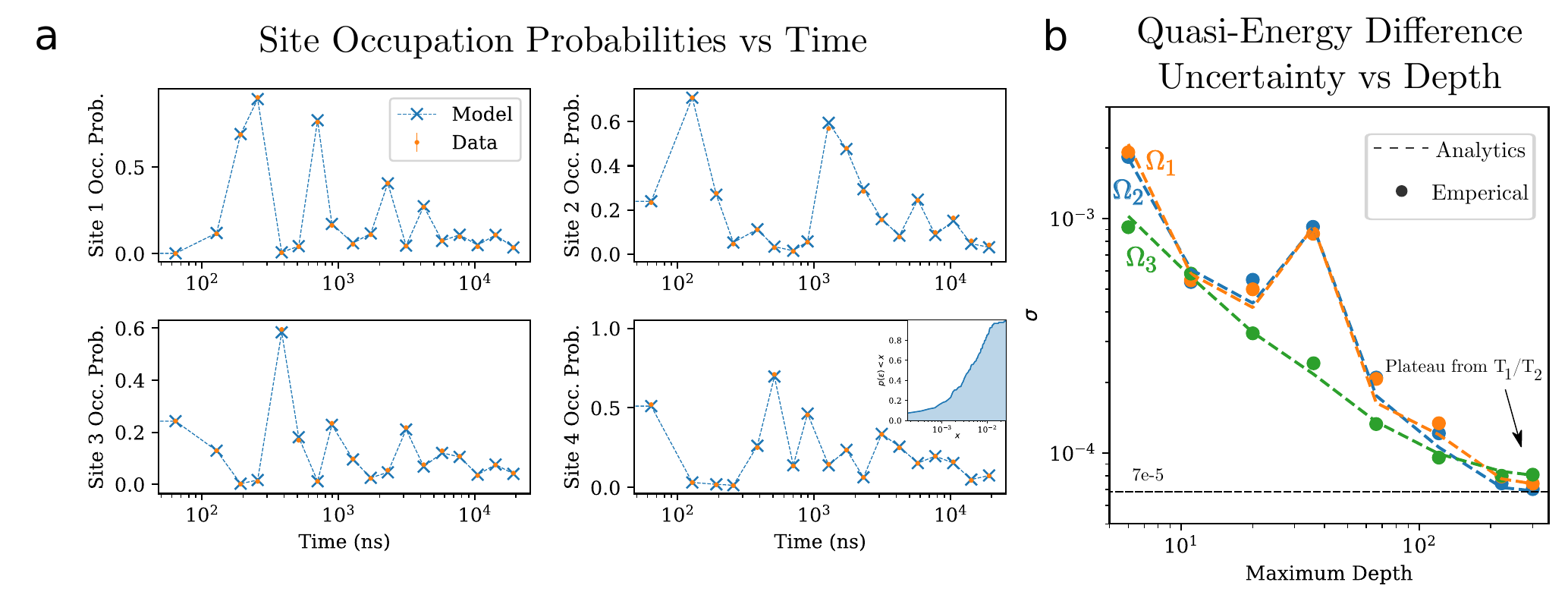}
    \caption{\textbf{Estimating open chain unitary parameters with Heisenberg scaling.} \textbf{a} Fit of a Lindblad equation open system model to log-spaced data. The nominal circuit is a 4-qubit $\sqrt{\mathrm{iSWAP}}$ chain, with all swap angles equal to $\pi/4$ and all single qubit phases equal to $0$, as described in section \ref{sec:chain_nominal}. One cycle has a real time duration of 64ns; each layer of two qubit gates is completed in 32ns. Error bars on the data points are $3 \sigma$ in each direction and were found via statistical bootstrapping of both the data and readout measurements. 50000 shots were taken per package. Data collection was completed over approximately 30 seconds of real time. \textbf{b}. Scaling of the uncertainty in the 3 fundamental quasi-energy differences with maximum number of cycles $D$. As expected, we achieve Heisenberg scaling out until a depth of approximately 300 and then start to observe the saturation predicted by equation \ref{eqn:sec_ll_hessian}. The calculation of the standard deviation was completed using both equation \ref{eqn:sec_ll_hessian} and statistical bootstrapping with 10000 re-samples. The two techniques agree well at all depths. $10^{-5}$ level standard deviations are observed for all 3 quasi-energy differences, with the lowest standard deviation observed being approximately $7 \times 10^{-5}$.}
    \label{fig:var}
\end{figure}

The following sections will detail the procedure used to extract the open chain quasi-energy energy differences to this level of precision.

\subsection{Open System Model Parameterization}\label{sec:open_params}

The minimum set of independent parameters $\vec{\theta}$ of the secular approximation model  must be identified to achieve a unique fit of the model to the data.

Through the quasi-energy difference dependence in equation \ref{eqn:secular_off_diag} it is clear that the secular approximation solution is dependant on all of the $2(N-1)$ parameters given in table \ref{eq:tab-param-O}.

At first glance it also seems that the solution would be sensitive to pulse shape parameters through the averages over pulses in equations \ref{eqn:trans_rate} and \ref{eqn:decay_rate}. However, in practice it was found that this dependence is extremely weak, and pulse shapes can fit using the nominal circuit unitary parameter values without being changed later in the fitting. Re-fitting pulse shapes to the corrected unitary parameters post-fitting typically only changes the population data on the $10^{-4}$ scale, which is an order of magnitude smaller than other known sources of error and therefore practically irrelevant. Therefore, the pulse-dependant terms can be considered to be fixed during fitting and do not add any free parameters to the model.

Equations \ref{eqn:trans_rate} and \ref{eqn:decay_rate} also contain $2N$ values of $\Gamma_1^m$ and $\Gamma_{2\phi}^{m}$. One may naively assume that they are all independent, but they generally are not. To see this, we analyze the sensitivity of the transition and decay rates to the vector $\Vec{x} = \begin{pmatrix}  \Vec{\Gamma}_1 \\ \Vec{\Gamma}_{2\phi} \end{pmatrix}$ via the least-squares problem

\begin{equation}\label{eqn:gamma_lstq}
    \mathcal{L} = \sum_{j \neq i} \left( Y_{ij} - \sum_{k=1}^{2n} A_{ij}^k x_k \right)^2 + \sum_{i \neq j} \left( W_{i \to j} - \sum_{k=1}^{2n} C_{ij}^k x_k \right)^2 \: ,
\end{equation}

where the coeffecient tensors $A_{ij}^k$ and $C_{ij}^k$ come from looking at structure of equations \ref{eqn:trans_rate} and \ref{eqn:decay_rate}. Differentiating with respect to each element in $x$ and setting the result equal to zero, we can form the system of equations $\Vec{L} = G \Vec{x}$, where $L$ is a length $2N$ vector with elements $L_l$ and $G$ is a $(2N, 2N)$ matrix with elements $G_{lk}$ given below.

\begin{equation}\label{eqn:sec_lstq_L}
    L_l = \sum_{j \neq i} Y_{ij} A_{ij}^l + \sum_{j \neq i} W_{i \to j} C_{ij}^l
\end{equation}

\begin{equation}\label{eqn:sec_lstq_G}
    G_{lk} = \sum_{j \neq i} A_{ij}^k A_{ij}^l + \sum_{j \neq i} C_{ij}^k C_{ij}^l
\end{equation}

G is a block matrix consisting of averages over pulses, 

\begin{align}\label{eqn:sec_g_block}
\begin{split}
    G =& \begin{pmatrix} \frac{1}{2} + \left( \frac{n}{2} + \frac{1}{2} \right)A & (n-3)A + 2B + 1 \\\left( (n-3)A + 2B + 1 \right)^T & 2 + (2n - 10) A + 4B + 4B^T + 8C   \end{pmatrix} \\
    A_{gh} =& \sum_{\alpha=0}^n \left(  \overline{|U_{h \alpha}(t)|^2} \: \overline{|U_{g \alpha}(t)|^2} \right)\\
    B_{gh} =& \sum_{i=0}^n \overline{| U_{g \alpha}(t)|^2} \: \overline{|U_{h \alpha}(t)|^4} \\
    C_{gh} =& \sum_{\alpha=0}^n \sum_{\beta \neq \alpha = 0}^n \overline{|U_{h \alpha} (t)|^2 |U_{h \beta}(t)|^2} \: \overline{| U_{g \alpha}(t)|^2 | U_{g \beta}(t)|^2} \: ,
\end{split}
\end{align}
for $g,h \in [1, N]$. G is a symmetric matrix, and therefore it's spectral decomposition, $G = \sum_{k=1}^{2N} \eta_k \Vec{g}_k \Vec{g}_k^T$ ,  describes the sensitivity of the secular approximation to different components of $\Vec{x}$. The linear combinations of $\Gamma_1$ and $\Gamma_{2\phi}$ values that the solution is sensitive to are given by the eigenvectors of G that are associated with nonzero eigenvalues. The rank of G (or more loosely, how many eigenvalues of G are a substantial fraction of the largest eigenvalue) tells us in total how many free parameters are associated with the incoherent evolution in the secular approximation. Evaluating this eigendecomposition numerically, for the 4 qubit $\sqrt{\mathrm{iSWAP}}$ nominal circuit we find that $G$ has rank 4 (but dimension 8), with the numerical values of the four nonzero eigenvalues and associated eigenvectors given below. The four other eigenvalues were at least 5 orders of magnitude smaller than the largest one, and are likely only nonzero due to numerical noise.

\begin{align}\label{eqn:4_q_gamma_eig}
\begin{split}
    \eta_k =& \begin{bmatrix} 13.21 & 0.94 & 0.71 & 0.14 \end{bmatrix}\\
    \Vec{g}_k =& \begin{bmatrix} \begin{pmatrix} 0.26 \\ 0.26 \\ 0.26 \\0.26 \\ 0.42 \\ 0.42 \\0.42\\0.42\end{pmatrix} & \begin{pmatrix}0.42 \\ 0.42\\0.42\\0.42\\-0.26\\-0.26\\-0.26\\-0.26\end{pmatrix}& \begin{pmatrix} 0.28 \\ -0.28\\-0.28\\0.28\\0.41\\-0.41\\-0.41\\-0.41\end{pmatrix}& \begin{pmatrix} -0.41 \\ 0.41\\0.41\\-0.41\\0.28\\-0.28\\-0.28\\0.28\end{pmatrix} \end{bmatrix} \\ 
\end{split}
\end{align}

There is significant structure in the spectrum of G. The largest 2 eigenvectors span the two dimensional space where $\Gamma_1^m = c_1$ and $\Gamma_{2\phi}^{m} = c_2$, which is very physically intuitive, as the overall strength of the incoherent effects should be the most important thing. The last two eigenvectors allow the edge and center values of $\Gamma_1^m$ and $\Gamma_{2\phi}^{m}$ to be different, which is an extremely interesting symmetry. 

In general, for $\sqrt{\mathrm{iSWAP}}$ chains on more than 4 qubits, the rank of G is still always exactly n. Changing the swap angles and single qubit phases away from nominal circuit values increases the rank of G. The spectrum of G should be examined numerically for every different circuit one wants to characterize.

In summary, for the nominal circuit open-system fitting problem on $N$ qubits, we will have $2(N-1)$ coherent parameters and $N$ incoherent parameters. The 4 qubit instance therefore has 10 total free parameters.

\subsection{Analysis of Drift}\label{sec:drift}

In this work, we generally rely on Markovian models to fit experimental data and extract unitary parameters. In the main text we fit an exponentially decaying oscillation, which is the correct form for a ring seeing a Markovian enviroment in the limit of large disorder, as justified in section \ref{sec:open_sys-ring}. In the case of the open chain, we directly fit a secular approximation to the Lindblad equation to data. The validity of these Markovian models when fitting to the Sycamore device must be studied, as one typically expects time-correlated noise to appear that would break this assumption.

\begin{figure}
    \centering
    \includegraphics[width=0.9\textwidth]{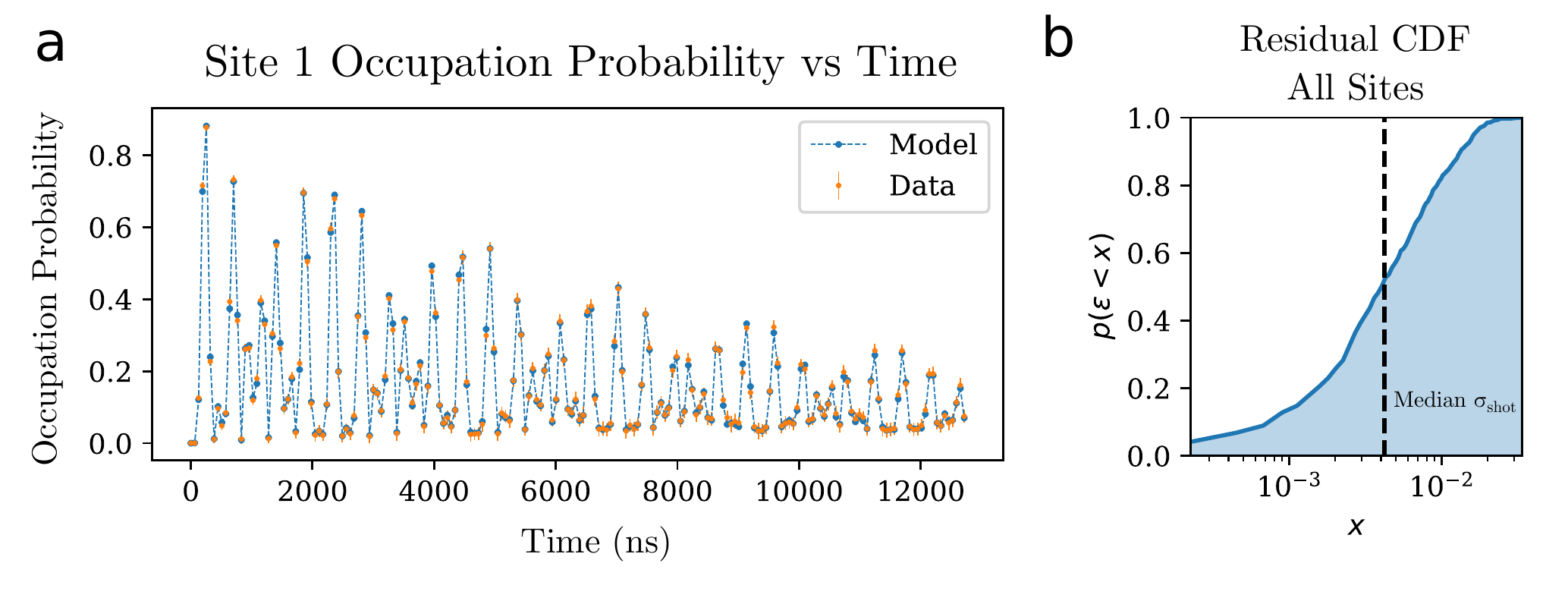}
    \caption{\textbf{Fitting the open system model to 4 qubit population data.} \textbf{a} Visualization of the fitting results for site 1 of 4. Data was collected at every depth between 0 and 200, meaning there are in total 800 elements in $p^{\text{data}}$ to fit the 10 parameter model to. 10000 shots were taken per data package.  Each data package was collected in a few seconds, implying the total time to take this data was 5 - 10 minutes. \textbf{b} Cumulative distribution of the fitting residuals for all of the 4 sites. We see that approximately $55\%$ of the residuals are smaller than the median value of the standard error of the data points. The error in the secular approximation is also generally on the $10^{-3}$ level, suggesting that this fit is practically perfect and that the Markovian model well represents the performance of this device during this experiment.}
    \label{fig:fit}
\end{figure}

There are two approaches one could take in studying this. The first is to analyze the fit of the Markovian model to the data in detail; if the fit is good there cannot be signifigant drift in the data, as the model has no way to adjust for time-variance. To study this, an experiment similar to that shown in Fig.~\ref{fig:var} is completed, with the data being taken with linear instead of logarithmic scaling to increase density and collection time. Results of fitting to this data are shown in Fig.~\ref{fig:fit}.

The fit is very high quality. Approximately $60\%$ of the residuals are at or below the level of shot noise, which happens to be of similar scale to the average error introduced by the secular approximation in the 4 qubit case, as shown in Fig.~\ref{fig:sec_error} b. The implication of this is that the Lindblad equation \ref{eqn:general_linbald} is an excellent model for the Sycamore device running this experiment. This is quite striking, as this implies that the environment truly appears Markovian to the device during this experiment, and that time-correlated noise sources are not effecting it on a scale we can resolve. 

Drift can also be analyzed on the timescale of individual data packages. Over a single package, the circuit parameters are nominally constant, and each bitstring in the timeseries should look like it is sampled from the same underlying distribution. Along these lines, rigorous techniques have been developed for the analysis of bitstring timeseries, such as the frequency domain hypothesis test developed in Ref.~\cite{proctor2020detecting}. However, these techniques are complicated to implement and ultimately provide much more information than we desire. Currently, we only want to test for the existence of drift, and we are not yet interested in the details of its frequency content or otherwise. As such, we propose a very basic protocol here that is capable of answering this simple question.

Consider the timeseries of $M$ bitstrings $y = [b_0, b_1, ..., b_M]$. This timeseries can be split into $k$ equal-sized chunks of size $m < M$. We can then ask the statistical question if all $k$ chunks appear to have been sampled from the same distribution. The chunk size m serves as a psuedo-frequency; using a smaller chunk size will allow us to look for drift at higher frequencies, however statistical uncertainty will be larger, lowering resolution. 

\begin{figure}
    \centering
    \includegraphics[width=\textwidth]{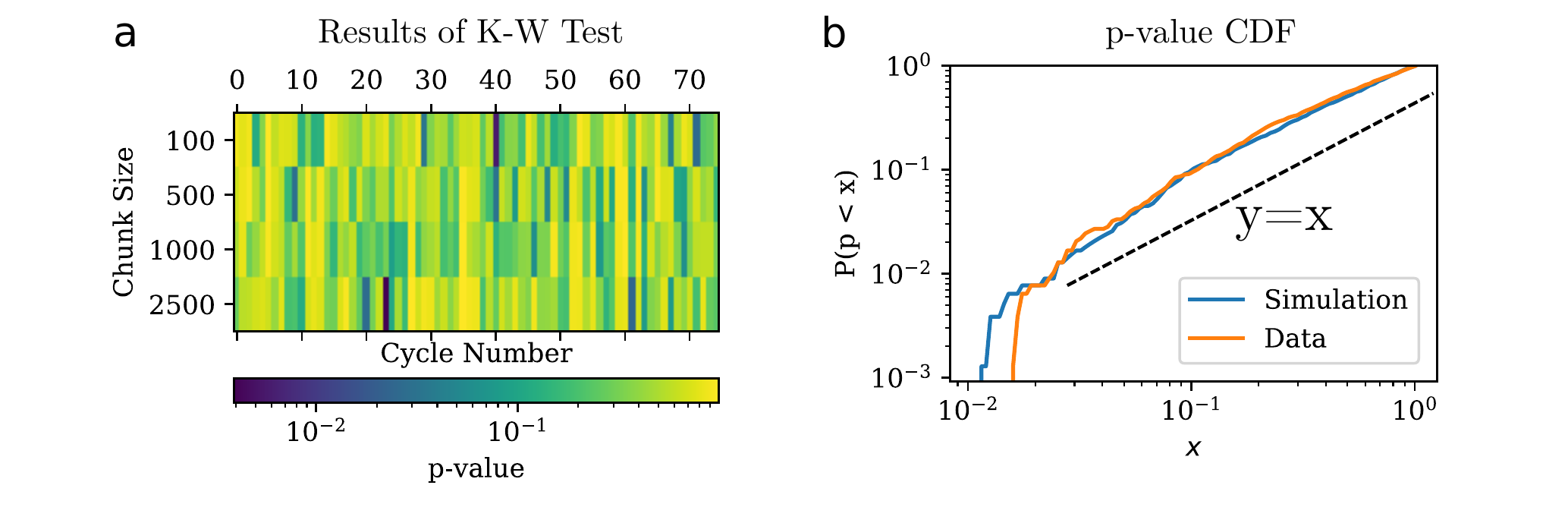}
    \caption{\textbf{Analysis of drift on the timescale of seconds.} \textbf{a}. Raw results of the Kruskal-Wallis test applied to experimental data. Results are shown out to depth 75, but the test was run on the full dataset out to depth 200. From this we see that no exceptionally small p-values are reported. \textbf{b} Plot of the cumulative distribution of the obtained p-values, for both the real experimental data set and a truly drift-less open system simulation. The test was completed for 200 depths and 4 different chunk sizes, so there are in total 800 points in the distribution. We see that the p-values generated using both the real and simulated datasets form a uniform distribution, which suggests that the null hypothesis is true and there is not evidence of drift in the data. The data and simulation distributions are also nearly coincident, which further supports this claim.}
    \label{fig:drift}
\end{figure}

There are several statistical tests that could be applied to this problem. A non-parameteric test is required, as the sampling distribution for a quantum computer does not at all resemble a normal distribution or other simple forms. The Kruskal-Wallis (KW) test \cite{kruskal} is an obvious candidate. The null hypothesis of this test is that all of the groups of samples come from distributions with the same median, which is a reasonable thing to look at when studying drift. The outcome of applying this test for various values of $m$ to the first 75 depths of the linearly spaced data show in Fig.~\ref{fig:fit} is shown in Fig.~\ref{fig:drift} a.

The way these p-values are distributed tells us if there is drift in the bitstrings or not. If the null is true and a statistical test has been used correctly (it's assumptions have been met), by the definition of a p-value one expects p-values to be uniformly distributed. As a test of this, we plot the distribution of p-values found by applying the previously described procedure to data generated by an open system simulation. The results of this are shown by the blue curve in Fig.~\ref{fig:drift} b. We see that the CDF of the p-values is nearly exactly linear, as expected. We then plot the distribution of p-values obtained from running the procedure on real data, as shown by the orange curve. we find that this distribution is also uniform and is almost indistinguishable from the distribution generated using simulated data (up to shot noise at very small p-values). This very strongly suggests that the null is true and that there is no statistically significant drift on the timescale of a single package. Combining this with the excellent fitting results shown in Fig.~\ref{fig:fit}, it seems that for one reason or another this experiment does not experience drift on the timescale of several minutes.

\subsection{Statistical Analysis of Parameter Estimates}\label{sec:stat_anal}

In any parameter estimation task, there are always two independent measures of success; bias and variance. Bias describes the difference in the mean value of an estimator and the actual parameter. It is desirable for an estimator $\hat{\theta}$ of the underlying parameter $\theta$ to be unbiased, eg $E(\hat{\theta} - \theta) = 0$. In practice, it is of course impossible to measure bias. Doing this would require knowledge of the true underlying distribution of $\theta$, which is generally what we are trying to measure in the first place. Bias can be avoided by choosing a model that is a good fit for the data being considered. The fit achieved with the open system model in Fig.~\ref{fig:fit} provides strong evidence that the parameter estimates achieved using this technique are unbiased on the $10^{-3}$ scale, as it it hard to imagine how the bias in parameter estimation could be substantially larger than the average residual in the fit of a minimal physical model.

The variance of a parameter estimate measures only statistical uncertainty and has nothing to do with the underlying distribution. As such, it can be rigorously analyzed and quantified. The variance of an estimator is given by $\sigma^2 = E\left((\hat{\theta} - \Bar{\hat{\theta}})^2\right)$. By the central limit theorem, the standard deviation $\sigma$ of a parameter estimate will typically scale with $1/\sqrt{n}$, where $n$ is the number of samples used in making the estimate. As mentioned in the main body and in previous work \cite{zhang2020}, the novelty of floquet calibration is that parameters are extracted at the Heisenberg limit; the scaling of the standard deviation with number of samples should exceed the central limit theorem and scale with $1/n$ instead of $1/\sqrt{n}$. This scaling will be carefully established for this open chain control sequence in the following work.

First, it is important to establish that we actually expect to be able to extract the quasi-energy differences, and not the parameter values themselves, with Heisenberg scaling. To see this, it is simplest to consider the populations that would result from the unitary evolution of some initial state $\ket{k}\bra{k}$, as in equation \ref{eq:pmks}. The populations are tied to the parameter values through explicit dependence on both the floquet eigenvectors $ \ket{\psi^{\alpha}}$ and the quasi-energy differences $\omega_{\alpha \beta}$. However, only the dependance on the quasi-energy differences scales with the depth of the circuit $d$. Therefore, the sensitivity of the populations to perturbations in the parameters only grows with depth though the quasi-energies. Noting that the entire quasi-energy difference matrix can be constructed from $N-1$ "fundamental" quasi-energy differences $\omega_{\beta 1}$, we should be able to learn $N-1$ linear combinations of the $2(N-1)$ unitary parameters with Heisenberg scaling from a single set of populations. This same argument can be made looking at the open system model presented in equation \ref{eqn:secular_diag}.

To solidify this argument we consider evaluating the variance of the parameter estimate though the inverse of the information matrix. Given a vector of parameters $\Vec{\theta}$, one may obtain the variance-covariance matrix of a maximum likelihood estimate of $\Vec{\theta}$ through the inverse of the hessian of the log likelihood function. The log-likelihood estimator for this problem is given by

\begin{equation}\label{eqn:ll_estimator}
    \mathcal{L}(\Vec{\theta}) = \sum_{d=0}^D N_{d} \sum_{k=0}^n p^{\text{data}}_{dk} \ln\left(  p^{\text{model}}_{dk}(\Vec{\theta})\right) \: ,
\end{equation}
with $N_d$ being the number of bitstrings collected at depth $d$ and the matrix $P_{dk}$ as defined in section \ref{sec:fitting}. $p^{\text{model}}_{dk}$ in the secular approximation is given explicitly

The variance-covariance matrix $\mathrm{var}(\Vec{\theta})$ is then given by 
\begin{align}\label{eqn:fisher_var}
    \begin{split}
        \mathrm{var}(\Vec{\theta})_{kj} = \left(-E\left(\pdv{\mathcal{L}(\Vec{\theta})}{\theta_j,\theta_k}\right)\right)^{-1} \: .
    \end{split}
\end{align}

In practice, it is easiest to simply evaluate this expression directly in TensorFlow to calculate the variance of the obtained parameter estimates. However, further analytical analysis is useful for gaining an understanding of the general behavior of this function. A simple expression can be developed by realizing that Eq.~\ref{eqn:psim} is only strongly coupled to the quasi-energy differences via the oscillating term $e^{-i \omega_{\alpha \beta} d}$; coupling through the transition and decay coeffecients is approximately $10^{-5}$ times weaker due to the $\Gamma$ prefactors and is therefore irrelevant when considering second derivatives.  Following this logic, one can arrive at the below expression for the Hessian of the log-likelihood function with respect to the fundamental quasi-energy differences in the open system case.

\begin{align}\label{eqn:sec_ll_hessian}
\begin{split}
    \pdv{\mathcal{L}(\Vec{\omega})}{\omega_{\beta 1}, \omega_{\alpha 1}} \approxeq & -4 \sum_{d=0}^D d^2 N_d \sum_{k=1}^n \frac{f(\alpha, \beta, k, d)}{p^{\text{model}}_{dk}}  e ^ {-d t_{\mathrm{cycle}} (Y_{\alpha 1} + Y_{\beta 1})}  \\
    \end{split}
\end{align}

Where $f(\alpha, \beta, k, d)$ is a function whose magnitude does not scale with d,

\begin{equation}\label{eqn:hessian_constant}
    f(\alpha, \beta, k, d)=  \mathrm{Im}\left( \bra{\psi^{\alpha}} \rho(0) \ket{\psi^1} \braket{k}{\psi^{\alpha}} \braket{\psi^1}{k} e^{-i \omega_{\alpha 1} d t_{\mathrm{cycle}}}\right) \mathrm{Im}\left( \bra{\psi^{\beta}} \rho(0) \ket{\psi^1} \braket{k}{\psi^{\beta}} \braket{\psi^1}{k} e^{-i \omega_{\beta 1} d t_{\mathrm{cycle}}}\right) \: .
\end{equation}

For data collected with sparsely sampled depths, equation \ref{eqn:sec_ll_hessian} scales with $D^2$ when the depth is sufficiently small such that $e ^ {-d t_{\rm cycle} (Y_{\alpha 1} + Y_{\beta 1})} \approxeq 1$. In this regime, the variance-covariance matrix will scale with $\frac{1}{D^2}$ and the standard deviations of the quasi energies will scale with $\frac{1}{D}$. This is the desired Heisenberg scaling. Beyond this regime, the variance will begin to increase due to incoherent effects. There is therefore an optimal depth that will allow for estimation of the parameters at the quantum limit, as observed in the two qubit case in section \ref{sec:two_q_floquet} and Ref. \cite{zhang2020}.

If one repeats the above analysis but takes the hessian with respect to the $2(N-1)$ unitary parameters directly, they will find that only $N-1$ of it's eigenvalues have the desired $D^2$ scaling. The eigenvectors associated with these eigenvalues represent the $N-1$ linear combinations of the $2(N-1)$ unitary parameters that can be learned from a single experiment with Heisenberg scaling. Alternatively, these linear combinations can be found more simply as the rows of the Jacobian of the fundamental quasi-energy differences with respect to the parameters.

\subsection{Readout Correction}

A readout correction procedure is applied to the bitstrings in each package to estimate the probabilities. As discussed in the main  text, this is not strictly necessary as the parameter information is encoded in the frequency domain. However, completing this process removes an imperfection from the open system model, reducing the number of experimental imperfections that need to be studied if the fit to the data is imperfect. We assume that readout error is a classical process, and it can therefore be inverted by solving the constrained quadratic problem given below.

\begin{equation}\label{eqn:readout_correct}
\begin{aligned}
    \mathcal{L} =& |A \Vec{v}^{prep} - \Vec{v}^{meas}|^2 \\
    v^{prep}_k > & 0 \\
    \sum_k v^{prep}_k =& 1\\
\end{aligned}
\end{equation}

$\Vec{v}^{prep}$ is a length $n+1$ vector containing the probabilities $\bra{m} \rho \ket{m}$ for $m \in [0, n]$. $\Vec{v}^{meas}$ is a sparse vector of computational basis measurement probabilities with at most $N$ entries given a package of $N$ bitstrings. $A$ is therefore a sparse matrix which can be measured efficiently by sequentially preparing the ground state and the single excitation states and measuring in the computational basis. Unlike most NISQ experiments, because we work in the 0-and-1 excitation subspace the dimension of $\Vec{v}^{prep}$ grows linearly instead of exponentially and the matrix $A$ can be measured without any assumptions about spatial correlation in the readout error. Therefore, this process will be optimal up to statistical error and quanutum effects in the readout, such as AC stark shifts. 

\newpage

\section{Band-structure demonstration with 30 qubits}\label{sec:thirty_qubit_data}

\begin{figure}[H]
\includegraphics{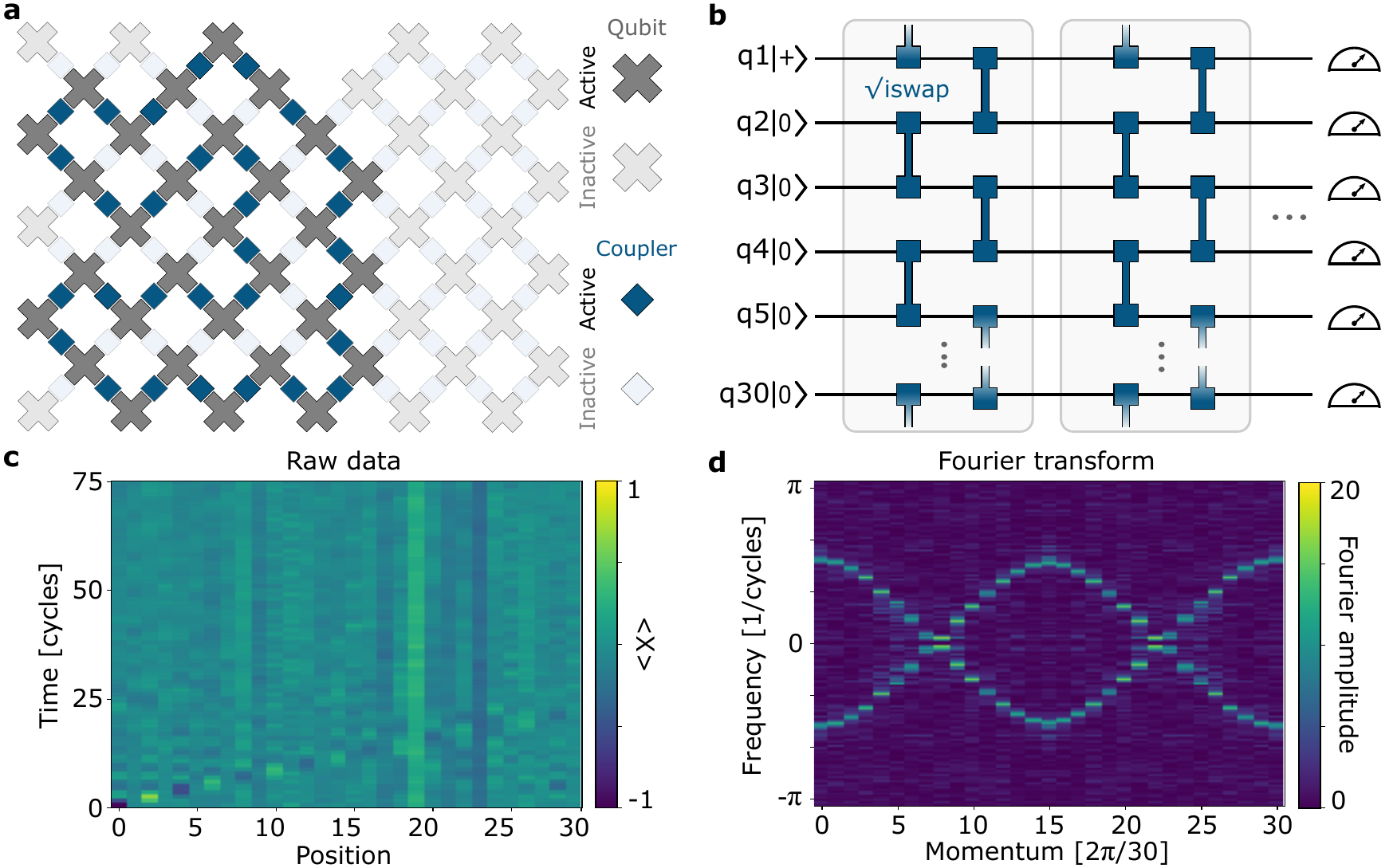}
\caption{\textbf{Measuring the single-particle band-structure with 30 qubits.  a}  Schematic of the 54-qubit Sycamore processor.   Qubits  are  shown  as  gray  crosses  and  tunable  couplers as blue squares.  Thirty of the qubits are isolated to form a one-dimensional ring. \textbf{b}  Schematic showing the control sequence used in this experiment. Each large vertical gray box indicates a cycle of evolution which we repeat many times.  Each cycle contains two sequential layers of $\surd{\text{i}}\text{SWAP}$ gates (blue).  \textbf{c} Raw data measured in this experiment.  Each qubit is measured in the x-basis and the data is plotted as a function of the position of the qubit along the ring and the number of cycles. \textbf{d} A Fourier-transform of the data is taken along both the x-axis and the y-axis revealing the single-particle band-structure.  By taking a Fourier-transform in position, we arrive at a signal with only a single frequency (each vertical column).  This greatly simplifies the data analysis over the method in the main text where all of the frequencies needed to be extracted from a single signal.}
\label{fig:30q_data}
\end{figure}


\end{document}